\documentclass[useAMS,usenatbib]{mn2e}
\usepackage{psfig}
\usepackage{epsfig}
\usepackage{longtable}
\usepackage{times}
\usepackage{lscape}
\usepackage{array}

\newbox\grsign \setbox\grsign=\hbox{$>$} \newdimen\grdimen
\grdimen=\ht\grsign
\newbox\simlessbox \newbox\simgreatbox \newbox\simpropbox
\setbox\simgreatbox=\hbox{\raise.5ex\hbox{$>$}\llap
     {\lower.5ex\hbox{$\sim$}}}\ht1=\grdimen\dp1=0pt
\setbox\simlessbox=\hbox{\raise.5ex\hbox{$<$}\llap
     {\lower.5ex\hbox{$\sim$}}}\ht2=\grdimen\dp2=0pt
\setbox\simpropbox=\hbox{\raise.5ex\hbox{$\propto$}\llap
     {\lower.5ex\hbox{$\sim$}}}\ht2=\grdimen\dp2=0pt

\def\mjup{\mbox{$M_\textrm{\tiny Jup}$}}
\def\msun{\mbox{$M_\odot$}}
\def\rsun{\mbox{$R_\odot$}}

\def\ga{\mathrel{\hbox{\rlap{\hbox{\lower4pt\hbox{$\sim$}}}{\raise2pt\hbox{$>$}}}}}\def\la{\mathrel{\hbox{\rlap{\hbox{\lower4pt\hbox{$\sim$}}}{\raise2pt\hbox{$<$}}}}}

\begin{document}

\title[White dwarfs in UKIDSS LAS DR8]{White dwarfs in the UKIDSS Large Area Survey: The Substellar Companion Fraction}

\author[P.\,R. Steele et~al. ]
{
P.\,R. Steele$^1$, M.\,R. Burleigh$^1$, P.\,D. Dobbie$^2$, R.\,F. Jameson$^1$ M.\,A. Barstow$^1$ \newauthor  and  R.\,P. Satterthwaite$^1$ \\
$^1$ Department of Physics and Astronomy, University of Leicester, University Rd., Leicester LE1 7RH, UK\\
$^2$ Anglo-Australian Observatory, PO Box 296, Epping, NSW 1710, Australia\\
}

\maketitle

\begin{abstract}
We present a near-infrared photometric search for unresolved substellar companions and debris disks around white dwarfs in the UKIRT Infrared Deep Sky Survey Large Area Survey. We cross-correlate the SDSS DR4 and McCook \& Sion catalogues of white dwarfs with the UKIDSS DR8 producing 3109 and 163 unique matches respectively. Cooling models are fitted to the optical photometry of a subsample of DA white dwarfs and extended to the near-infrared. A comparison is then made with the observed photometry to identify those stars with a near-infrared excess consistent with the presence of a cool companion or debris disk. Where present, we have estimated the approximate spectral type of any putative companion, or an upper limit on the temperature of a debris disk. In total we identify 14-16 new candidate white dwarf $+$ very low mass stellar systems, 9-11 candidate white dwarf $+$ brown dwarf systems, and 3 candidate white dwarf $+$ debris disks. We place lower limits on the unresolved ($<2\arcsec$) companions to all DA white dwarfs and thus assess the sensitivity of UKIDSS to such objects.  We use this result to estimate unresolved binary fractions of $f_{\rm WD+dL}\geq0.4\pm0.3$\%, $f_{\rm WD+dT}\geq0.2$\% and $f_{\rm WD+BD}\geq0.5\pm0.3$\%.

\end{abstract}


\section{Introduction}
\label{intro}
Near-infrared (NIR) surveys of white dwarfs enable the detection and study of late stellar and substellar companions, and circumstellar dust disks. A typical white dwarf is $10^3 - 10^4$ times fainter than its main sequence progenitor, significantly reducing the brightness contrast problem when searching for cool, low mass secondaries and dust. In addition, the spectral energy distributions of most white dwarfs (blue) and their low mass companions (red) are markedly different, facilitating easy separation of the components in broadband photometry and enabling straightforward spectroscopic follow-up (e.g.~\citealt{dobbie05}). Previous NIR photometric surveys of white dwarfs include those by \citet{probst83}, \citet{zuckerman87}, \citet*{green00}, \citet*{farihi05}, and the analysis of white dwarfs in the Two-Micron All Sky Survey (2MASS; \citealt{skrutskie06}) by \citet{wachter03}, \citet{wellhouse05}, \citet{holberg05}, \citet{tremblay07} and \citet{hoard07}. 

Many white dwarfs are known to have low mass M dwarf companions, and analysis of their population statistics  allows the investigation of binary formation and evolution. In particular, consideration of stellar evolution suggests the possible existence of two distinct populations: close systems in which the secondary has survived a phase of common envelope evolution and which may eventually lead to the formation of a cataclysmic variable (CV), and wide pairs where the secondary has migrated outwards in response to mass-loss from the white dwarf's progenitor \citep{farihi05b}. One particularly interesting case are the magnetic white dwarfs. Although hundreds of magnetic CVs (Polars) are known, no convincing progenitor system has been identified \citep{liebert05} despite exhaustive analysis of the 2MASS dataset for these objects \citep{wellhouse05}. 

Searches for substellar companions to white dwarfs additionally allow the investigation of the known deficit of brown dwarf companions to main sequence stars ($< 10$AU: $0.5\%$; \citealt{marcy00}) The closest brown dwarf $+$ white dwarf binaries might also represent either another channel for CV evolution or the end state of CV evolution, in which the secondary has become highly evolved through mass transfer. In close detached binaries, the brown dwarf is expected to be irradiated by the white dwarf's high UV flux, possibly leading to substantial temperature differences between the ``day'' and ``night'' hemispheres. Such systems can provide laboratories for testing models of irradiated "hot Jupiter" atmospheres (e.g.~HD\,189733b; \citealt{knutson07}). However, detached brown dwarf companions to white dwarfs are rare. \citet*{farihi05} suggest a binary fraction of $f_{\rm WD+dL}<0.5$\% for L dwarfs (We re-assess this limit later in this work).  Radial velocity and proper motion surveys, and searches for NIR excesses have so far found only five spectroscopically confirmed examples: GD\,165 (DA$+$L4, \citealt{becklin88}), GD\,1400 (DA$+$L$6-7$; \citealt{farihi04}; \citealt{dobbie05}), WD\,$0137-349$ (DA$+$L8;~\citealt{maxted06,burleigh06}), PHL\,5038 (DA$+$L8; \citealt{steele09}) and LSPM\,1459$+$0857 (DA$+$T4.5; \citealt{dayjones11}). GD\,165, PHL\,5038 and LSPM\,1459$+$0857 are wide orbit binaries with projected separations of 120\,AU, 55\,AU and 16,500-26,500\,AU respectively, whereas GD\,1400 and WD\,0173$-$349 are in much closer orbits with periods of 10 hours (Burleigh et al. 2011, in prep) and 116 mins respectively.

Infra-red observations have recently revealed the presence of metal-rich debris dust disks around a small number of white dwarfs (e.g.~\citealt{becklin05}; \citealt{kilic05}; \citealt{kilic06}; \citealt{kilic07}; \citealt{vonhippel07}; \citealt*{jura07}). Indeed, the large NIR excess in G\,29$-$38 was first identified twenty years ago by \citet{zuckerman87}, and {\it Spitzer} mid-infrared spectroscopy has now revealed that this dust disk is composed largely of silicates \citep{reach05}. The favoured explanation for the origin of this material is the tidal disruption of an asteroid that has wandered into the Roche radius of the white dwarf \citep{jura03}. With one notable exception, these disks have been found around relatively cool white dwarfs whose atmospheres are polluted with heavy elements. These metals should sink from the photospheres on a timescale of days, so the presence of close, orbiting dust disks provides a reservoir for on-going low level accretion. The exception is the central star of the Helix nebula. This hot, extremely young white dwarf has a mid-IR excess emission identified with {\it Spitzer} by \citet{su07}, which might be explained by the presence of a ring of colliding cometary material at an orbital distance of $\approx 30$\,AU. Therefore, dust disks around white dwarfs may have a variety of origins linked to the post-main sequence evolution of both the central stars and surrounding planetary systems, and large IR surveys will help establish how common they are. 

In this paper we present the first results from studying white dwarfs detected in the eigth data release (DR8) of the UKIRT Infrared Deep Sky Survey, the largest and deepest survey of the near-infrared sky yet undertaken. 

\section{The UKIDSS surveys}
\label{survey}

The UKIRT Infrared Deep Sky Survey (UKIDSS) is the NIR counterpart of the Sloan Digital Sky Survey (SDSS; \citealt{york00}) and is several magnitudes deeper than the previous 2MASS survey \citep{skrutskie06}, although unlike the latter UKIDSS will cover only $\sim20\%$ of the sky. UKIDSS actually consists of a set of five complementary wide/shallow, narrow/deep (extra-galactic), Galactic cluster and Galactic Plane surveys. A full description of these surveys, the science goals, observations and data processing and products are given in \citet{lawrence07}. Briefly, UKIDSS uses the 0.21\,deg$^2$ field of view Wide Field Camera (WFCAM; \citealt{casali07}) on the 3.8m UKIRT telescope on Mauna Kea, Hawaii. The UKIDSS surveys began in May 2005 and are expected to take 7~yr to complete. The survey of most relevance to this work is the Large Area Survey (LAS; PI: S.~Warren), which aims to cover $\sim4000$\,deg$^2$ of the Northern Sky coincident with the SDSS. The LAS makes observations in the $YJHK$ filters (see \citealt{hewett06} for a description of the UKIDSS photometric system) to a $5\sigma$ depth for point sources of $Y \approx 20.2$, $J \approx 19.6$, $H \approx 18.8$ and $K \approx 18.2$. At $K$ the LAS is 2.7~magnitudes deeper than 2MASS. UKIDSS data are being released to the community in stages and are public immediately to astronomers in all ESO states, before being made available to the rest of the world after 18 months. Uniformly processed images and catalogues are available through the WFCAM Science Archive\footnote{http://surveys.roe.ac.uk/wsa}. Following an the release of science verification data and an Early Data Release (EDR; \citealt{dye06}), UKIDSS made the First Data Release (DR1; \citealt{warren07}) in July 2006 and the Second Data Release (DR2; \citealt{warren07b}) in March 2007. A further 6 data releases haven subsequently been released (DR3-DR8) approximately once every 6 months. The most recent release, DR8, is inclusive of all previous data releases, and covers 2445~deg$^2$ in all four $YJHK$ filters, representing $\approx$64\% of the final survey.

The LAS observations are carried out in the $YJHK$ filters with the goal of reaching a uniform depth in all fields. The minimum schedulable blocks (MSBs) for the LAS use a pair of filters; either $YJ$ or $HK$. If an area has been covered by one pair an effort is made to follow up with the remaining pair within one month. However, conditions do not always allow for this and observations depend on a number of constraints (Dye et al. 2006). Hence, before completion a small number of observations in each data release will only have measurements in either the $YJ$ or $HK$ bands only. It is also entirely possible that any single measurement will not pass data quality tests and be omitted from the database. Therefore, a number of objects may have a selection of $Y$, $J$, $H$ or $K$ magnitudes, but not all four.

\section{Identification of White Dwarfs in the UKIDSS LAS DR8}

We have cross-correlated the UKIDSS LAS DR8 with two catalogues of white dwarfs: the on-line August 2006 version of the McCook \& Sion catalogue (hereafter MS99) of Spectroscopically Identified White Dwarfs\footnote{http://www.astronomy.villanova.edu/WDCatalog/index.html} (see \citealt{mccook99} for a fuller description) which contains 5557 stars, and the 9316 spectroscopically confirmed white dwarfs catalogued by  \citet{eisenstein06} (hereafter EIS06) from the SDSS Data Release 4. The MS99 catalogue includes the brightest white dwarfs known. The white dwarfs identified in the SDSS are all fairly hot ($T_{\rm eff}>8000-9000$\,K), and are constrained by the upper brightness limit ($g' \approx 15$) and lower sensitivity limit ($g' \approx 21$) of that survey. 

 Inevitably, there is considerable overlap between the catalogues. However, we note that the majority of SDSS white dwarfs are too faint to have been detected in the 2MASS survey (for example, \citealt{hoard07} excluded all objects $J \le 15.8$, $H \le 15.1$, $K_{\rm s} \le 14.3$). Therefore, UKIDSS allows us for the first time to investigate the NIR properties of the large number of new white dwarfs discovered in the SDSS, as well as significantly reducing the measurement errors on the NIR photometry of the brighter white dwarfs detected by 2MASS. 

Many of the relatively bright white dwarfs have large and unmeasured proper motions, meaning their astrometry is often unreliable. For the MS99 white dwarfs, we have made use of a set of improved co-ordinates kindly supplied by Jay Holberg (private communication) in our cross-correlation with UKIDSS DR8. The SDSS stars, being generally fainter and more distant, have small proper motions and generally reliable astrometry. We require all matches have an S/N $\ge 5$, corresponding to $J \ge 19.56$, $H \ge 18.81$ {\it and} $K \ge 18.23$ \citep{warren07b}. We performed the cross-correlation using the on-line ``CrossID'' tool at the WFCAM Science Archive. We used a pairing radius of $2\arcsec$, and took the nearest NIR source to the white dwarf's co-ordinates as being the best match. In general, the UKIDSS and white dwarf co-ordinates agree to better than $2\arcsec$ and we are confident that the correct match has been made. A few SDSS ``white dwarfs'' had very red UKIDSS colours: an inspection of the corresponding Sloan spectra revealed these to be quasar contaminants of the EIS06 catalogue. 

The cross correlation produced 3109 matches with the EIS06 catalogue, and 163 exclusive matches with the MS99 white dwarf catalogue. The corresponding spectral types for the white dwarfs (as assigned by EIS06 for the SDSS DR4 white dwarfs, and from literature for the MS99 white dwarfs) are listed in Table~\ref{spectypes}. It should be noted that the EIS06 catalogue is not a complete sample of white dwarfs, but the result of the placement of SDSS spectroscopic fibers according to complex and largely non-stellar considerations. Likewise, the MS99 catalogue is a heterogeneous sample. Thus, a number of white dwarfs outside of this sample may still be present in the area covered by the UKIDSS surveys.

\begin{table}
\caption{Number of each white dwarf spectral type detected in the cross correlation of UKIDSS LAS DR8.}
\begin{center}
\begin{tabular}{cc|cc|cc}
\hline
Spectral Type & No. & Spectral Type & No. & Spectral Type & No. \\
\hline\hline
DA      &  2306  & DO       & 10 & DBAZ     & 1 \\
DA$+$M  &   350  & DBZA     & 3  & DAO      & 1 \\
DC      &    94  & PG1159   & 5  & DBZ      & 1 \\
DB      &   153  & DZA      & 6  & DO$+$M   & 2 \\
DQ      &    31  & DBH      & 3  & DAZ      & 2 \\
DZ      &    37  & DBZ      & 4  & DZ$+$M   & 1 \\
DAH     &    19  & DAB      & 5  & DBA$+$M  & 1 \\
DBA     &    37  & DH       & 2  & DAH$+$DA & 1 \\
WD$+$M  &    13  & DAOB     & 2  & DAH$+$M  & 1 \\
DB$+$M  &     6  & DA CV    & 2  & DBO      & 1 \\
DC$+$M  &     5  & DZA$+$M  & 1  & DQ$+$M   & 1 \\ 
        &        &          &    & DZB      & 1 \\
\hline
\end{tabular}
\end{center}
\label{spectypes}
\end{table}

\section{Analysis}

\subsection{The Near-Infrared Two Colour Diagram}

\begin{figure}
\centering
\caption{$J-H$ v $H-K$ colour-colour diagram 
for the EIS06 white dwarfs (top) and the MS99 (bottom) white dwarfs with complete $JHK$ photometry present in the UKIDSS archive. The error bars have been omitted from the EIS06 diagram for the sake of clarity. }
\psfig{file=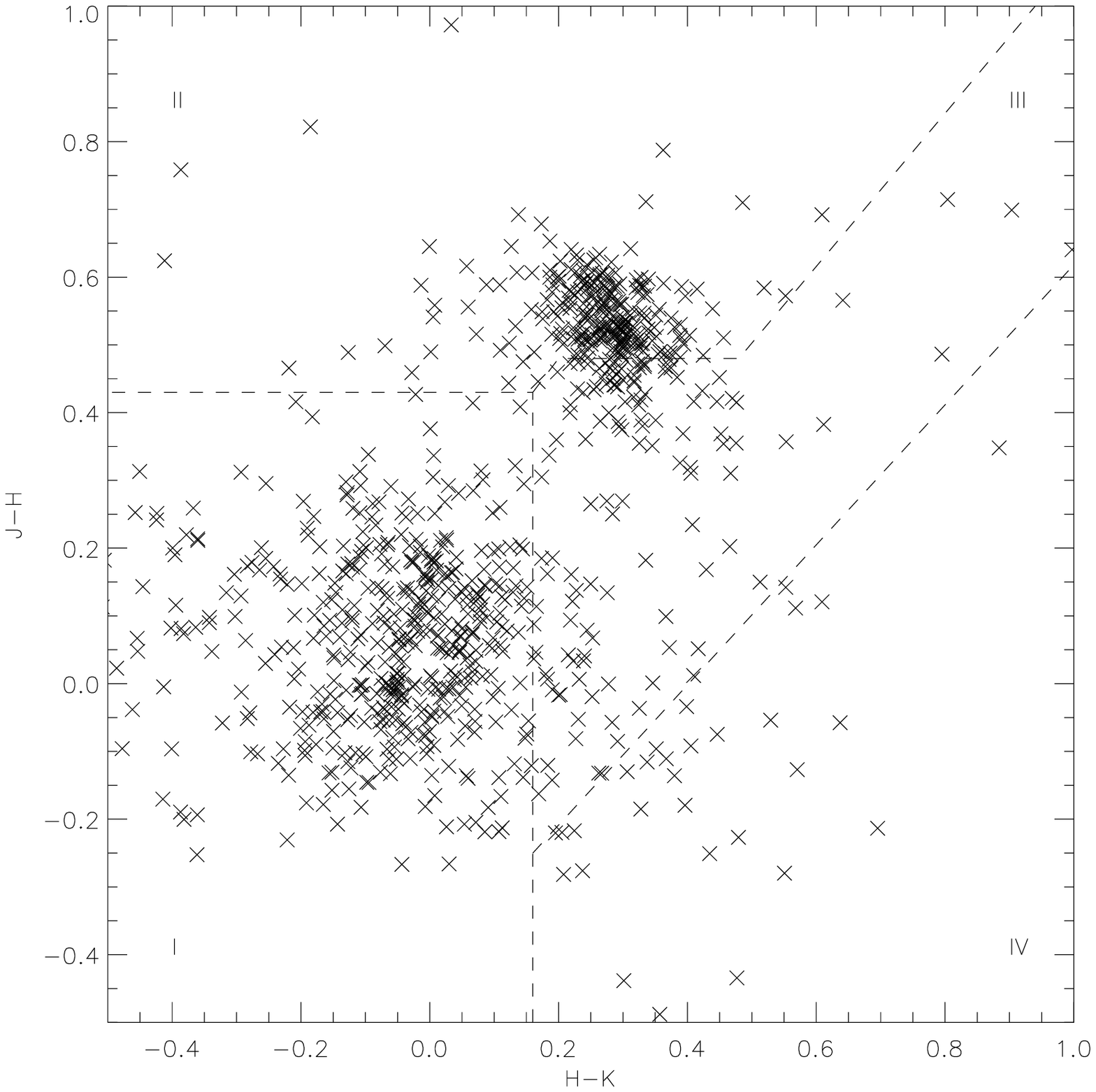,width=8.0cm}
\psfig{file=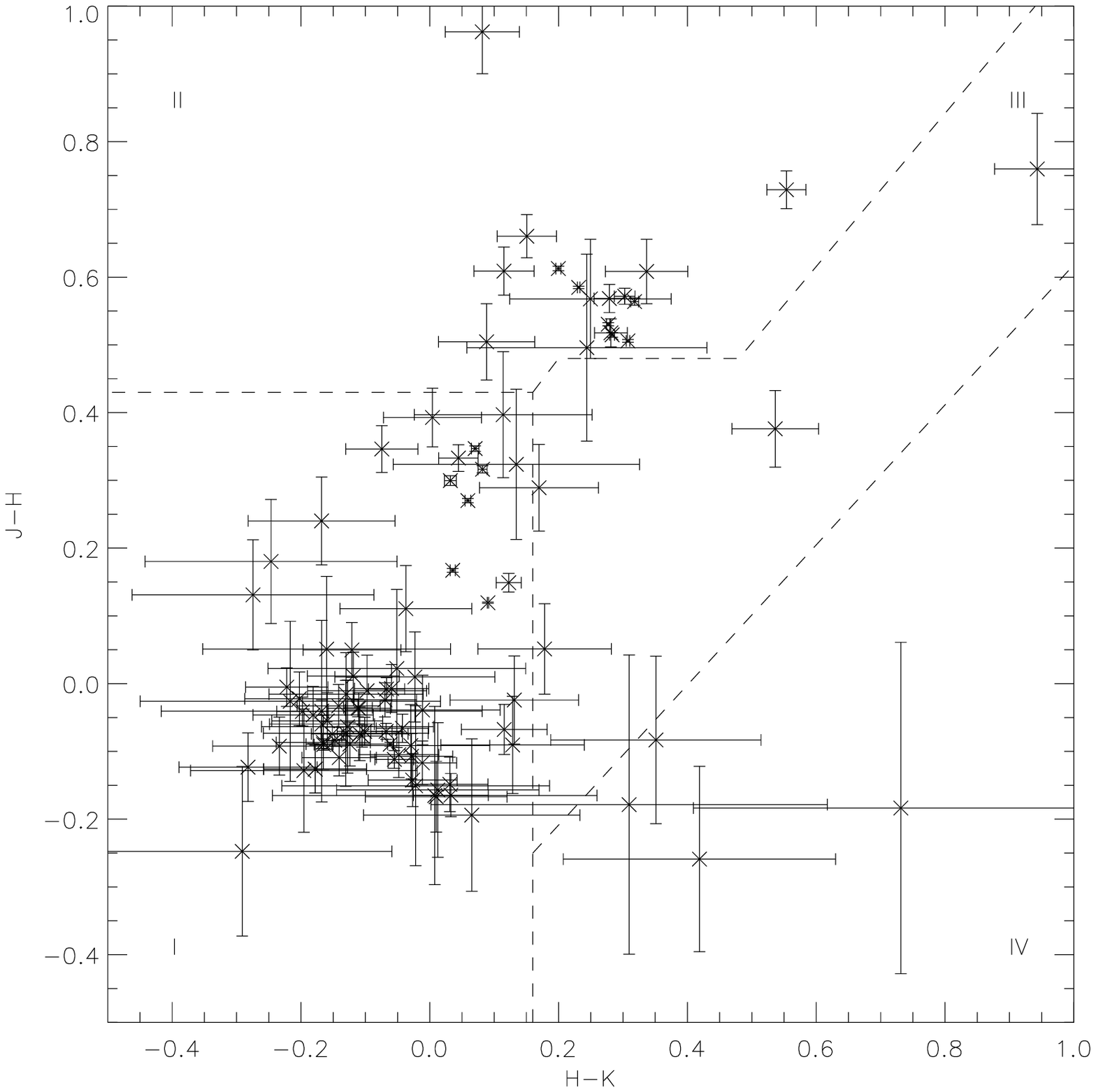,width=8.0cm}
\label{colour1}
\end{figure}

In order to make an immediate and rapid search for white dwarfs with a NIR excess, indicative of a low mass companion or dust disk, the UKIDSS DR8 white dwarfs were plotted on the ($J-H$,$H-K$) two-colour diagram shown in Figure~\ref{colour1}. This method was first applied by \citet{wachter03} in their analysis of white dwarfs detected by 2MASS. Then, based on Wachter et al.'s work and theoretical colour simulations, \citet{wellhouse05} further developed the technique by splitting the two-colour diagram into four distinct regions; (I) single white dwarfs, (II) white dwarfs with late-type main-sequence companions, (III) white dwarfs with lower mass, possible brown dwarf companions and (IV) white dwarfs contaminated with circumstellar material. Wellhouse et al.'s regions have been plotted in the two-colour diagram in Figure~\ref{colour1}.

\citet{hoard07} argue this colour selection method allows a simple, fast and efficient means to exploit large, homogeneous survey data to provide an initial list of candidate stars with an interesting NIR excess. Hoard et al. (2007) confirmed 80\% of the white dwarf $+$ main-sequence binaries reported by Wachter et al. (2003) with a false positive rate (i.e. stars rejected as misidentification by Wachter et al. 2003) of 20\%. However, they failed to identify any new candidate brown dwarf companions or dust disks. This is probably in part due to the excess emission from additional unresolved sources being small compared to the errors on the 2MASS data. However, it could also be due to some white dwarfs with companion brown dwarfs or dust disks occupying regions (I) and (II) of the two-colour diagram, making them indistinguishable from single white dwarfs or the large population of white dwarfs with main sequence K and M dwarf companions.

\begin{figure}
\begin{center}
\psfig{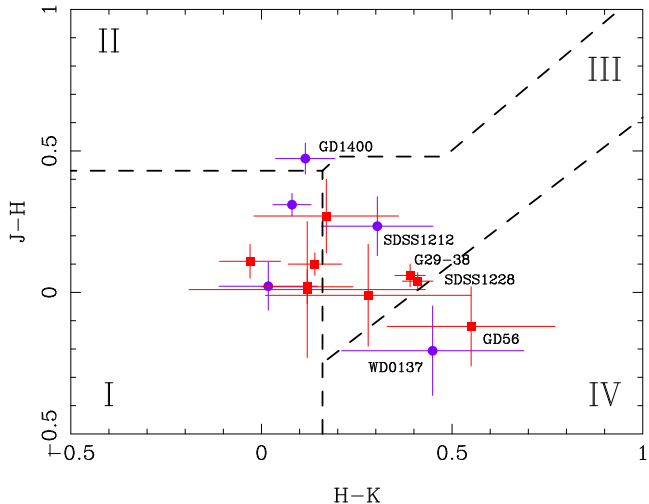}
\end{center}
\caption{NIR colour-colour diagram of white dwarfs with confirmed low-mass companions (circles) and circumstellar disks (squares). The data are given in Tables~\ref{knownbd} and \ref{knowndd}. The four regions defined by \citet{wachter03} and \citet{wellhouse05} and described in the text are shown for comparison  with Figure~1.}
\label{colour2}
\end{figure}

To investigate this further, white dwarfs with known brown dwarf companions and dust disks have been plotted in a NIR two-colour diagram in Figure~\ref{colour2} using photometry from 2MASS and other sources in literature (See Tables~\ref{knownbd} and \ref{knowndd}). Clearly only eight of the fourteen white dwarfs plotted lie within regions where further investigation would have been solicited using Wellhouse et al.'s (2003) regions. Indeed only a single dust disk (GD56) and a single white dwarf brown dwarf binary (SDSS\,J121209.31$+$013627.7) lie within the regions expected of such systems. In this case we can crudely conclude that the colour selection method is likely to easily distinguish $\sim50$\% of such systems. However, more examples would be required to accurately constrain such statistics.

\begin{table*}
\caption{Near infra-red magnitudes of known white dwarfs
  with unresolved brown dwarf companions}
\begin{center}
\begin{tabular}{lcccccl}
\hline
Name & $J$ & $H$ & $K$  & Source \\ 
\hline
\hline
WD\,$0137-349$  & $15.681 \pm 0.055$ & $15.887 \pm 0.148$ &  $15.438
\pm 0.188$ &   2MASS$^1$ \\
GD\,1400  & $14.923 \pm 0.032$ & $14.450 \pm 0.045$ & $14.335 \pm
0.064$ &   2MASS$^1$ \\
SDSS\,J$121209.31+013627.7$ & $17.91 \pm  0.05$ & $17.676 \pm 0.091$ & 
$17.372 \pm 0.114$ &   Farihi et al. (2008)\\
PG\,$1234+482$ & $14.977 \pm 0.043$ &  $14.955 \pm 0.073$ & $14.937
\pm 0.106$  &  2MASS$^1$ \\
PHL\,5038$^{2}$ & $16.75 \pm 0.02$ & $16.44 \pm 0.03$ & $16.36 \pm 0.04$  & Steele et a. (2009) \\
\hline
\end{tabular}
\end{center}
$^1$ Skrutskie et al. (2006) \\
$^{2}$ Partially resolved in UKIDSS
\label{knownbd}
\end{table*}

\begin{table*}
\caption{Near infra-red magnitudes of known white dwarfs
  with circumstellar debris disks}
\begin{center}
\begin{tabular}{lcccccl}
\hline
Name & $J$ & $H$ & $K$ & Source \\ 
\hline
\hline
G\,$29-38$  &     $13.132 \pm 0.029$ &  $13.075 \pm 0.022$ & $12.689 \pm
0.029$  &  2MASS$^1$\\
GD\,362     & $16.09 \pm  0.03$    & $15.99 \pm  0.03$ &   $15.85 \pm
0.06$    & Becklin et al. (2003) \\
GD\,56 &        $15.870 \pm 0.061$   & $15.991 \pm 0.129$ &  $15.440
\pm 0.175$ &   2MASS$^1$\\
WD\,$2115-560$ &   $14.110 \pm 0.029$ &  $13.996 \pm 0.055$ &  $14.022
\pm 0.061$ &   2MASS$^1$\\
GD\,40  &       $15.885 \pm 0.072$ &  $15.897 \pm 0.161$ &  $15.621
\pm 0.214$  &   2MASS$^1$\\
PG\,$1015+161$ & $16.131 \pm 0.085$ & $16.120 \pm 0.222$ & $16.003 \pm 0.216$ &    2MASS$^1$\\
GD\,133   &     $14.752 \pm 0.039$ &   $14.730 \pm 0.051$ & $14.611
\pm 0.105$ &   2MASS$^1$\\
GD\,16   &     $15.799 \pm 0.076$ &  $15.528 \pm 0.101$ &  $15.358 \pm
0.163$  &  2MASS$^1$\\
SDSS\,J$122859.93+104032.9$ &       $16.897 \pm 0.016$ &   $16.856 \pm
0.023$ &   $16.442 \pm 0.038$  &   UKIDSS DR5\\
\hline
\end{tabular}
\end{center}
$^1$ Skrutskie et al. (2006)
\label{knowndd}
\end{table*}

It should also be noted that this method can only be used to analyse white dwarfs which have complete $JHK$ photometry. In Section~\ref{survey} we noted that some stars will only have either $YJ$ or $HK$ photometry at this time in UKIDSS and so can not be included in this analysis. A massive 70\% of the white dwarfs in this sample are missing one of the required $J$, $H$ or $K$-band magnitudes (Table~\ref{magnumber}) needed for this type of analysis. 

Other limitations of the colour selection method are discussed by \citet{tremblay07}, who critically analysed  the 2MASS photometric data for white dwarfs and the analyses of Wachter et al. (2003) and Wellhouse et al. (2005). They concluded that the method is appropriate but should be interpreted with caution near the detection threshold where larger errors are to be expected. They demonstrated that there exists a number of single stars with large photometric uncertainties that contaminate regions (III) and (IV) of the colour-colour diagram. Indeed, It can be seen in Figure~\ref{colour1} that there are a number of stars with errors bars that extend from regions (III) and (IV) back into regions (I) or (II).

Tremblay \& Bergeron (2007) demonstrate that a more thorough technique is to compare observed NIR photometric fluxes with those predicted by model atmospheres (as also adopted by ~\citealt{holberg05}). This method first requires accurate effective temperatures and surface gravities, which are determined from high S/N optical spectroscopy. The model spectral energy distribution (SED) is then normalised to the observed $V$ or $g'$ magnitude of the white dwarf. Predicted  photometry can then be calculated by folding the synthetic SED through the appropriate filter transmission profiles and a direct comparison made. However, this technique is limited to those stars with known effective temperatures and surface gravities. Many of the MS99 white dwarfs do have accurately determined temperatures and gravities. The SDSS DR4 white dwarfs do have published atmospheric parameters (EIS06), however these are determined in an automated fashion and users of the catalogue are warned to be cautious when adopting these values.

\begin{table}
\caption{The combinations of UKIDSS data available for the entire white dwarf sample}
\begin{center}
\begin{tabular}{cccc}
\hline
Bands Available & No. & Bands Available & No.\\
\hline\hline
$YJHK$ & 926 & $JH$   &  28 \\
$YJH$  & 477 & $JK$   &   1 \\
$YHK$  &  28 & $HK$   &  67 \\
$YJK$  &  43 & $Y$    & 708 \\ 
$JHK$  &  56 & $J$    &  81 \\
$YJ$   & 790 & $H$    &  29 \\
$YH$   &  22 & $K$    &   4  \\
$JK$   &   2 &        &      \\
\hline
\end{tabular}
\end{center}
\label{magnumber}
\end{table}

\begin{figure}
\begin{center}
\psfig{file=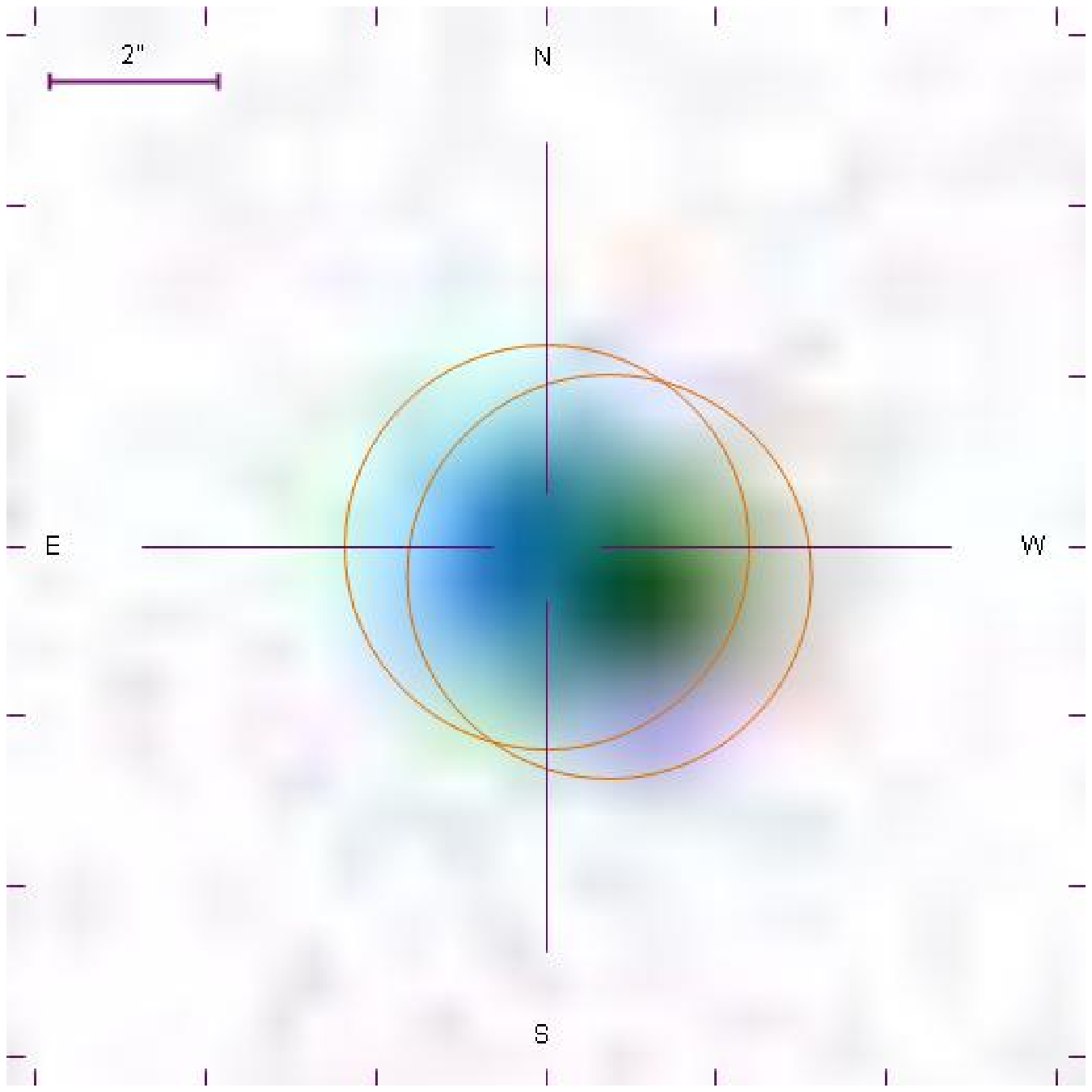,height=5.0cm,angle=0}
\end{center}
\caption{SDSS finding chart for SDSS\,J020538.12$+$005835.3 clearly showing both system components (circled). The colours have been inverted from the original finding chart.}
\label{020538_spec}
\begin{center}
\psfig{file=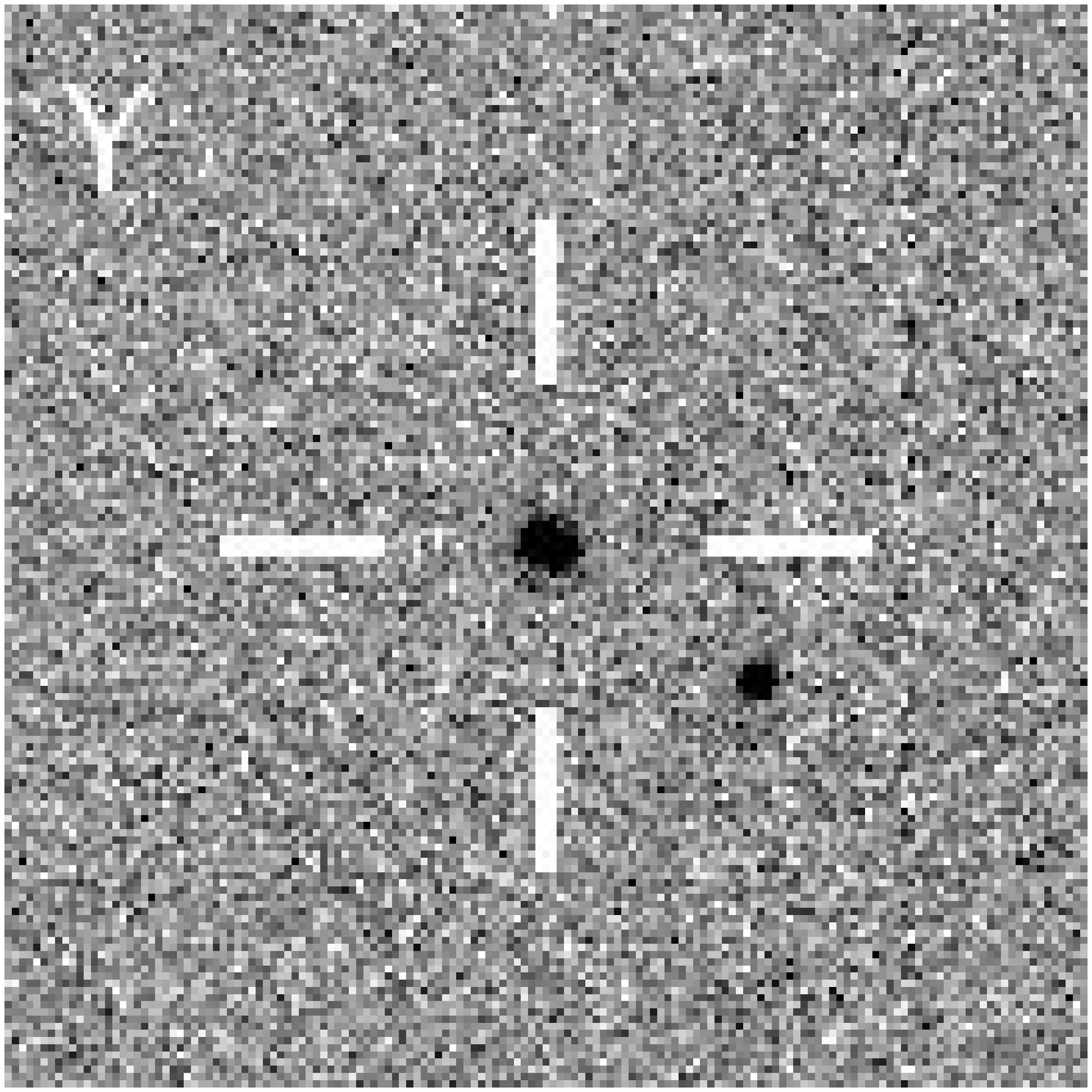,height=2.8cm,angle=0}
\psfig{file=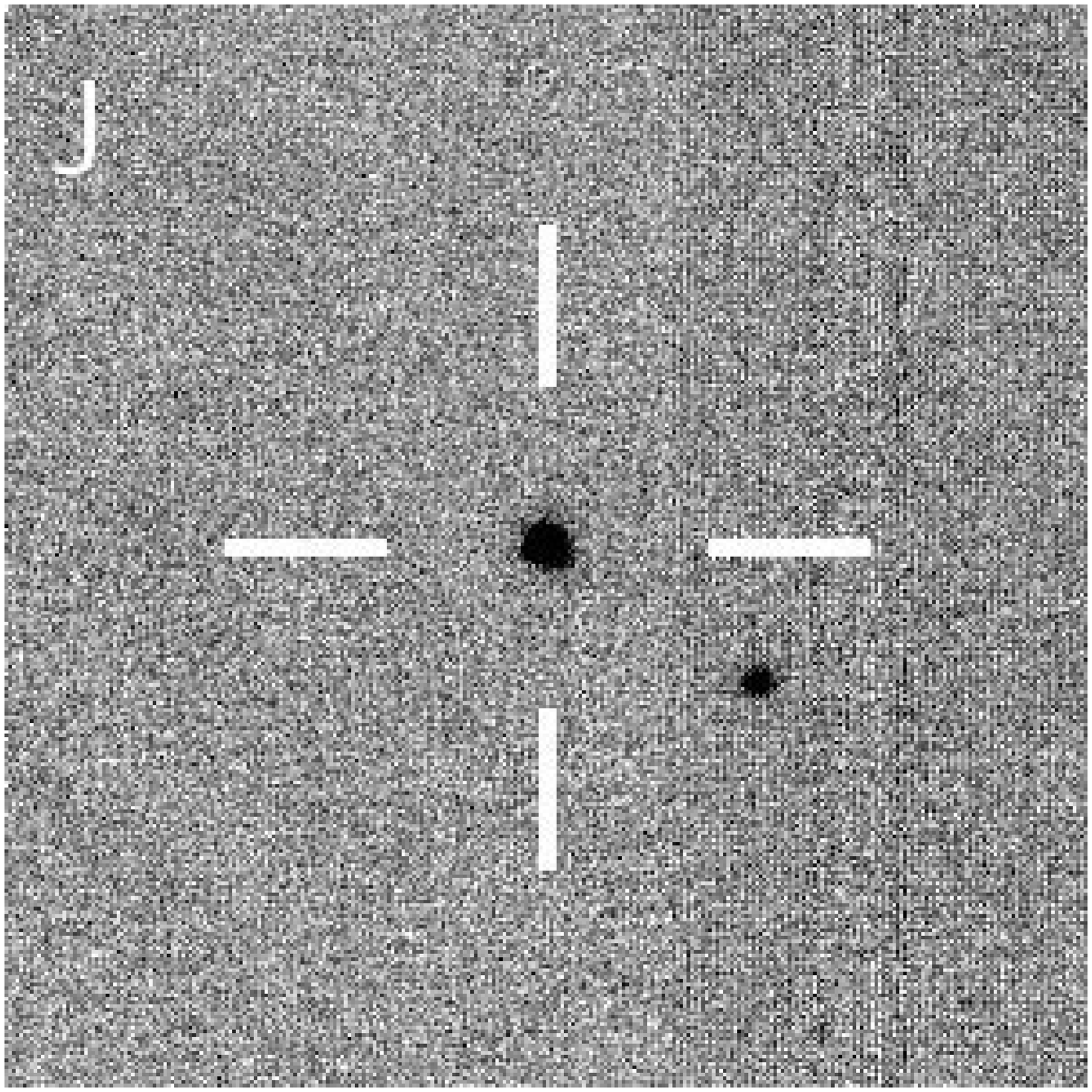,height=2.8cm,angle=0}
\psfig{file=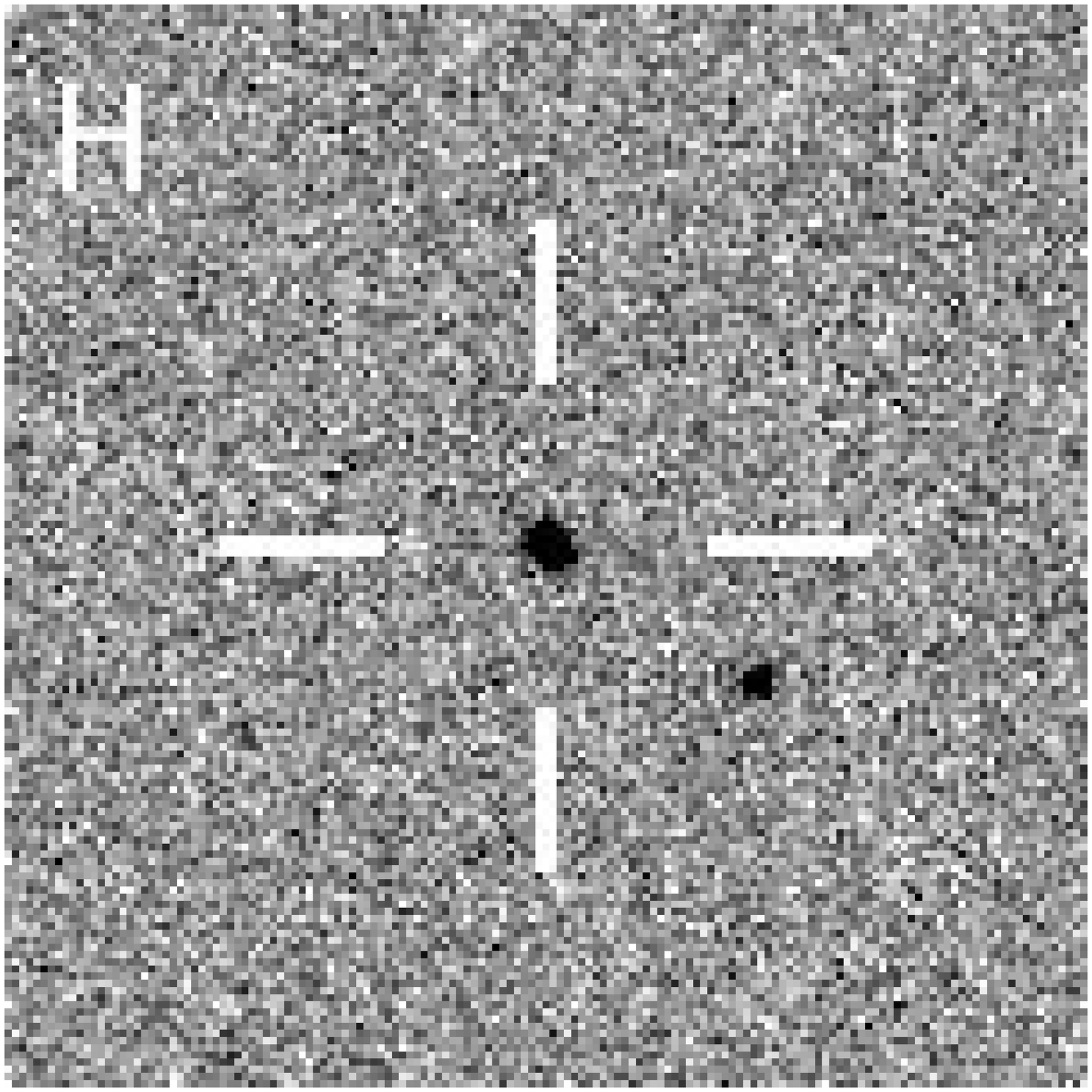,height=2.8cm,angle=0}
\psfig{file=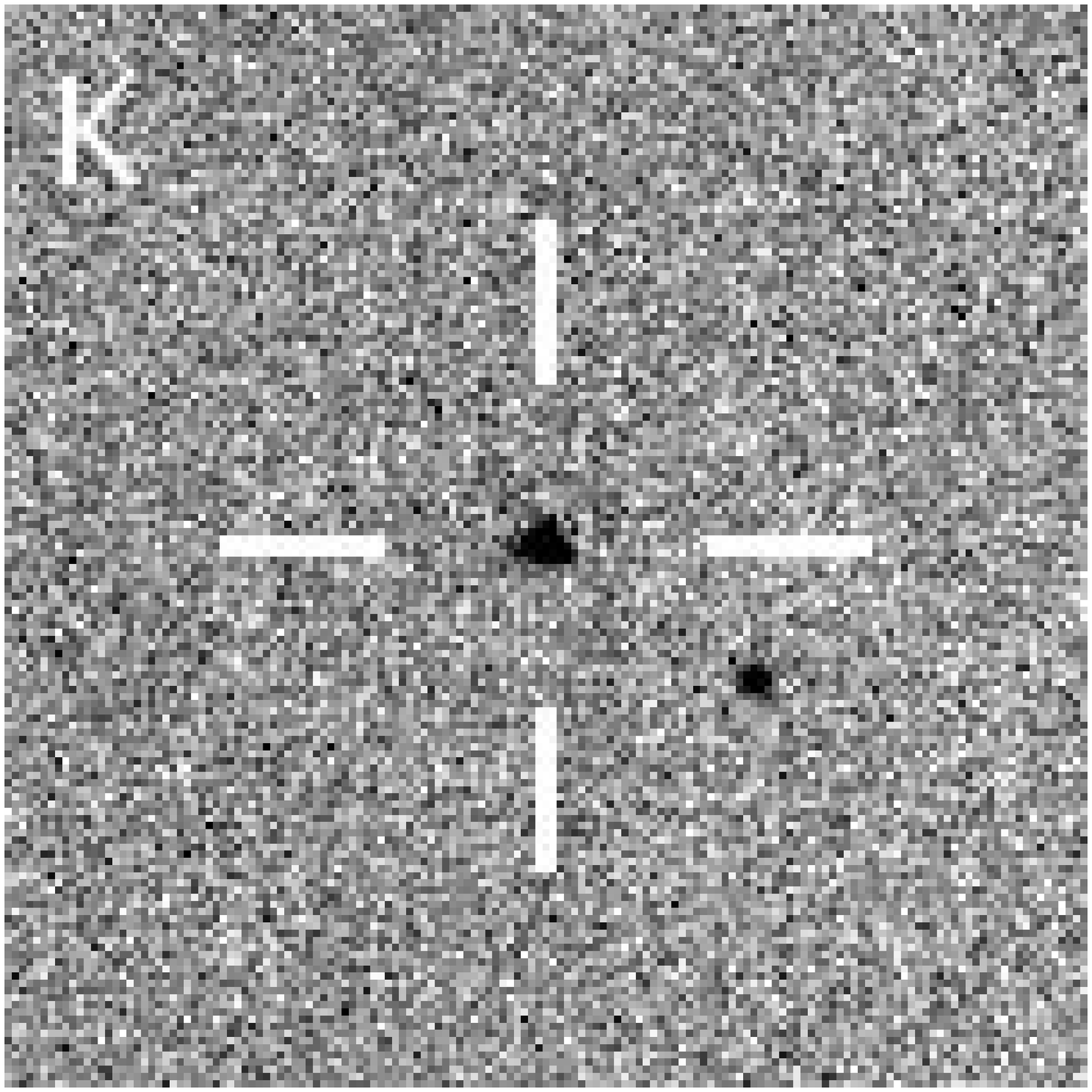,height=2.8cm,angle=0}
\end{center}
\caption{UKIDSS $YJHK$ images of SDSS\,J020538.12$+$005835.3.}
\label{020538_UKIDSS}
\end{figure}

The two colour diagram of the UKIDSS white dwarfs (Figure~\ref{colour1}) clearly separates out those stars with optical companions into the bottom of region~(II). The majority of stars are in region (I) potentially indicating that these are single white dwarfs. There are then a small number of stars in regions (III) and (IV) which would warrant further investigation. It should be noted that a number of the clear outliers (e.g. the MS99 object near the top edge of the diagram) are classified in the UKIDSS database as merged sources. This could indicate contamination, perhaps by an background galaxy or star. It can also indicate that a companion is becoming resolved in the NIR, as is most likely the case for e.g. SDSS\,J020538.12$+$005835.3 which clearly has a companion in the optical (Figure~\ref{020538_spec}) and UKIDSS classes as a merged source, as can be seen from the science images (Figure~\ref{020538_UKIDSS}). The UKIDSS \it{mergedClass} \rm statistic (-1 for a point source, +1 for an extended source) was checked for each candidate object to assess these possibilities (Table~\ref{results}).

It is clear there are significant advantages to using the modelling technique over the two colour diagram; a direct comparison can be made between observed and expected photometry determined from UKIDSS filter transmission profiles, and those objects with photometry in only one pair of filters can be analysed. Therefore we adopt this method for our work. 

\subsection{White Dwarf Modelling}
\label{modelling}
In order to efficiently process the cross-matched catalogue of white dwarfs we decided to model each white dwarf independently and then inspect each for evidence of a NIR excess. We also decided that only those DA white dwarfs with both $H$ and $K$-band photometry were to be modelled, as a substellar companion will give only a small or non-detectable excess at $Y$ and $J$, depending on the temperature of the components. Also, $H$ and $K$-band photometry together allow candidate substellar companions and blackbody-like dust disk emissions to be distinguished from each other. 

We have used the online version of Pierre Bergeron's synthetic colours and evolutionary sequences for DA white dwarfs\footnote{http://www.astro.umontreal.ca/~bergeron/CoolingModels/} \citep{bergeron95,holberg06}. This grid is built from a local thermodynamic equilibrium (LTE) model atmosphere code \citep{bergeron92}, which incorporates convective energy transport and hydrogen molecular opacity, up to an effective temperature ($T_{\rm eff}=20,000$\,K) where non-LTE effects are still negligible and the stellar atmosphere is completely radiative.  For pure hydrogen white dwarfs with $T_{\rm eff}>20,000$\,K, the grid then switches to using the plane-parallel, hydrostatic, non-LTE atmosphere and spectral synthesis codes \textsc{TLUSTY} (v200,~\citealt{hubeny88,hubeny95}) and \textsc{SYNSPEC} (v49; Hubeny \& Lanz, ftp:/tlusty.gsfc.nasa.gov/synsplib/synspec). The original model grid extends from $1500<T_{\rm eff}<100,000$\,K and $7.0<$log\,$g<9.0$ for DA white dwarfs. This was then linearly extrapolated to $6.5<$log\,$g<9.5$ to include the lower and higher mass white dwarfs, but to preclude any possible sub-dwarf contamination of the sample. 


This provided theoretical absolute magnitudes for the SDSS $i'$ and the 2MASS $JHK_{\rm S}$ filters, as well as an estimated distance to the white dwarf and cooling age of the system. The 2MASS magnitudes were converted to the corresponding UKIDSS filters using the colour transformations of \citet{carpenter01}. Out of the 2677  white dwarfs classified by EIS06 and MS99 as DA white dwarfs, 811 had both $H$ and $K$-band photometry and fell within the extended DA model grid. 


We identified candidate stars as those showing at least a $>3\sigma$ excess in the UKIDSS $K$-band when compared to the model predicted values. 313 such excesses were identified, with 274 of these accounted for as previously identified binaries from an optical excess in SDSS (or other optical data). This left 39 candidate white dwarfs with an apparent NIR excess. The UKIDSS photometry for these is given in Table~\ref{candidates} with additional parameters given in Table~\ref{candidates2}.

\begin{table*}
\caption{UKIDSS $YJHK$ magnitudes of the candidate white dwarfs showing near-infrared excesses.}
\begin{center}
\begin{tabular}{lccccc}
\hline
SDSS\,J & WD & $Y$ & $J$ & $H$ & $K$ \\
\hline\hline
003902.47-003000.3 &    -      & 18.70$\pm$0.06 & 18.37$\pm$0.08 & 17.86$\pm$0.11 & 17.41$\pm$0.10 \\
003923.04+003534.7 &    -      & 18.44$\pm$0.04 & 17.87$\pm$0.03 & 17.48$\pm$0.05 & 17.04$\pm$0.06 \\ 
012032.27-001351.1 &    -      & 19.25$\pm$0.08 & 18.34$\pm$0.07 & 17.82$\pm$0.07 & 17.42$\pm$0.08 \\
013532.98+144555.8 & 0132+145  & 16.49$\pm$0.01 & 16.24$\pm$0.01 & 15.86$\pm$0.01 & 15.56$\pm$0.02 \\ 
032317.00-002612.7 &    -      & 15.03$\pm$0.00 & 14.63$\pm$0.00 & 14.08$\pm$0.00 & 13.81$\pm$0.01 \\
033209.25+010722.1 &    -      & 17.45$\pm$0.02 & 17.47$\pm$0.02 & 17.75$\pm$0.10 & 17.65$\pm$0.10 \\
034221.55+005345.6 &    -      & 16.48$\pm$0.01 & 16.05$\pm$0.01 & 15.45$\pm$0.01 & 15.31$\pm$0.01 \\
075327.05+244705.5 &    -      & 19.17$\pm$0.06 & 19.22$\pm$0.07 & 18.65$\pm$0.18 & 18.01$\pm$0.16 \\
082705.04+284402.7 & 0824+288  & 13.06$\pm$0.00 & 12.72$\pm$0.00 & 12.14$\pm$0.00 & 11.84$\pm$0.00 \\
085956.47+082607.5 &    -      & 18.33$\pm$0.03 &     -          & 17.49$\pm$0.03 & 17.09$\pm$0.08 \\
090759.59+053649.7 &    -      & 18.79$\pm$0.08 & 18.66$\pm0.08$ & 18.62$\pm0.16$ & 18.40$\pm$0.19 \\
092648.84+102828.8 &    -      & 19.57$\pm$0.12 & 19.19$\pm$0.16 & 18.74$\pm$0.19 & 18.29$\pm$0.19 \\ 
093821.34+342035.6 &           & 18.48$\pm$0.05 & 18.08$\pm$0.03 & 17.65$\pm$0.08 & 17.35$\pm$0.10 \\
100259.88+093950.0 &    -      & 19.30$\pm$0.08 & 19.19$\pm$0.13 & 18.55$\pm$0.18 & 18.55$\pm$0.22 \\ 
101642.93+044317.7 &    -      &        -       &         -      & 17.73$\pm$0.09 & 17.59$\pm$0.09 \\ 
103220.19+011227.0 &    -      &        -       &         -      & 17.76$\pm$0.08 & 17.15$\pm$0.07 \\ 
103448.92+005201.4 & 1032+011  &        -       &         -      & 18.20$\pm$0.11 & 18.03$\pm$0.14 \\ 
103736.75+013912.2 &    -      &        -       &         -      & 17.25$\pm$0.06 & 17.12$\pm$0.07 \\ 
110826.47+092721.5 &    -      &        -       &         -      & 16.44$\pm$0.02 & 16.21$\pm$0.04 \\
113416.09+055227.2 &    -      &        -       &         -      & 18.10$\pm$0.12 & 17.94$\pm$0.20 \\ 
115808.44-012312.9 & 1155-011  & 18.26$\pm$0.03 & 17.83$\pm$0.03 & 17.22$\pm$0.03 & 16.89$\pm$0.06 \\
120503.99+070031.0 &    -      & 18.47$\pm$0.03 & 18.31$\pm$0.05 & 17.76$\pm$0.08 & 17.70$\pm$0.13 \\
121209.31+013627.7 &    -      & 18.04$\pm$0.04 & 17.85$\pm0.05$ & 17.21$\pm0.05$ & 16.21$\pm$0.03 \\
122625.93+002700.2 & 1223+007  & 19.11$\pm$0.08 & 19.13$\pm$0.14 & 18.54$\pm$0.14 & 18.02$\pm$0.18 \\
122859.92+104033.0 &    -      & 16.86$\pm$0.01 & 16.89$\pm$0.02 & 16.84$\pm$0.02 & 16.42$\pm$0.04 \\ 
124455.15+040220.6 &    -      &        -       &         -      & 18.59$\pm$0.12 & 17.70$\pm$0.13 \\
125044.42+154957.4$^{1}$ & 1248+161  &        - & 16.61$\pm$0.02 & 16.14$\pm$0.02 & 15.00$\pm$0.01 \\
131955.04+015259.5 & 1317+021  &        -       &         -      & 17.79$\pm$0.05 & 17.57$\pm$0.10 \\
132044.68+001855.0 & 1318+005  & 17.61$\pm$0.02 & 17.62$\pm$0.04 & 17.76$\pm$0.11 & 17.38$\pm$0.11 \\ 
132925.21+123025.4 &    -      & 17.07$\pm$0.01 & 16.66$\pm$0.01 & 16.27$\pm$0.02 & 15.98$\pm$0.02 \\ 
134154.39+005600.8 & 1339+011  & 18.30$\pm$0.03 & 17.82$\pm$0.04 & 17.33$\pm$0.05 & 16.95$\pm$0.06 \\
141448.25+021257.7 & 1412+024  & 17.77$\pm$0.03 & 17.87$\pm$0.04 & 17.76$\pm$0.08 & 17.19$\pm$0.09 \\
154431.47+060104.3 &    -      & 17.96$\pm$0.03 & 17.83$\pm$0.04 & 17.64$\pm$0.05 & 17.64$\pm$0.10 \\ 
155720.77+091624.7 &    -      & 18.82$\pm$0.04 & 18.82$\pm$0.06 & 19.03$\pm$0.14 & 18.34$\pm$0.15 \\ 
162514.88+302610.8 &    -      & 18.09$\pm$0.02 & 17.95$\pm$0.03 & 17.66$\pm$0.07 & 17.72$\pm$0.13 \\
220841.63-000514.5 &    -      & 18.16$\pm$0.04 & 17.48$\pm$0.03 & 17.06$\pm$0.05 & 16.59$\pm$0.04 \\ 
222030.68-004107.9 &    -      & 16.95$\pm$0.01 & 16.75$\pm$0.02 & 16.44$\pm$0.03 & 16.36$\pm$0.04 \\ 
222551.65+001637.7 &    -      & 18.59$\pm$0.05 & 18.51$\pm$0.07 & 18.15$\pm$0.13 & 17.60$\pm$0.13 \\
231133.88+144119.9 &    -      & 19.51$\pm$0.11 & 19.17$\pm$0.15 & 18.68$\pm$0.16 & 17.88$\pm$0.17 \\
233345.97-000843.0 &    -      & 19.36$\pm$0.08 & 18.80$\pm$0.10 & 18.22$\pm$0.13 & 17.67$\pm$0.13 \\ 

\hline
\end{tabular}
\end{center}
$^{1}$Identified in Section~\ref{ukidss:magnetic}
\label{candidates}
\end{table*}

\begin{table*}
\caption{Physical parameters of the candidate white dwarfs showing near-infrared excesses. Temperatures and surface gravities are from EIS06 unless otherwise stated.}
\begin{center}
\begin{tabular}{lcccc}
\hline
SDSS\,J & WD & $g'$ & $T_{\rm eff}$ (K) & log\,$g$  \\
\hline\hline
003902.47-003000.3 &   -        & 19.11$\pm$0.02 & 12314$\pm$268    & 7.39$\pm$0.08 \\ 
003923.04+003534.7 &   -        & 19.20$\pm$0.02 & 10080$\pm$81     & 7.99$\pm$0.10 \\ 
012032.27-001351.1 &   -        & 20.23$\pm$0.04 & 10487$\pm$194    & 7.91$\pm$0.22 \\ 
013532.98+144555.8 &   0132+145 & 16.96$\pm$0.01 & 8083$\pm$22      & 8.08$\pm$0.04 \\ 
032317.00-002612.7 &   -        & 20.72$\pm$0.07 & 22193$\pm$1446   & 7.10$\pm$0.19 \\
033209.25+010722.1 &   -        & 17.90$\pm$0.01 & 16700$\pm$89     & 7.89$\pm$0.02 \\      
034221.55+005345.6 &   -        & 19.73$\pm$0.03 & 23035$\pm$684    & 6.97$\pm$0.08 \\
075327.05+244705.5 &   -        & 19.47$\pm$0.02 & 13256$\pm$321    & 7.80$\pm$0.07 \\     
082705.04+284402.7 &   0824+288 & 14.26$\pm$0.00 & -                & -             \\
085956.47+082607.5 &   -        & 19.13$\pm$0.02 & 15623$\pm$237    & 7.44$\pm$0.06 \\      
090759.59+053649.7 &   -        & 18.85$\pm$0.01 & 19474$\pm$293    & 7.82$\pm0.05$ \\
092648.84+102828.8 &   -        & 20.47$\pm$0.04 & 13214$\pm$935    & 8.11$\pm$0.29 \\ 
093821.34+342035.6 &   -        & 19.00$\pm$0.01 & 8130$\pm$59      & 8.02$\pm$0.14 \\
100259.88+093950.0 &   -        & 19.57$\pm$0.02 & 22331$\pm$435    & 7.87$\pm$0.06 \\ 
101642.93+044317.7 &   -        & 18.99$\pm$0.01 & 27039$\pm$276    & 7.43$\pm$0.04 \\ 
103220.19+011227.0 &   -        & 19.17$\pm$0.02 & 8901$\pm$93      & 7.71$\pm$0.18 \\ 
103448.92+005201.4 &   1032+011 & 19.16$\pm$0.02 & 9904$\pm$109     & 8.13$\pm$0.15 \\ 
103736.75+013912.2 &   -        & 18.82$\pm$0.01 & 16300$\pm$221    & 7.73$\pm$0.05 \\ 
110826.47+092721.5 &   -        & 18.48$\pm$0.01 & 11873$\pm$176    & 8.97$\pm$0.05 \\ 
113416.09+055227.2 &   -        & 19.59$\pm$0.03 & 24184$\pm$650    & 7.85$\pm$0.08 \\ 
115808.44-012312.9 &  1155-011  & 19.63$\pm$0.03 & $\sim$13300$^{2}$&   -           \\
120503.99+070031.0 &   -        & 18.97$\pm$0.07 & 8688$\pm$61      & 8.21$\pm$0.12 \\
121209.31+013627.7 &   -        & 18.24$\pm$0.01 & $\sim$10000$^{3}$&$\sim$8.0$^{3}$\\
122625.93+002700.2 &  1223+007  & 20.17$\pm$0.04 & 8744$\pm$146     & 8.97$\pm$0.04 \\
122859.92+104033.0 &   -        & 16.90$\pm$0.01 & 22125$\pm$136    & 8.22$\pm$0.02 \\ 
124455.15+040220.6 &   -        & 19.25$\pm$0.02 & 67380$\pm$4504   & 7.95$\pm$0.27 \\
125044.42+154957.4 &  1248+161  & 18.36$\pm$0.01 & $\sim$10000$^{4}$&   -           \\
131955.04+015259.5 &  1317+021  & 18.71$\pm$0.01 & 16800 $^{5}$     &   -           \\
132044.68+001855.0 &  1318+005  & 17.69$\pm0.01$ & 19649$\pm$130    & 8.36$\pm$0.02 \\ 
132925.21+123025.4 &   -        & 17.70$\pm$0.01 & 13537$\pm$138    & 7.48$\pm$0.02 \\ 
134154.39+005600.8 &  1339+011  & 19.14$\pm$0.02 & 18032$\pm$226    & 7.91$\pm$0.05 \\
141448.25+021257.7 &  1412+024  & 18.18$\pm$0.01 & 8296$\pm$41      & 8.45$\pm$0.07 \\
154431.47+060104.3 &   -        & 18.51$\pm$0.01 & 7946$\pm$66      & 8.39$\pm$0.15 \\
155720.77+091624.7 &   -        & 19.11$\pm$0.01 & 21990$\pm$403    & 7.67$\pm$0.06 \\ 
162514.88+302610.8 &   -        & 18.32$\pm$0.03 & 74667$\pm$3097   & 7.88$\pm$0.13 \\
220841.63-000514.5 &   -        & 19.19$\pm$0.03 & 9224$\pm$81      & 8.28$\pm$0.12 \\ 
222030.68-004107.9 &   -        & 17.38$\pm$0.01 & 8037$\pm$31      & 8.28$\pm$0.07 \\ 
222551.65+001637.7 &   -        & 18.95$\pm$0.02 & 10640$\pm$94     & 8.16$\pm$0.09 \\ 
231133.88+144119.9 &   -        & 19.90$\pm$0.03 & 39161$\pm$1369   & 7.40$\pm$0.18 \\
233345.97-000843.0 &   -        & 20.65$\pm$0.05 & 11198$\pm$433    & 8.35$\pm$0.29 \\ 

\hline
\end{tabular}
\end{center}
$^{1}$ Helium fit. \\
$^{2}$ Blackbody fit of SDSS photometry. \\
$^{3}$ \citet{farihi08}\\
$^{4}$ \citet{vanland05}\\
$^{5}$ \citet{vennes02}
\label{candidates2}
\end{table*}

\subsection{Identifying Sub-Stellar Companions}
\label{companions}
For each candidate white dwarf we first examined the SDSS optical spectrum for any previousley undetected evidence of a low mass main-sequence companion. An increase in flux of the spectral energy distribution towards longer wavelengths or spectral features indicative of M dwarfs (e.g. H$\alpha$ emission and/or TiO absorption bands) were searched for by eye. Empirical models for low mass sub-stellar objects were then added to a white dwarf synthetic spectrum and these composites were compared to UKIDSS photometry to obtain an approximate spectral type for the putative companion.

The white dwarf pure-hydrogen synthetic spectra were calculated at the effective temperatures and surface gravities given in Table~\ref{candidates2} using the models of \cite{koester08}. The synthetic spectra were then normalised to the Sloan $i'$ magnitude of the corresponding white dwarf (Table~\ref{candidates2}). The $i'$-band was chosen for this work to avoid the strongest absorption lines in the white dwarf's optical spectrum.


The empirical models for the candidate secondary stars were constructed using the NIR spectra of late M dwarfs from the IRTF spectral library (\citealt{cushing05}, Rayner et al. in prep.), and the L and T dwarfs in Sandy Leggett's L and T dwarf archive \citep{chiu06,golimowski04,knapp04}.

\begin{figure}
\begin{center}
\psfig{file=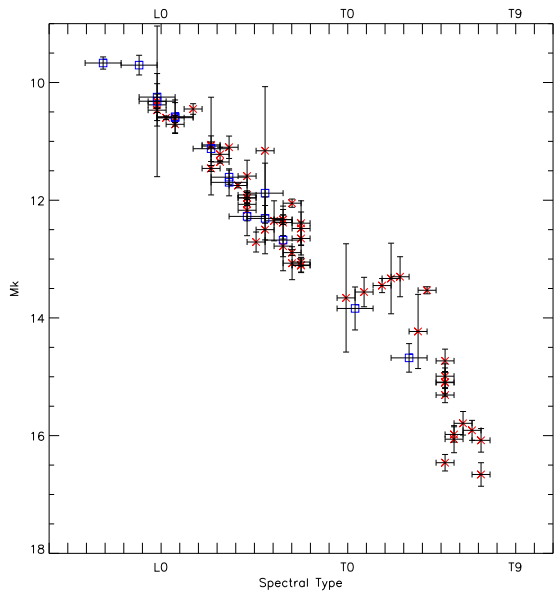,width=8.0cm,angle=0}
\end{center}
\caption{Absolute $K$-band photometry verses spectral type for L and T dwarfs (Crosses: Reproduced from Vrba et al. 2004). Also plotted are the predicted absolute magnitudes of the candidate companions in this survey (Squares). }
\label{absK}
\end{figure}

In order to obtain an initial estimate on an approximate spectral type of a putative companion, we calculated an absolute $K$-band magnitude for each candidate system. This was achieved by subtracting the model predicted $K$-band flux from the observed UKIDSS $K$-band flux, and then scaling to the estimated distance of the white dwarf (Table~\ref{results}). This value was compared to observed absolute $K$-band magnitudes of M, L and T dwarfs, which show a trend distinctive enough to separate the boundary regions of these 3 spectral types (Figure~\ref{absK}; \citealt{vrba04}). Due to errors in temperature, surface gravity and therefore distance, estimates of the companion spectral types are likely within $\pm$1 spectral types of the best fitting composite white dwarf $+$ companion model spectrum. 

\subsection{Identifying Debris Disks}
A debris disk around a white dwarf can be identified through a substantial excess emission in the $K$-band, whilst lacking a significant excess in the $J$ and $H$-band. Using the following method, blackbody models were used to obtain an upper limit on the temperature of the potential disk: a blackbody model was generated at the temperature of white dwarf and scaled appropriately. This model was then added to the white dwarf synthetic spectrum to create a combined white dwarf + blackbody model (WD$+$BB). This combined model represents a white dwarf with a cool ($<1000$\,K) disk. Starting at a temperature of 1000\,K, the temperature of the secondary blackbody was reduced in steps of 100\,K until the combined WD$+$BB model matched the $H$ and $K$-band photometry. Here we assumed there is no excess  in the $H$-band. This temperature can then be adopted as a upper limit as lower temperature disk combinations can also produce models which match the $K$-band photometry alone. NIR $K$-band spectroscopy or mid-infrared photometry would be needed to more accurately constrain the temperature of any debris disk containing system.  

\section{Follow-up NIR Photometric Observations}
\label{UKIDSS:followup}
For a small number of sources, we obtained follow-up NIR photometry to independently measure their $JHK$ magnitudes. SDSS\,J222551.65$+$001637.7 was observed on 2007 October 2 (at $K_{\rm s}$) and 2007 October 3 with the 3.5m New Technology Telescope (NTT) and the Son-of-ISAAC (SOFI) Imager. Individual exposure times were 15\,secs, 12\,secs and 8\,secs for the $J$, $H$ and $K_{\rm s}$-bands respectively, with a total exposure time in each band of 150\,secs, 288\,secs and 240\,secs.  The weather was clear with seeing of $J = 1.3\arcsec$ on the first night and $K_{\rm s} = 1.4\arcsec$ on the second night. 

SDSS\,J003902.47$-$003000.3 was observed also with the NTT$+$SOFI on 2007 October 25 in $J,H$ \& $K_{\rm s}$. Individual exposure times were 15\,secs, 12\,secs and 12\,secs for the $J$, $H$ and $K_{\rm s}$-bands respectively, with a total exposure time in each band of 75\,secs, 144\,secs and 144\,secs. The weather was clear and photometric with seeing at $K_{\rm s} = 0.7\arcsec$. 

WD\,1318$+$005 was observed at the 3.9m Anglo-Australian Telescope (AAT) with the IRIS2 NIR imager on 2007 August 27. Individual exposures times were 5\,secs in the $K$-band with a total exposure time of 75\,secs. The average seeing over the course of the observations was $=2.4\arcsec$.   \\

Standard nodding was used in all observations to ease te removal of NIR background. We reduced all the data automatically in the standard manner with the ORAC-DR package. The combined frames were photometrically calibrated using UKIDSS sources in each field. Aperture photometry was performed interactively using the Starlink Gaia package. 

\section{Candidate UKIDSS Photometry}

For each candidate system with a putative stellar/substellar companion, we have estimated an upper limit on the projected orbital separation. This was achieved by measuring the FWHM of the system in the bandpass where the contributions of flux from each object would be roughly equal (usually the $z'$ or $J$-band depending on the availability of the data), and equating this to an orbital seperation at the estimated distance to the white dwarf. In the few cases where the candidate system has been resolved, the separation between components has been measured directly and equated to a projected orbital seperation.

For the few resolved systems, we have also calculated the probability of a chance alignment ($\rm P_{al}$) by referring to the spatial number densities of M, L and T-dwarfs \citep*{caballero08}. The expected number of stars with the spectral type assigned to the putative companion is then calculated within a region of space defined by the errors on the white dwarf's distance. This then has to be multiplied by the total sample size in order to take the complete volume of the survey into account.

We discuss a number of examples in detail in the following sections. A complete summary of results is given in Table~\ref{specresults} and Figures inluded in Appendix~\ref{spectra}.


\subsection{Examples I: White Dwarfs with Putative Low Mass Companions}

\subsubsection{SDSS\,J003902.47$-$003000.3: A White Dwarf with an Unresolved Low Mass Stellar Companion}
SDSS\,J0039$-$003's optical spectrum shows no obvious evidence for an M dwarf companion. EIS06 give the white dwarf physical parameters as $T_{\rm eff} = 12,314\pm268$K and log~$g = 7.39\pm0.08$ from an automated fit to the optical spectrum. When compared to an atmospheric model at this temperature and gravity, a small excess emission is visible in the SDSS $z'$-band (Figure~\ref{003902}), with a clear strong excess emission in the UKIDSS $YJHK$-band photometry. The predicted absolute $K$-band photometry of the potential companion suggests a spectral type of M7. However, we found that a composite WD$+$L0 spectrum is the best fit to the UKIDSS photometry.

We first identified this object in UKIDSS DR2 and so follow-up $JHK$-band photometry was obtained using SOFI on the NTT in October 2007 (Section~\ref{UKIDSS:followup}). The SOFI photometry confirms the presence of a NIR excess and that the companion is likely of spectral type L0. The UKIDSS \it{mergedClass} \rm statistic for SDSS\,J0039$-$003 indicates the system is a point source, with a measured FWHM of 0.73'' in the $J$-band. This equates to a projected orbital separation of $<284$\,AU for the secondary at the estimated distance to the white dwarf (Table~\ref{results}). However, the low mass of the white dwarf (0.34$\pm$0.02\msun) suggests an accelerated evolution caused by a secondary component, and so it is likely that the companion is close. If this is the case, a discrepancy in estimated spectral type of the secondary may arise from variability in the NIR photometry, due to night and dayside heating of the companions atmosphere. NIR time series photometry is needed to confirm this scenario.

\subsubsection{SDSS\,J012032.27$-$001351.1: A White Dwarf with a Resolved Low Mass Stellar Companion}
\begin{figure}
\begin{center}
\psfig{file=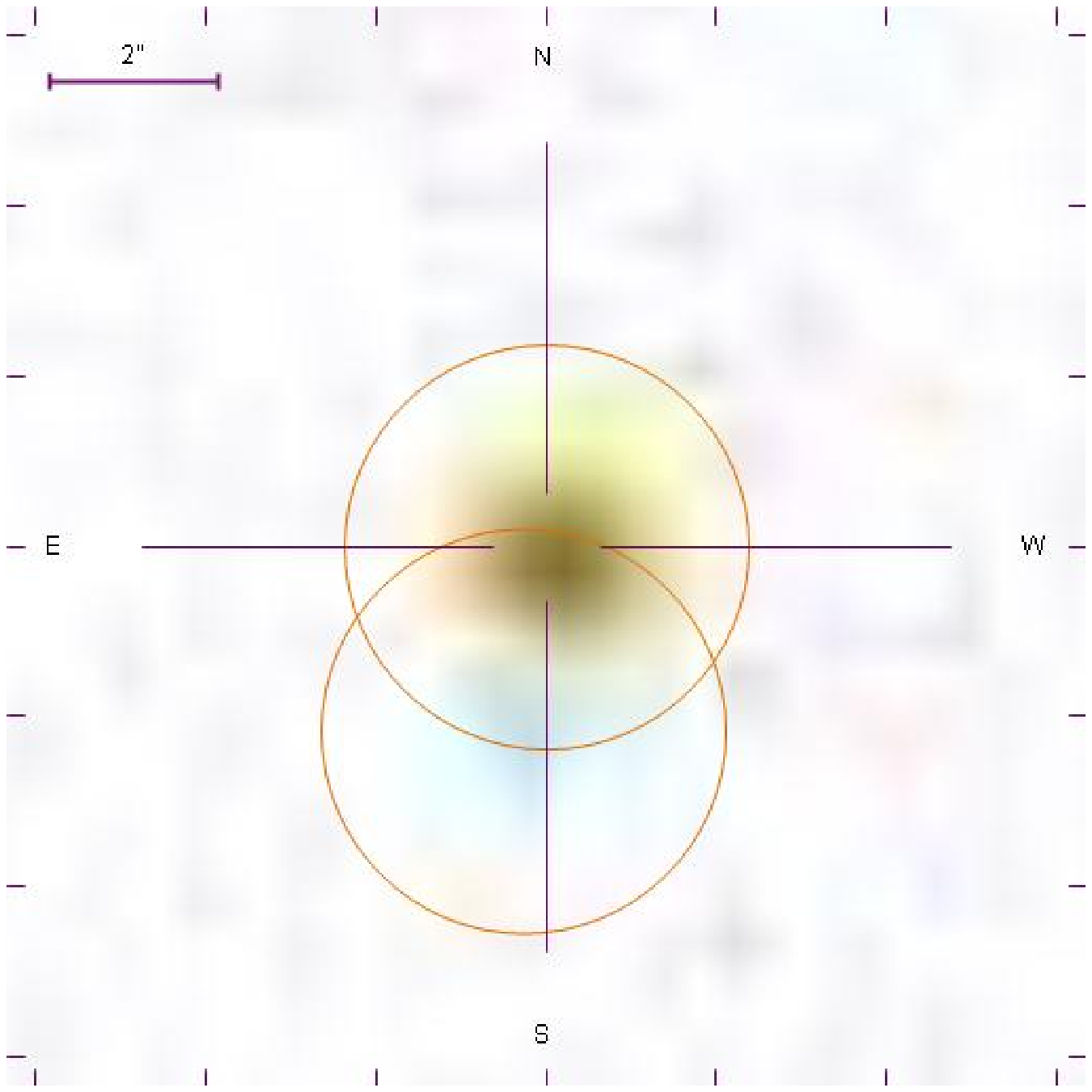,width=5.0cm,angle=0}
\end{center}
\caption{SDSS finding chart for SDSS\,J012032.27$-$001351.1. A secondary object can be seen just to the south of the central white dwarf. The colours have been inverted from the original finding chart.}
\label{012032_image}
\begin{center}
\psfig{file=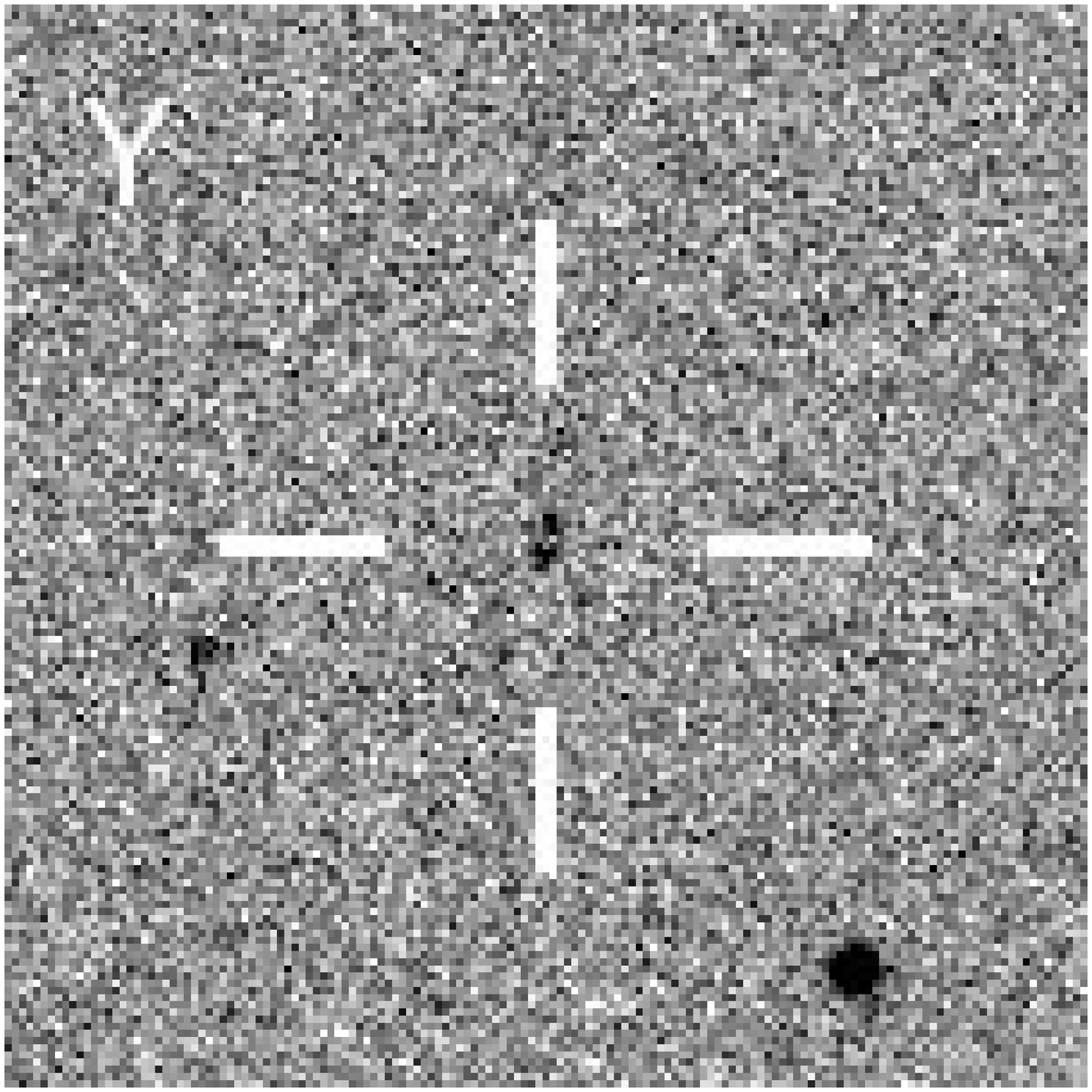,height=2.8cm,angle=0}
\psfig{file=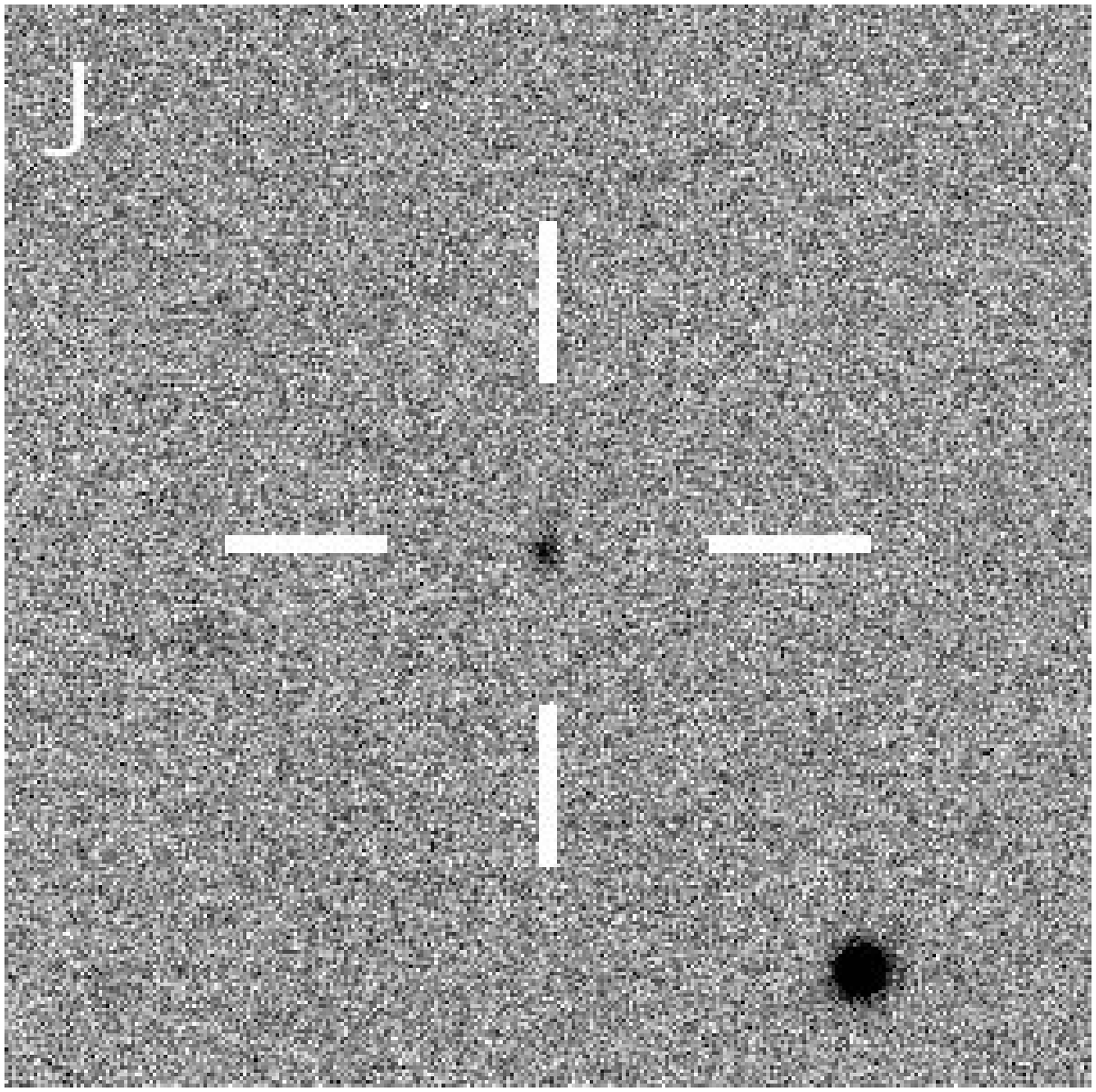,height=2.8cm,angle=0}
\psfig{file=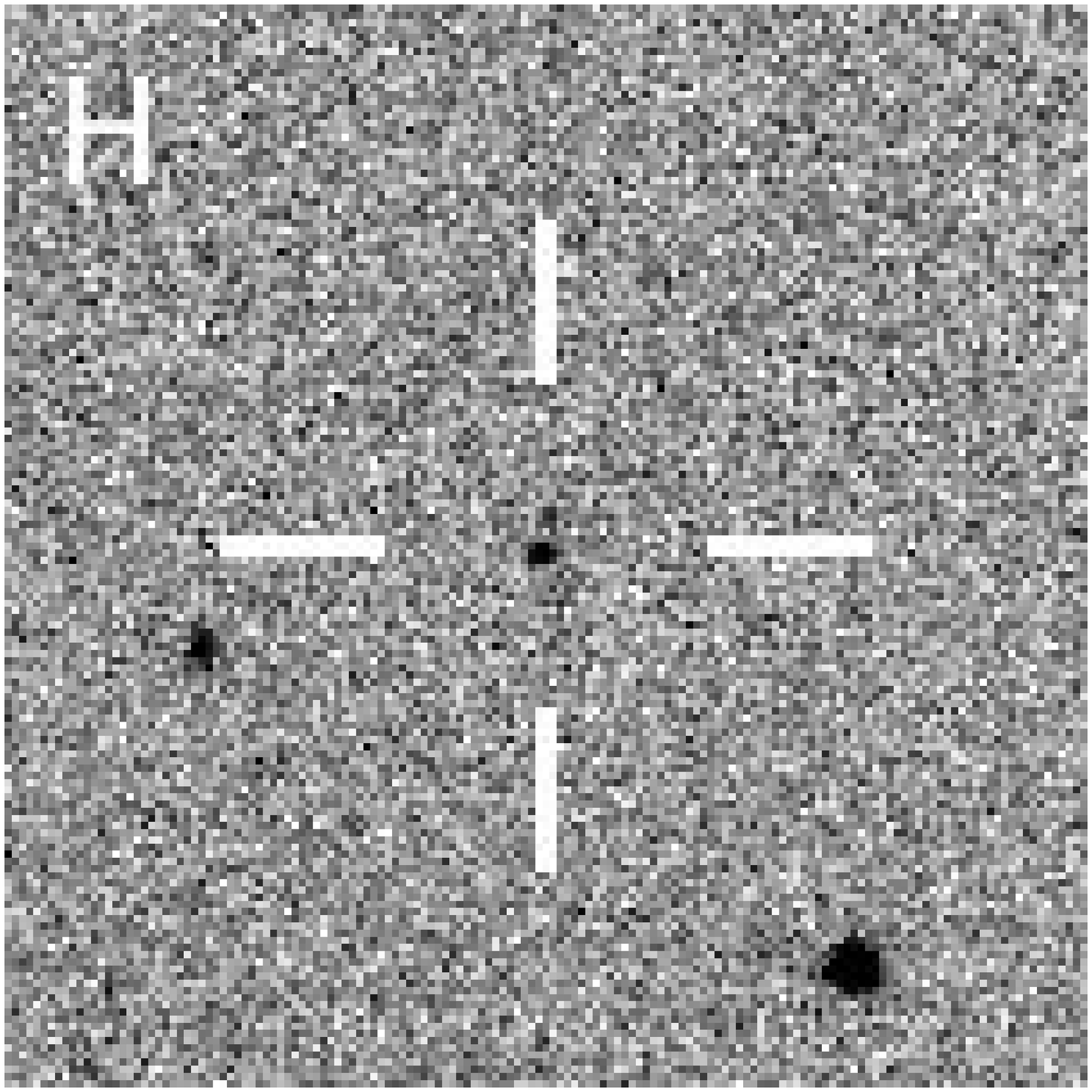,height=2.8cm,angle=0}
\psfig{file=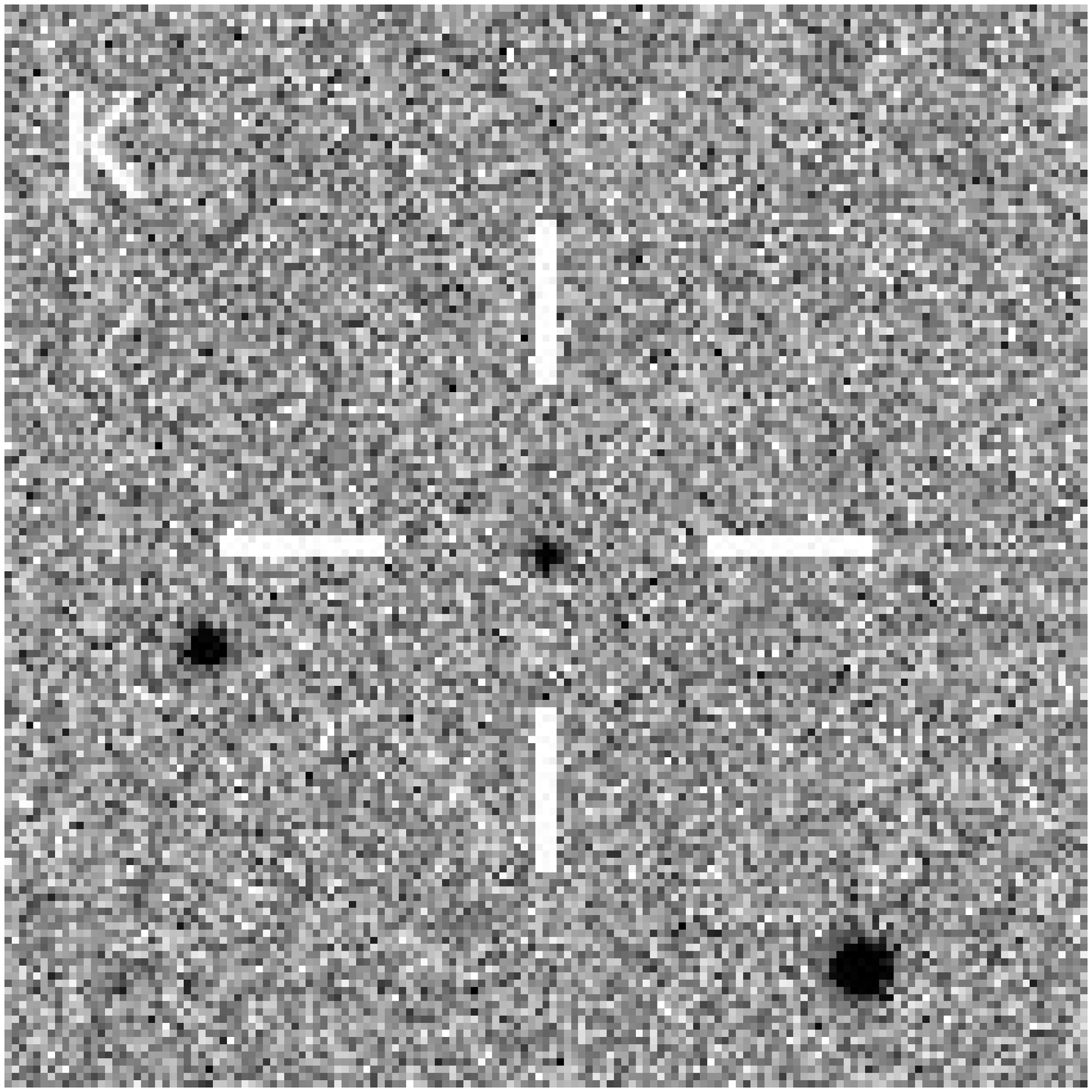,height=2.8cm,angle=0}
\end{center}
\caption{UKIDSS $YJHK$ images of SDSS\,J012032.27$-$001351.1. Both objects from the SDSS image can be seen in the UKIDSS $Y$-band, with only the southern most object appearing in the $JHK$ images.}
\label{012032_image2}
\end{figure}

SDSS\,J0120$-$001 was classified as a DA white dwarf by EIS06 with $T_{\rm eff} = 10,487\pm194$\,K and log~$g = 7.91\pm0.23$. However, the SDSS finder image shows a faint red background object which may have contaminated the optical spectrum to some extent (Figure~\ref{012032}. An excess is clearly seen in the UKIDSS photometry, most likely due to this background source. Indeed, a closer inspection of the SDSS and UKIDSS images (Figure~\ref{012032_image} \& \ref{012032_image2}) reveals that the white dwarf has not been detected in the the NIR and hence it is only the flux from the nearby red object that is being measured. The two components are separated by $\approx 2\arcsec$ equating to a projected orbital separation of $\approx910$\,AU. If the two are associated then the UKIDSS photometry is best matched by an M7 dwarf scaled to the estimated distance of the white dwarf. 

The expected number density of stars of spectral type M6-M8\,V is 6.8$\times10^{-3}$\,pc$^{-3}$ \citep{caballero08}. Therefore, assuming an error of $\pm$1 on the spectral type of the red object, the expected number of stars of spectral type M6-M8\,V in a volume of space defined by the errors of the white dwarf's distance (140\,pc) and the projected separation is 5.8$\times10^{-5}$.  Thus, this could be the widest white dwarf + M dwarf binary to be identified in this work. Follow-up proper motion measurements are required to assess the association of the two objects, as a background giant star cannot be discounted at this stage.

\subsubsection{SDSS\,J013532.98$+$144555.8 $=$ NLTT\,5306: A White Dwarf with an Unresolved Brown Dwarf Companion}
\label{NLTT5306}
NLTT\,5306 is classified by EIS06 as a DA white dwarf with $T_{\rm eff}=8100\pm20$\,K and log$g=8.08\pm0.04$. The SDSS optical photometry and spectrum (Figure~\ref{013532}) show no sign of an excess due to a stellar companion. However, an excess is clearly visible in the UKIDSS photometry, with the predicted absolute $K$-band magnitude of the companion suggesting a spectral type of L6. A best fit is achieved with a WD$+$dL5 spectrum. At the white dwarf age of 2.1\,Gyr (Section~\ref{UKIDSS:starordwarf}), the DUSTY atmospheric models \citep{chabrier00,baraffe02} predict a mass of 58$\pm$2\mjup.  Therefore, it is likely SDSS\,0135$+$144 harbours a brown dwarf companion. The UKIDSS \it{mergedClass} \rm statistic for NLTT\,5306 indicates the system is a point source, with a measured FWHM of 0.96'' in the $J$-band. This equates to a projected orbital separation of $<57$\,AU for the secondary at the estimated distance to the white dwarf (Table~\ref{results}).

\begin{figure}
\begin{center}
\psfig{file=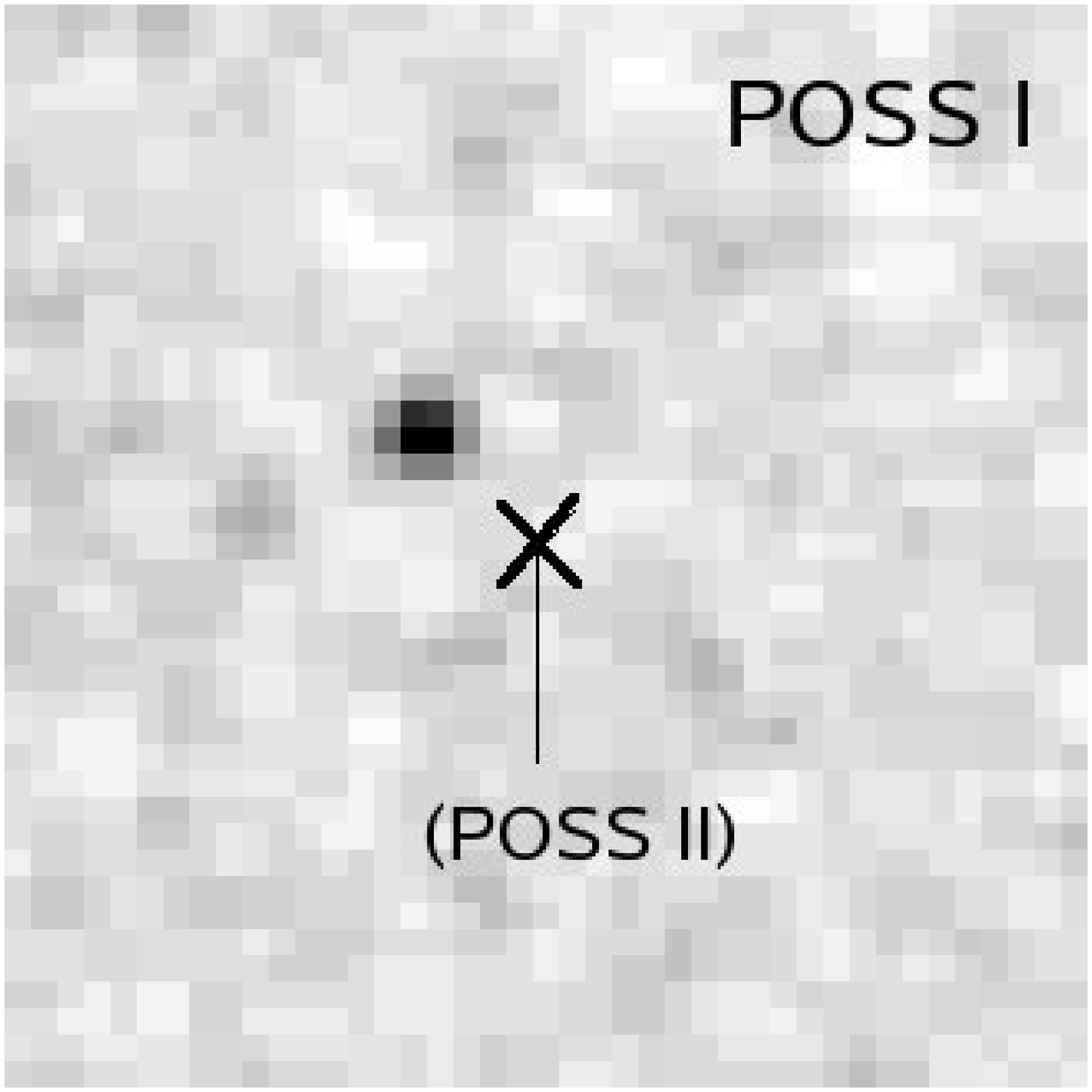,height=2.8cm,angle=0}
\psfig{file=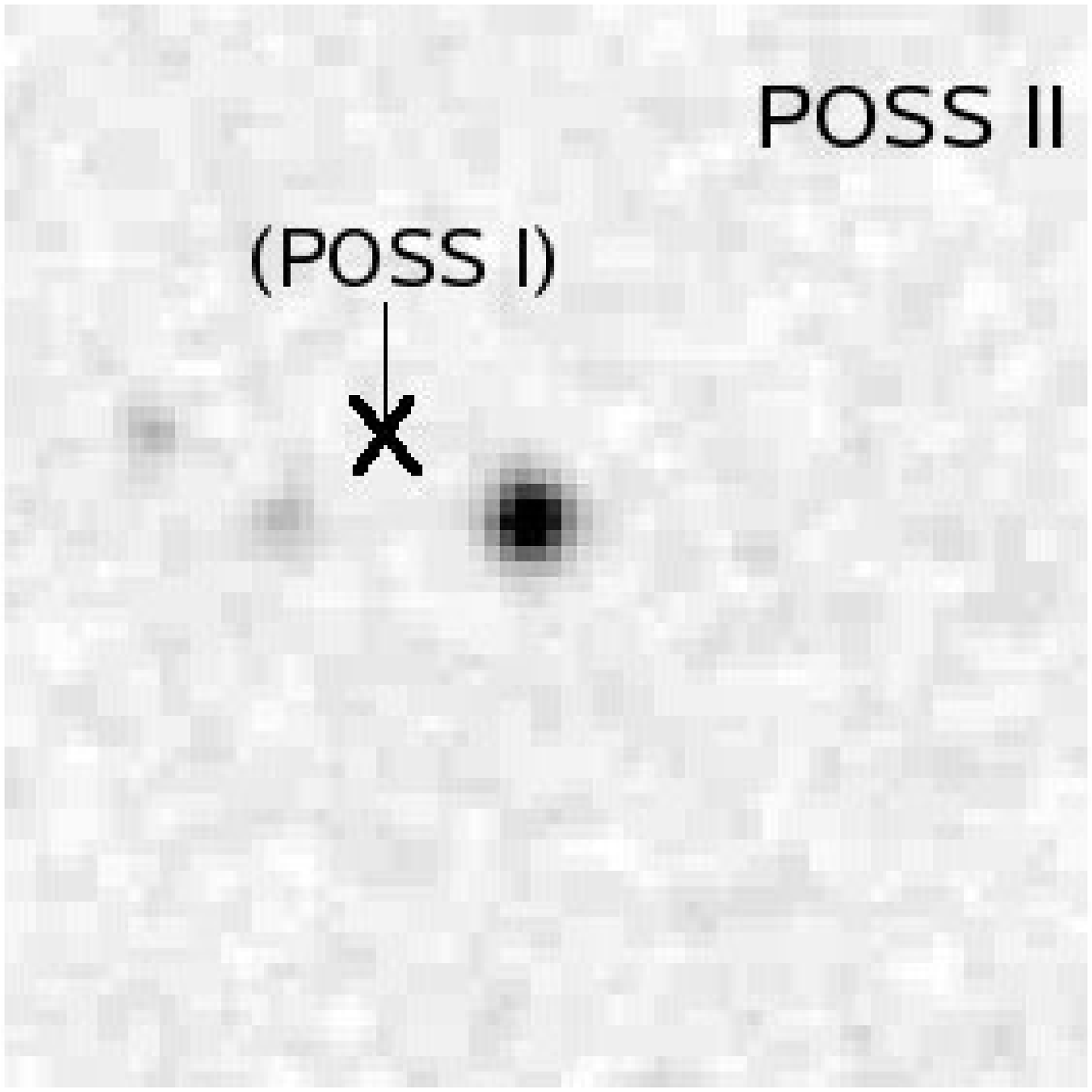,height=2.8cm,angle=0}
\psfig{file=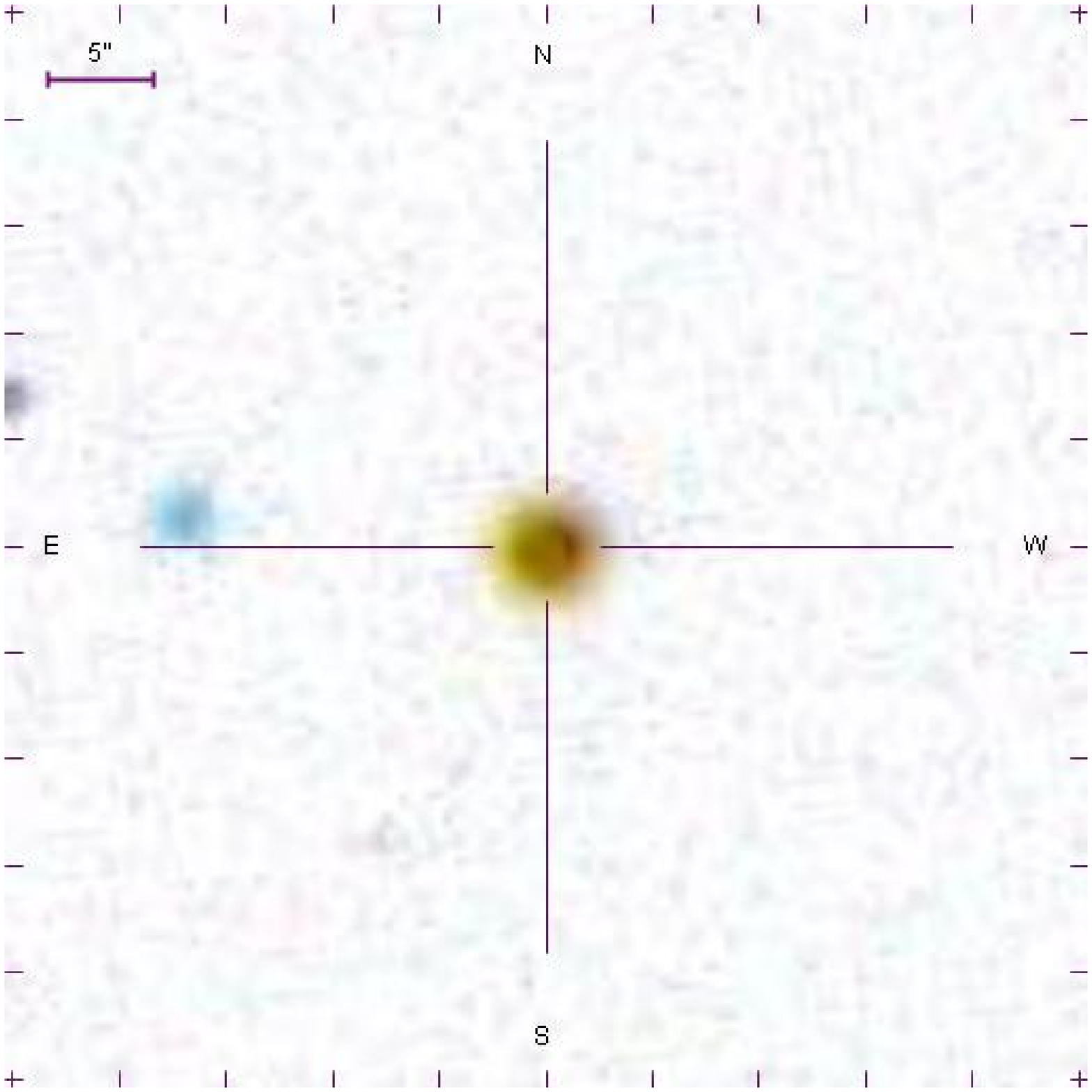,height=2.8cm,angle=0}
\psfig{file=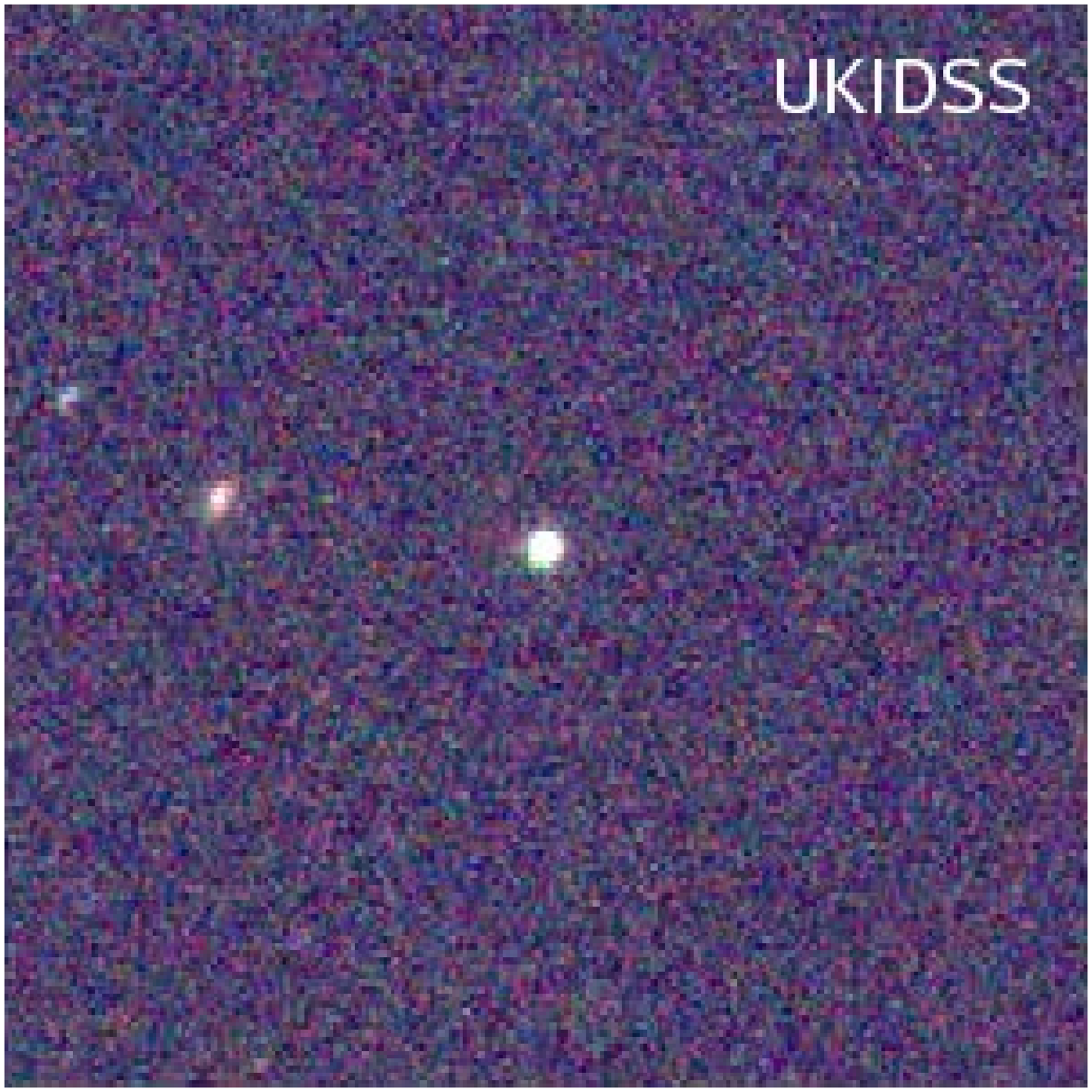,height=2.8cm,angle=0}
\end{center}
\caption{POSS\,I and POSS\,II images of NLTT\,5306, with the current and prior positions of the star respectively marked. Also shown is the current epoch finder image from SDSS, and a composite $JHK$ image from the UKIDSS DR8.}
\label{nltt5306}
\end{figure}

NLTT\,5306 is known to have a high proper motion ($\mu_{\rm abs}=0.194\arcsec$ yr$^{-1}$, \citealt{lepine05}). The star was identified in both the first and second Palomar Sky Surveys (POSS\,I and POSS\,II), and a composite image was produced from the data (Figure~\ref{nltt5306}). No background object can be seen to be present along the path of NLTT\,5306, increasing the likelihood that this star has a secondary low mass companion.

\subsubsection{SDSS\,J023247.50$-$003909.3: Contamination from a Foreground/Background Source}
\begin{figure}
\begin{center}
\psfig{file=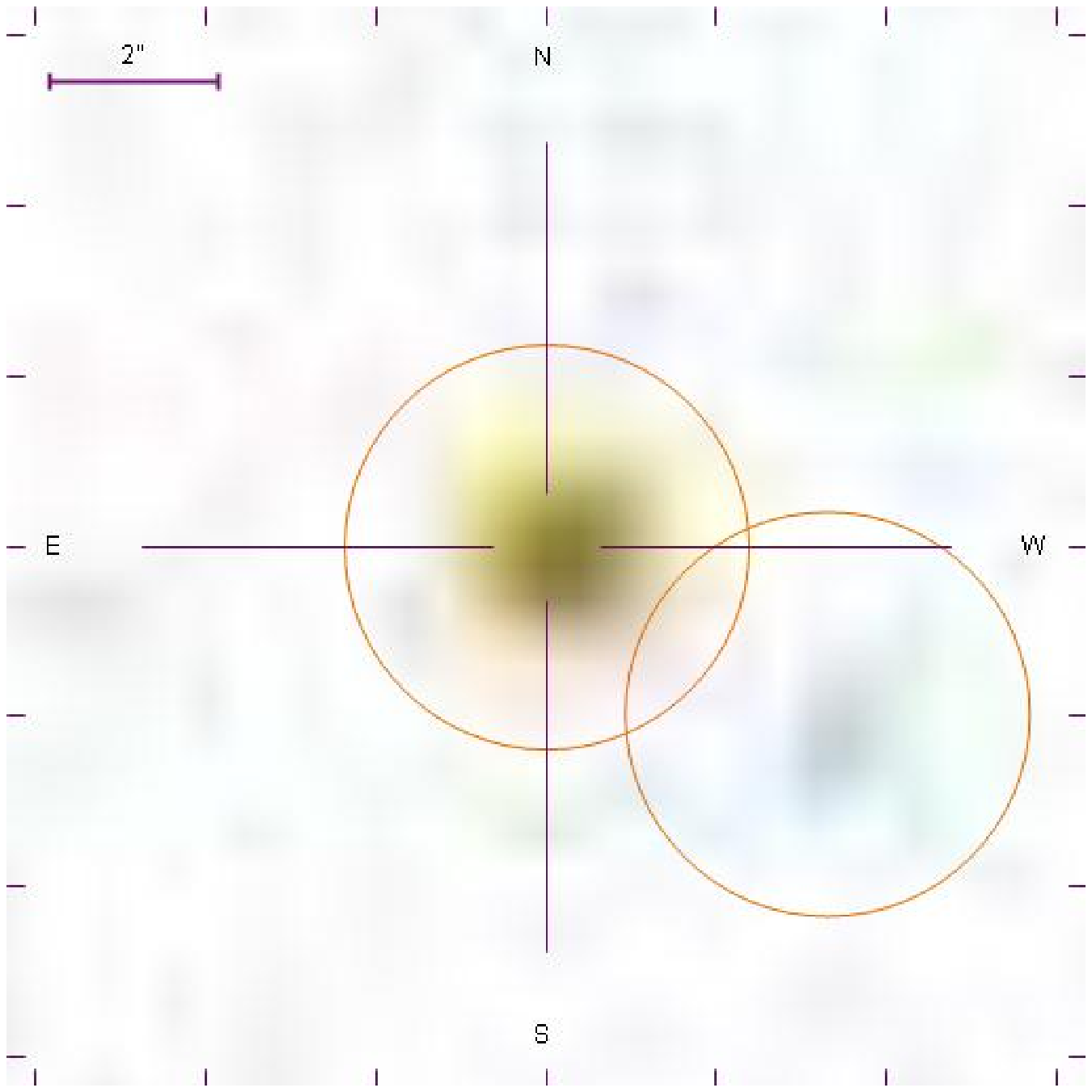,width=5.0cm,angle=0}
\end{center}
\caption{SDSS finder image of SDSS\,J023247.50$-$003909.3. A secondary object can be seen just to the south west of the central white dwarf. The colours have been inverted from the original finding chart.}
\label{023247_image}
\begin{center}
\psfig{file=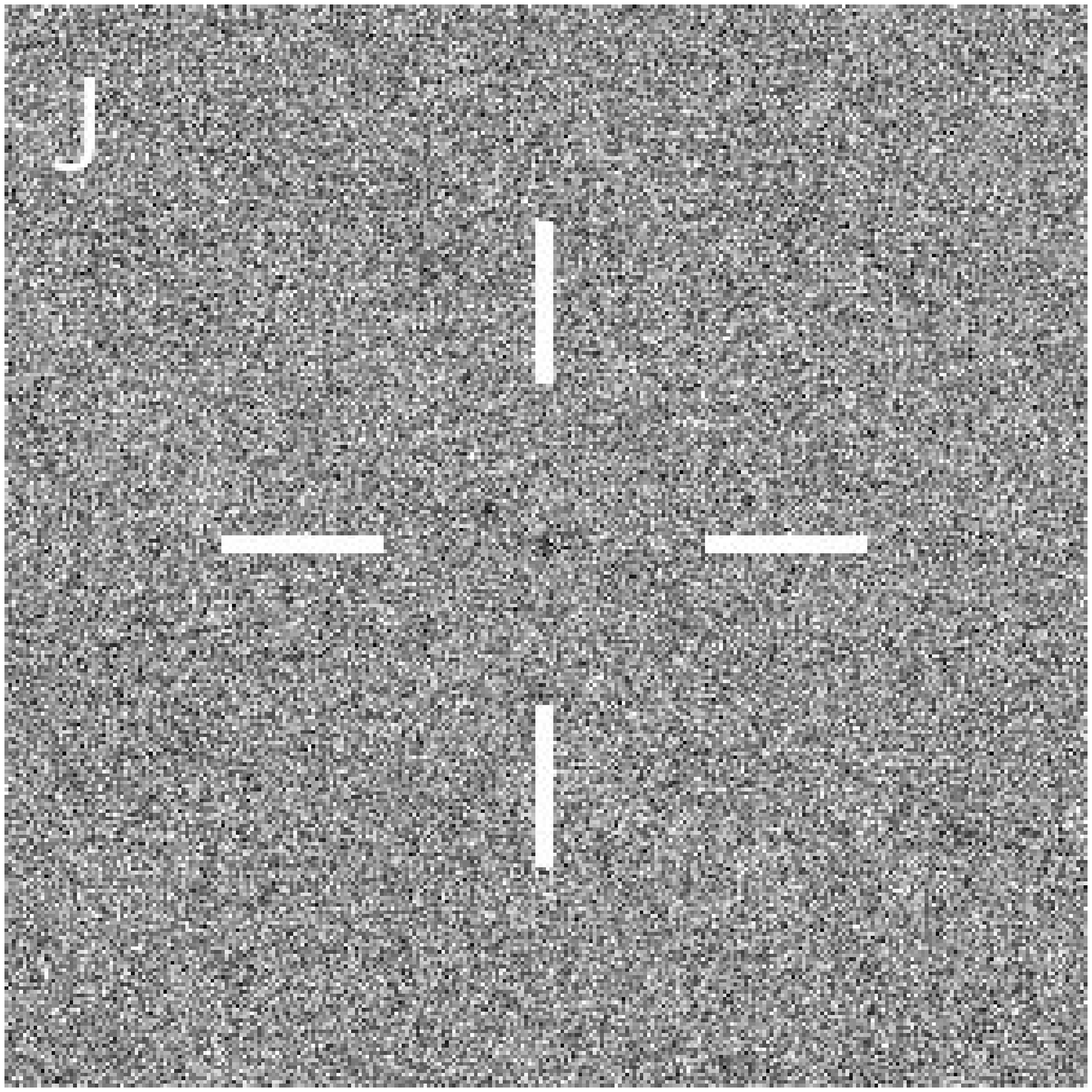,height=2.8cm,angle=0}
\psfig{file=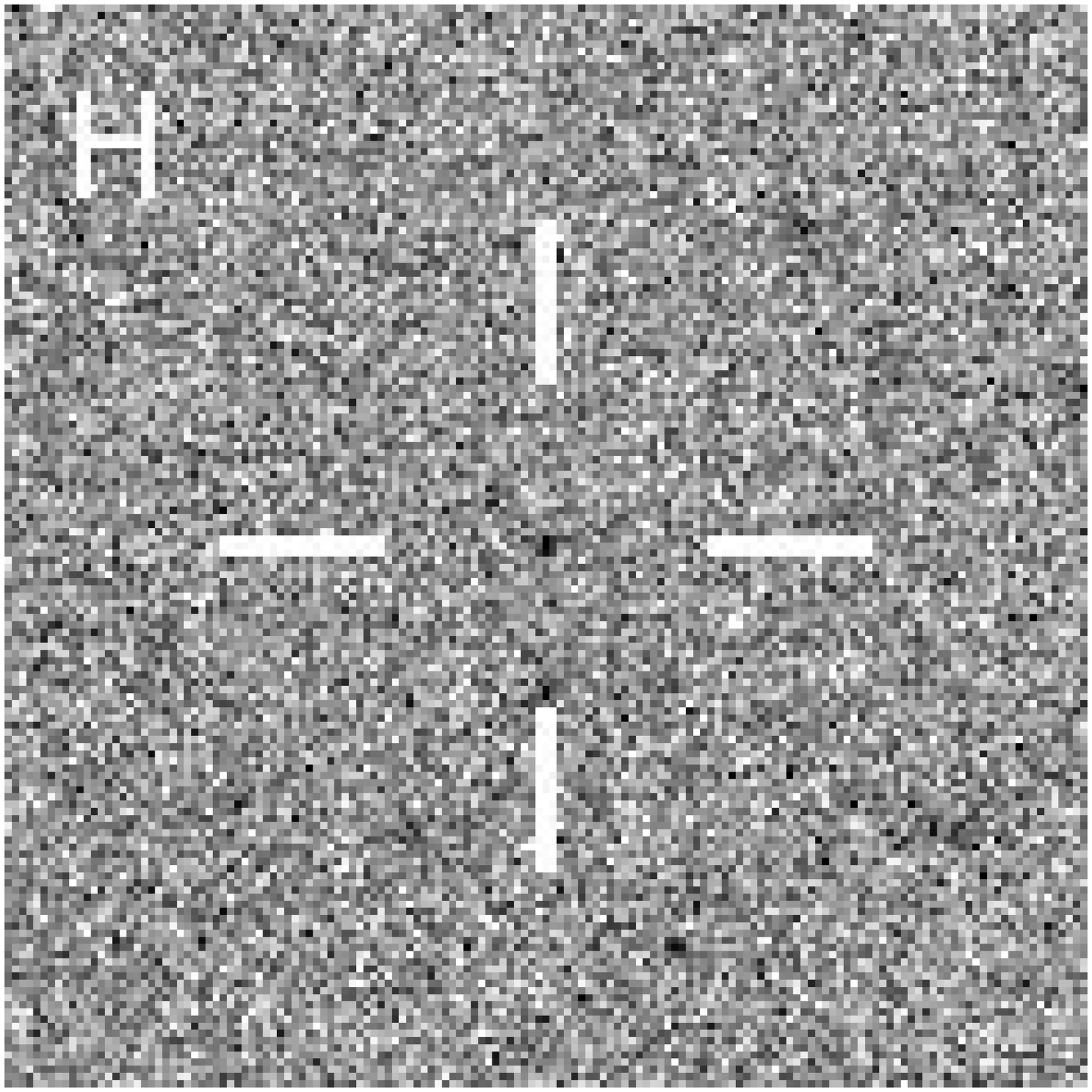,height=2.8cm,angle=0}
\psfig{file=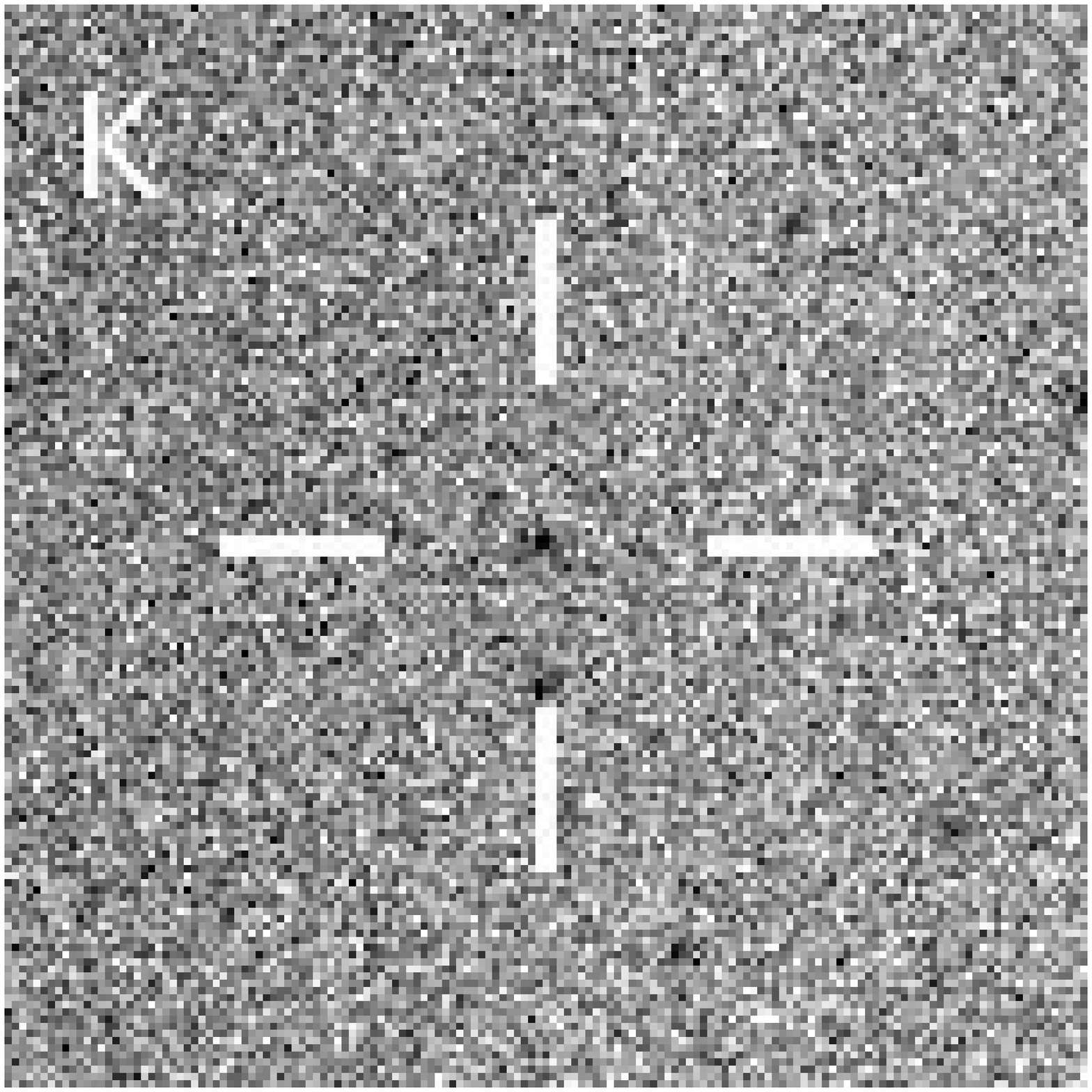,height=2.8cm,angle=0}
\end{center}
\caption{UKIDSS $JHK$ images of SDSS\,J023247.50$-$003909.3. The white dwarf has almost completely faded from view in the $J$-band image, with the red object completely dominating in the $H$ and $K$-bands.}
\label{023247_image2}
\end{figure}

SDSS\,J0232$-$003 is a DA white dwarf with EIS06 parameters of $T_{\rm eff}=9898\pm161$\,K and log$g=8.04\pm0.24$. The white dwarf has $i'=20.11\pm0.03$ and so would be too faint to be detected by UKIDSS in the NIR. However, a secondary object was detected with a separation of 2.0'' from SDSS\,J0232$-$003 (Figures~\ref{023247_image} \& \ref{023247_image2}), equating to a projected separation of 700\,AU. At the estimated distance to the white dwarf of $334\pm83$\,pc, the $K$-band magnitude of the potential secondary indicates a spectral type of M8-L2. However, it was found that a star of spectral type L3 was the best fit to the UKIDSS photometry. This difference in estimates could be due to the large uncertainties in the UKIDSS magnitudes and therefore distance estimate to the white dwarf. However, it is highly unlikely that an object of this spectral type would have been seen in the optical images. Therefore, it is more likely that this is a foreground or background object and so will be excluded from our statistical analysis. 

\subsubsection{SDSS\,J154431.47$+$060104.3: A White Dwarf with an Unresolved T Dwarf Companion}
SDSS\,1544$+$060 shows no optical excess when compared to a blackbody model at $T_{\rm eff}=7946\pm66$\,K, the temperature assigned by the  EIS06 automated fit (Figure~\ref{154431}).  An excess is clearly visible in each of the UKIDSS bands with the predicted absolute $K$-band magnitude of the secondary suggesting a companion type of early T. Each data point matches a composite WD$+$T3 model particularly well with the $J$-band photometry possibly indicating the detection of the CH$_{4}$ absorption feature. If confirmed this would potentially be the first detection of a close WD$+$dT dwarf binary, and only the second detection of a WD+dT system. The UKIDSS \it{mergedClass} \rm statistic for SDSS\,J1329$+$123 indicates the system is a point source, with a measured FWHM of $2.12\arcsec$ in the $J$-band. This equates to a projected orbital separation of $<190$\,AU for the secondary at the estimated distance to the white dwarf (Table~\ref{results}). If the system were in a close orbit the secondary would be the lowest object to have survived common envelope evolution. We now urgently require spectroscopic confirmation for this potential new type of WD binary system, with follow up radial velocity measurements to assess the separation of components.
 
\subsubsection{SDSS\,J222551.65$+$001637.7: A White Dwarf with a Partially Resolved Brown Dwarf Companion}
SDSS\,J2225$+$001 shows no sign of an M dwarf in it's optical spectrum and an atmospheric model with parameters $T_{\rm eff}=10640\pm94$\,K and log\,$g=8.16\pm0.09$ matches the SDSS photometry (Figure~\ref{222551}). A clear excess is then seen in the $H$ and $K$-band photometry with the predicted absolute $K$-band magnitude of the secondary suggesting a spectral type of approximately L6. A composite WD$+$L7 model provides the best match in this case. The UKIDSS \it{mergedClass} \rm statistic for SDSS\,J2225$-$001 suggest the system is partially resolved. However, there is not a clear separation of the components in the UKIDSS images. The  measured FWHM of the star is 2.28'' in the $J$-band which equates to a projected orbital separation of $<435$\,AU for the secondary at the estimated distance to the white dwarf (Table~\ref{results}).

\begin{figure}
\begin{center}
\psfig{file=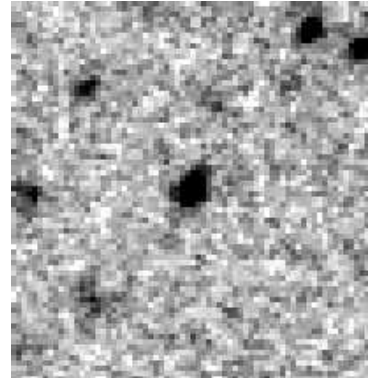,width=5.0cm,angle=-90}
\end{center}
\caption{$K$-band image of SDSS\,J222551.65$+$001637.7 taken with SOFI on the NTT. The star is in the centre and is clearly elongated, possibly indicating the system is becoming resolved.}
\label{222551_image}
\end{figure}

Since SDSS\,J2225$+$001 is a good candidate for a rare, previously unknown unresolved substellar companion to a white dwarf, independent $JHK$ photometry was obtained at the 3.5m NTT at La Silla in October 2007 (Section~\ref{UKIDSS:followup}). Indeed, the SOFI $JHK$ photometry confirms the NIR excess and is a good fit to the WD$+$L7 composite model. The SOFI $K$-band image of SDSS\,J2225$+$001 is shown in Figure~\ref{222551_image}, and shows that the system may indeed become resolved with a higher resolution instrument. The FWHM of this image is 1.8'' which equates to a projected orbital separation of $<350$\,AU.

\subsection{Examples II: Magnetic White Dwarfs}
\label{ukidss:magnetic}
 We detected 19 hydrogen atmosphere magnetic white dwarfs (DAHs) in the cross correlation between the EIS06 and MS99, and the UKIDSS DR8 database. All of these were found to fit their predicted NIR photometry to within 3$\sigma$ of their UKIDSS observed values, with the exception of SDSS\,J121209.31$+$013627.7 (See below). In an attempt to increase this sample size, we performed a further cross correlation between the UKIDSS DR8 database and the list of known magnetic white dwarfs from \citet{kawka07}. However, after accounting for duplicate entries between the catalogues, this only increased the sample by a single DAH star. This star turned out to exhibit a NIR excess and we discuss it below. 

\subsubsection{SDSS\,J121209.31$+$013627.7}
SDSS\,J1212$+$013 was first identified as a binary through an optical spectrum and a single photometric $J$-band detection, limiting the spectral type of the secondary to L5 or later \citep{schmidt05}. It was initially thought to be a detached system until \citet{debes06} detected strong cyclotron emission, an indication of mass transfer, in $K$-band time series photometry. Both \citet{koen06} and \citet{burleigh06b} detected optical variability (Period\,$\approx90$\,mins) consistent with that of a polar in a low accretion state. The NIR excess was successfully modelled by \citet{farihi08}, who modelled the system as a DAH $+$ a brown dwarf companion of spectral type L8. Although the evidence suggests that SDSS\,J1212$+$013 is a CV in a low state, the NIR observations do not rule out a detached system with accretion occurring via a wind. This system shall therefore be included in our statistical analysis.

\subsubsection{SDSS\,J125044.42$+$154957.4 $=$ WD\,1248$+$161}
WD\,1248$+$161 is a DAH white dwarf identified by \citet{vanland05} in the SDSS DR3. It has an estimated effective temperature of $T_{\rm eff}=10,000$\,K and a polar magnetic field strength $B_{p}=$20\,MG. The SDSS optical spectrum and photometry are a reasonable fit when compared to a blackbody model generated at the spectroscopic temperature of the white dwarf, with a clear excess in all four of the UKIDSS bands (Figure~\ref{125044}). We found that all of the UKIDSS data could not be matched effectively with any composite model, with a discrepancy existing between the $JH$ and the $K$-band photometry. This could be explained by the existence of strong cyclotron emission due to the magnetic field of the white dwarf, increasing the flux in the $K$-band. Therefore, the predicted absolute magnitude of the secondary was calculated using the $H$-band, which suggests a spectral type of M8. A composite WD$+$M8 model matches the UKIDSS $JH$ photometry. However, the $J$-band data must be interpreted with caution as it was flagged as 'noise' in the UKIDSS archive. We cannot rule out another source of this excess emission, such as the presence of a debris disk, and so NIR spectroscopy is required to discriminate between these possibilities.

The UKIDSS \it{mergedClass} \rm statistic for WD\,J1248$+$161 indicates the system is a point source, with a measured FWHM of $1.57\arcsec$ in the $H$-band. This equates to a projected orbital separation of $<245$\,AU for the putative secondary at the estimated distance to the white dwarf (Table~\ref{results}).

\subsection{Examples III: White Dwarfs with Debris Disks}

\subsubsection{SDSS\,J122859.92$+$104033.0}

SDSS\,J1228+104 was first identified by \citet{gansicke06} in DR6 of the SDSS, where they identified distinct emission lines of the Calcium 850-866\,nm triplet. The lines profiles of the Ca emission triplet show a double-peaked morphology, indicative  of a gaseous, rotating disk. The velocity of the Ca~II lines peaks indicated an outer radius of the disk of $\sim1.2$\rsun. The star was later investigated in the NIR by \citet{brinkworth09}, where it was discovered to have a large near- and mid-infrared excess, an indication of the presence of a dusty component in addition to that of the gas. This was the first evidence for the coexistence of a both a gaseous and dusty debris disk around a white dwarf.

We have recovered SDSS\,J1228+104 in this work as having a large $K$-band excess, as was expected from this system.

\subsubsection{SDSS~J132044.68$+$001855.0 = WD\,1318$+$005 }

The SDSS optical spectrum of WD\,1318$+$005 shows no evidence for a companion, and the UKIDSS $JH$ photometry can be matched with a white dwarf model for $T_{\rm eff} =19,649$K and log~$g =8.36$  (Figure~\ref{132044}). However, the UKIDSS $K$-band measurement is $3.1\sigma$ in excess of this model, and can be matched by the addition of a 600\,K blackbody to the white dwarf model. The UKIDSS \it{mergedClass} \rm statistic for WD\,1318$+$005 indicates the system is a point source, which rules out contamination due to a background galaxy. 

We obtained NIR photometry of WD\,1318$+$005 with IRIS2 at the 3.9m Anglo-Australian Telescope (Section~\ref{UKIDSS:followup}). These data are consistent with both the $K$-band excess emission and the white dwarf model within errors (Figure~\ref{132044}). WD\,1318$+$005 possibly possesses a warm debris disk, and if so then it is highly likely to be a DAZ white dwarf. Although it cannot be classified as a DAZ from the existing low resolution SDSS optical spectrum,  this star lies in the temperature regime of these objects. \cite{zuckerman03} suggest that $\approx25\%$ of all DAs in this temperature range are metal-polluted DAZs, and higher resolution and higher signal-to-noise data are now urgently required to search for  Ca\,II absorption.  

\subsubsection{SDSS\,J155720.77$+$091624.7}

SDSS\,J1557$+$091 shows no optical excess when fitted with an atmospheric model with physical parameters of $T_{\rm eff}=21990\pm403$\,K and log\,$g=7.67\pm0.06$ (EIS06). An excess is clearly evident in the $K$-band photometry which is best matched by the addition of a 700\,K blackbody, indicating a potential cool disk. However, we also note that there is a possible excess in the $Y$ and $J$-band photometry which can be best matched by the addition of a companion with spectral type L4. This is within 2$\sigma$ of the $H$-band flux and so a sub-stellar companion, or indeed another source, cannot be discounted as a possibility at this stage. The UKIDSS \it{mergedClass} \rm statistic for SDSS\,J1557$+$091 indicates the system is a probable point source (-2: 70\% probability that the source is stellar), with a measured FWHM of $0.51\arcsec$ in the $J$-band. This equates to a projected orbital separation of $<250$\,AU for a secondary star at the estimated distance to the white dwarf (Table~\ref{results}).

\begin{table*}[p]
\caption{Summary of proposed sources for the near-infrared excesses identified in UKIDSS. For the resolved systems, the probability of a chance alignment is given. Further information on systems with a putative companion is given in Table~8}
\begin{center}
\begin{tabular}{lclc}
\hline
SDSS\,J & WD & Proposed Source of Excess & Figure \\
\hline\hline
003902.47-003000.3 &    -      & Unresolved Companion                               & \ref{003902} \\
003923.04+003534.7 &    -      & Unresolved Companion                               & \ref{003923} \\
012032.27-001351.1 &    -      & Resolved Companion (P$_{\rm al}=5.8\times10^{-2}$) & \ref{012032} \\
013532.98+144555.8 & 0132+145  & Unresolved Companion                               & \ref{013532} \\ 
032317.00-002612.7 &    -      & Foreground/Background Contamination                & -            \\
033209.25+010722.1 &    -      & Foreground/Background Contamination                & -            \\
034221.55+005345.6 &    -      & Foreground/Background Contamination                & -            \\
075327.05+244705.5 &    -      & Debris Disk                                        & \ref{075327} \\
082705.04+284402.7 & 0824+288  & WD+dC+d2 \citep*{farihi10}				    & -            \\
085956.47+082607.5 &    -      & Unresolved Companion                               & \ref{085956} \\
090759.59+053649.7 &    -      & Unresolved Companion                               & \ref{090759} \\
092648.84+102828.8 &    -      & Unresolved Companion                               & \ref{092648} \\
093821.34+342035.6 &    -      & Unresolved Companion                               & \ref{093821} \\
100259.88+093950.0 &    -      & Unresolved Companion                               & \ref{100259} \\
101642.93+044317.7 &    -      & Unresolved Companion                               & \ref{101642} \\
103220.19+011227.0 &    -      & Foreground/Background Contamination                & -            \\
103448.92+005201.4 & 1032+011  & Unresolved Companion                               & \ref{103448} \\
103736.75+013912.2 &    -      & Unresolved Companion                               & \ref{103736} \\
110826.47+092721.5 &    -      & Unresolved Companion                               & \ref{110826} \\
113416.09+055227.2 &    -      & Unresolved Companion                               & \ref{113416} \\
115808.44-012312.9 & 1155-011  & Unresolved Companion                               & \ref{1155}   \\
120503.99+070031.0 &    -      & Foreground/Background Contamination                & -            \\
121209.31+013627.7 &    -      & DAH+L8 (Section~6.2.1)                                      & -            \\
122625.93+002700.2 & 1223+007  & Unresolved Companion                               & \ref{122625} \\
122859.92+104033.0 &    -      & Debris Disk (Section~6.3.1)                                 & -            \\
124455.15+040220.6 &    -      & Foreground/Background Contamination                & -            \\
125044.42+154957.4 & 1248+161  & Unresolved Companion (Magnetic WD)                 & \ref{125044} \\
131955.04+015259.5 & 1317+021  & Unresolved Companion                               & \ref{1317}   \\
132044.68+001855.0 & 1318+005  & Debris Disk                                        & \ref{132044} \\
132925.21+123025.4 &    -      & Unresolved Companion                               & \ref{132925} \\
134154.39+005600.8 & 1339+011  & Unresolved Companion                               & \ref{134154} \\
141448.25+021257.7 & 1412+024  & Debris Disk                                        & \ref{141448} \\
154431.47+060104.3 &    -      & Unresolved Companion                               & \ref{154431} \\
155720.77+091624.7 &    -      & Debris Disk                                        & \ref{155720} \\
162514.88+302610.8 & 	       & Unresolved Companion			            & \ref{162514} \\
220841.63-000514.5 &    -      & Resolved Companion (P$_{\rm al}=3.7\times10^{-4}$) & \ref{220841} \\
222030.68-004107.9 &    -      & Partially Resolved Companion (Steele et al. 2009)  & -            \\
222551.65+001637.7 &    -      & Partially Resolved Companion                       & \ref{222551} \\
231133.88+144119.9 &    -      & Foreground/Background Contamination                & -            \\
233345.97-000843.0 &    -      & Unresolved Companion                               & \ref{233345} \\

\hline
\end{tabular}
\end{center}
\label{specresults}
\end{table*}

\section{Low Mass Star or Brown Dwarf?}
\label{UKIDSS:starordwarf}
In order to assess whether or not each putative companion is a low mass main-sequence star or a brown dwarf an estimate of the mass of the secondary has been calculated. Firstly an age of the white dwarf was calculated by the addition of the white dwarf cooling age (Section~\ref{modelling}) and the main-sequence lifetime of the progenitor star. This was estimated using the initial-final mass relationship of \citet{dobbie06}, which is valid for initial masses $>1.6$\msun \citep{kalirai08}. An approximate main-sequence lifetime can then be calculated from the models of \citet{girardi00}. It should be noted that for white dwarfs where $M_{\rm WD}<0.5$\msun\, it is highly likely that the star has evolved through mass transfer and for these stars an age can not be calculated through this method. However, this can be seen as further evidence for the existence of a secondary star. For these stars, a lower limit on the mass of the secondary is estimated by using the cooling age of the white dwarf. 

\begin{figure*}
\begin{center}
\psfig{file=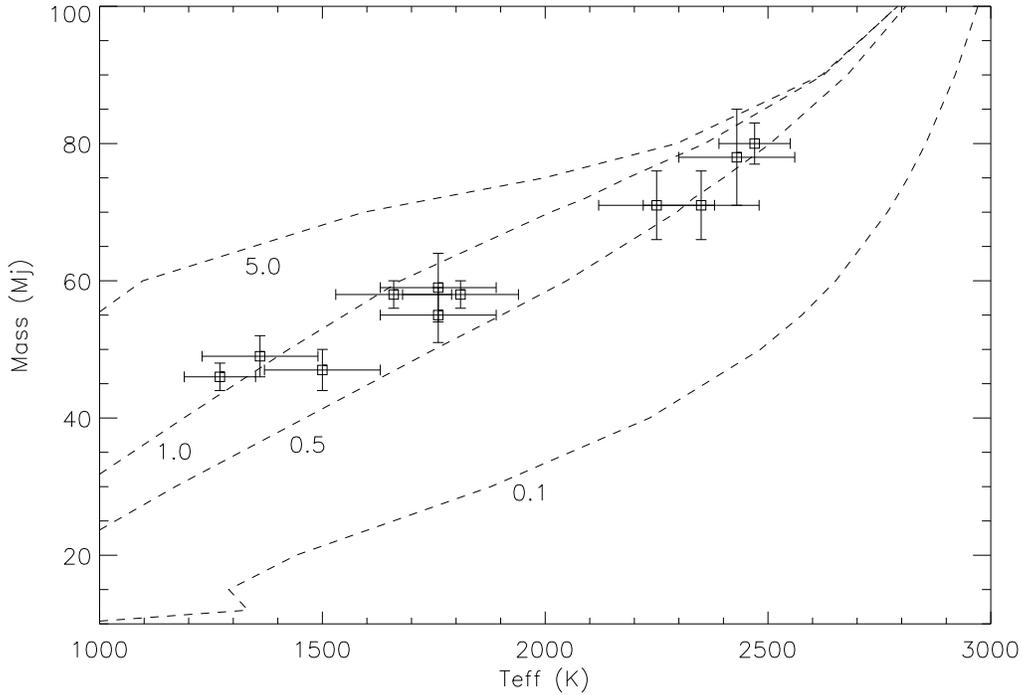,height=10.0cm}
\end{center}
\caption{Predicted masses of companions based on the Lyon group models. The dashed lines indicate constant age in Gyr.}
\label{mass_dusty}
\end{figure*}

We then estimate a mass for the secondary by interpolating the Lyon group atmospheric models \citep{chabrier00,baraffe02}, given the age of the white dwarf and an estimate of effective temperature of the companion. We estimated these temperatures by comparison with observed M, L and T dwarfs \citep{vrba04} and assuming an error of $\pm1$ spectral type. For the WD$+$dM binaries, we have estimated the masses by referring to the models of \citet{baraffe96}. The results are given in Table~\ref{results} and plotted in Figure~\ref{mass_dusty}.

If we assume a value of 80\mjup\, as a conservative cut-off mass for the stellar-substellar boundary, then 9-11 new DA white dwarf $+$ brown dwarf binaries (1 DA$+$L8 has been spectroscopically confirmed, see \citealt{steele09}). All of these have spectral types of M9 or later, with one candidate WD $+$ T dwarf.

\section{A Wide or Close Binary?}
\label{ukidss:wideorclose}
\cite{farihi05} suggest a bimodal distribution of separation for white dwarf + low mass secondary binaries. The UKIDSS \it{mergedClass} \rm statistic may provide us with evidence which supports this conclusion. The white dwarf PHL\,5038 was identified using this statistic as partially resolved, and was later fully resolved into a wide orbiting ($\approx55$\,AU) WD$+$L8 binary \citep{steele09}. Therefore, it may be possible to separate these stars into long and short period systems using this value. 

\begin{figure}
\begin{center}
\psfig{file=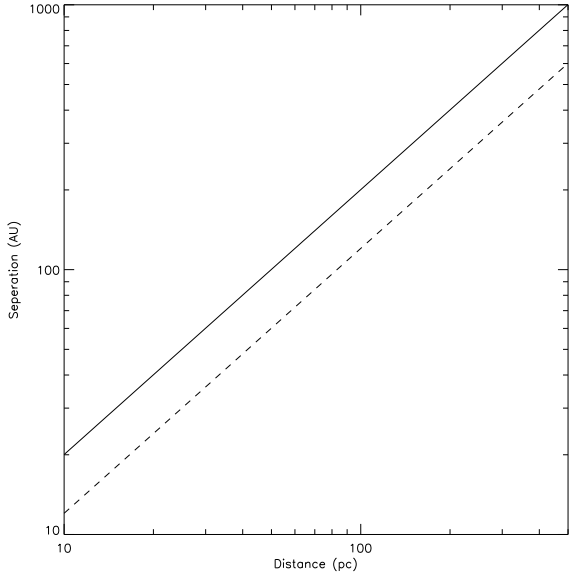,width=8.0cm,angle=0}
\end{center}
\caption{Distance verses projected orbital separation for a $2\arcsec$ (solid) and $1.2$\arcsec (dashed) radius.}
\label{resolve}
\end{figure}

In order to investigate this possibility, the resolving power of UKIDSS must be determined for the binary candidates in this survey. The UKIDSS photometry is measured using a $2\arcsec$ aperture in all bands. The projected orbital separation for this radius at distances from 10-100\,pc are plotted in Figure~\ref{resolve}. The same is also plotted for the average FWHM of the binary candidates ($\approx1.2$\arcsec). We see that for the majority of stars it will not be possible to distinguish between long and short period period systems if the system is classified as a point source, unless the binary has a particularly wide orbit. The maximum and projected orbital separations for all unresolved and resolved candidates in this survey are given in Table~\ref{results}.

\section{Limits on Unresolved Ultracool Companions to DA White Dwarfs}
Atmospheric modelling successfully fitted 521 DA white dwarfs to within 3$\sigma$ of both their UKIDSS $H$ and $K$-band photometry. We will use these stars to estimate the limits on the spectral types of unresolved companions to DA white dwarfs and hence determine the sensitivity of UKIDSS to both L and T dwarfs in white dwarf binaries. The method used to calculate these limits is detailed below. 


In order to place a limit on the spectral type of the coolest detectable unresolved companion to each DA white dwarf we required the absolute magnitudes for the spectral types of low mass objects ranging from L0-T8. \cite{patten06} lists known M, L and T dwarfs with 2MASS $JHK_{\rm S}$ photometry ranging from spectral types M5-T8 and also has measured parallaxes for many of these.  We converted the 2MASS $JHK_{\rm S}$ photometry for each object to the UKIDSS $JHK$ photometric system using the colour corrections of \citet{carpenter01}. Taking only those with a measured parallax, and thus an estimated distance, the $K$-band photometry was scaled to 10\,pc for each spectral type to obtain an estimated absolute $K$-band magnitude. For multiple stars of the same spectral type we used an average of the magnitudes. For the few spectral types that had no parallax measurement we performed a linear interpolation to predict an absolute magnitude. The $K$-band magnitude was used for the reasons discussed in Section~\ref{companions}. 

\begin{figure*}
\begin{center}
\psfig{file=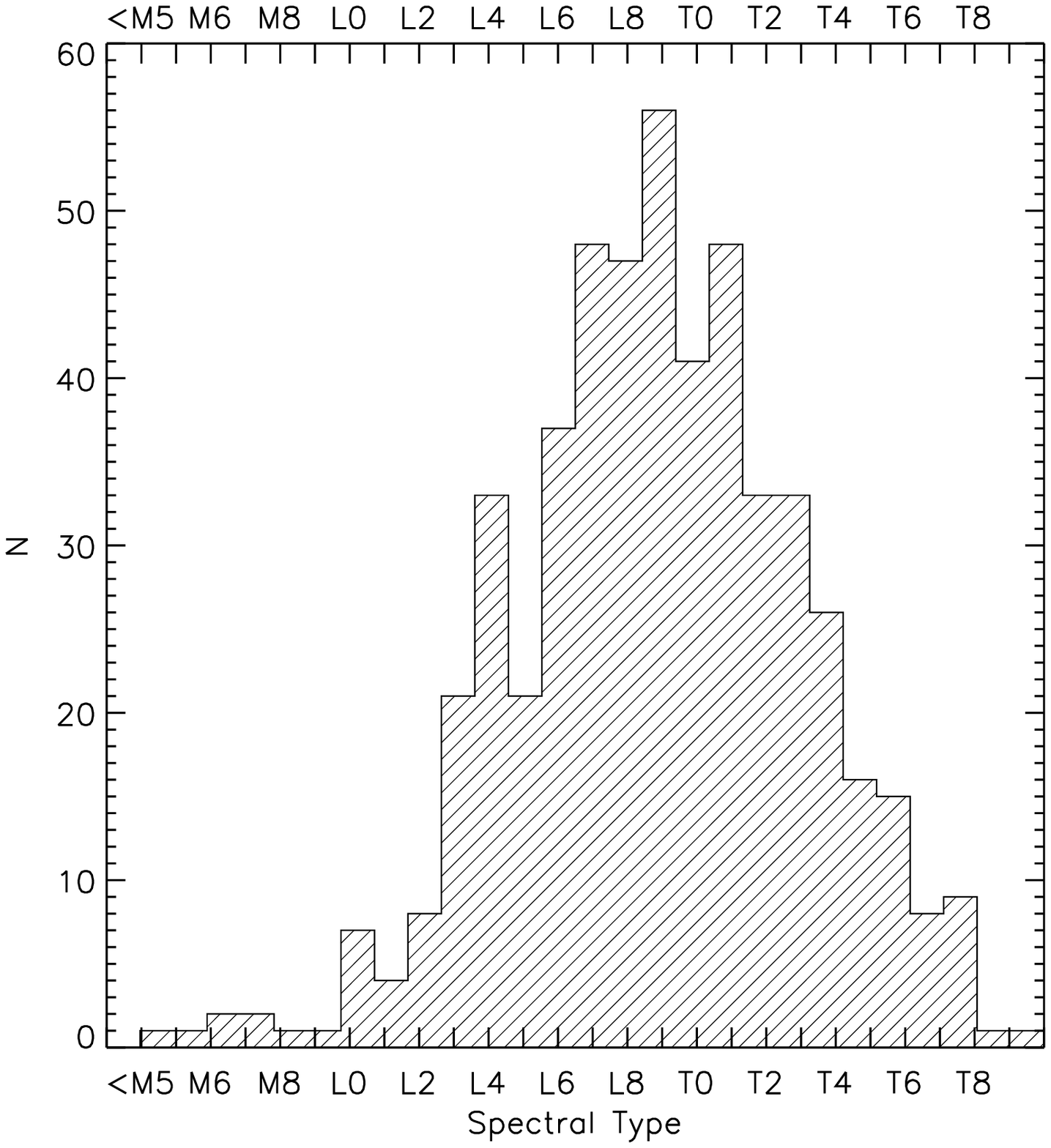,width=8.0cm,angle=0}
\psfig{file=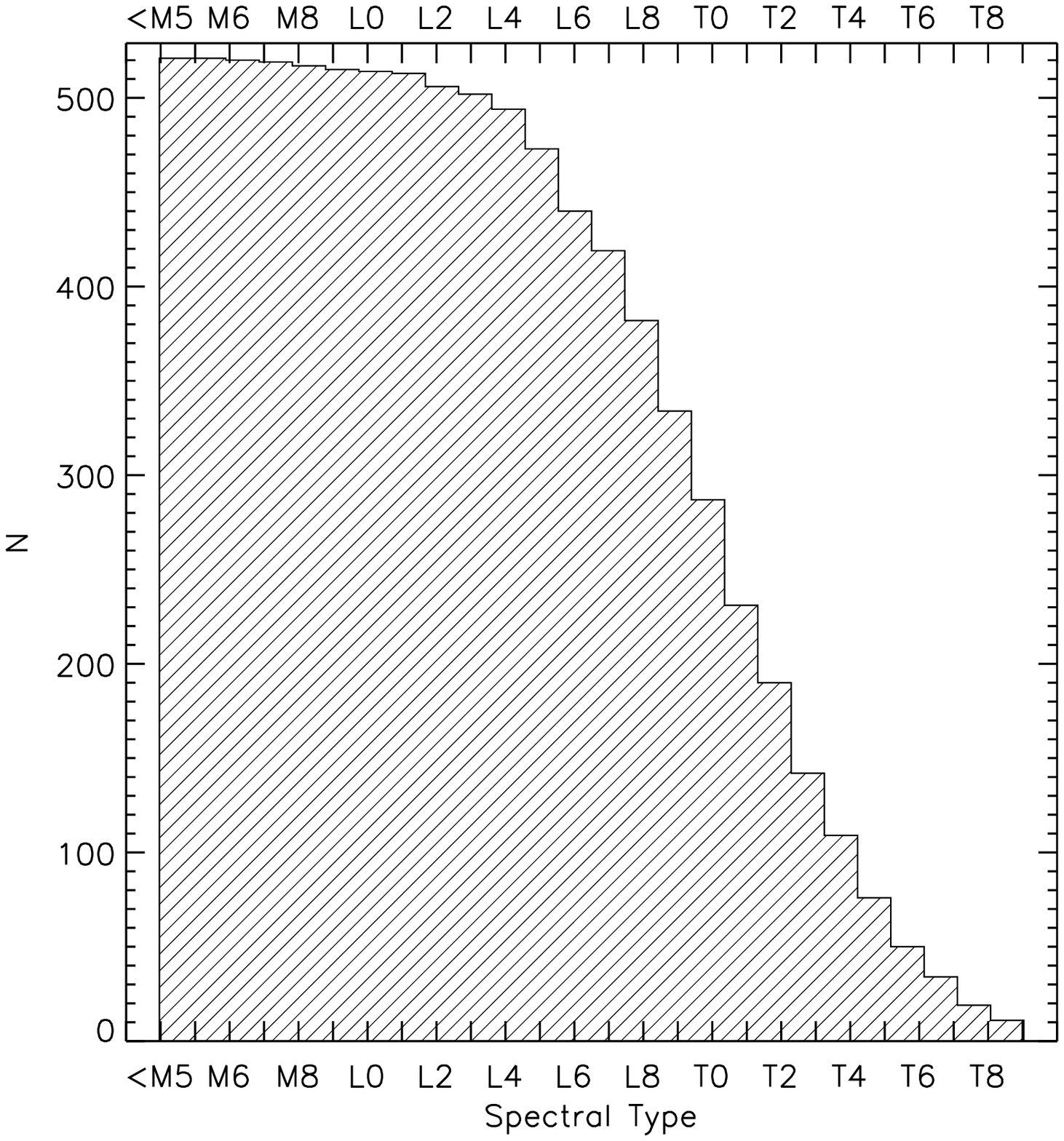,width=8.0cm,angle=0}
\end{center}
\caption{Left: Distribution of limits on unresolved companions to the 521 ``single'' DA white dwarfs (with well determined parameters) detected in UKIDSS DR8. Right: Plotted as a cumulative distribution.}
\label{limits}
\end{figure*}

For each apparently single DA white dwarf we calculated an expected $K$-band magnitude (Section~\ref{modelling}). The observed $K$-band magnitude was not utilised, as any of the white dwarfs could be in excess within the observed errors. This would have resulted in an over-estimate of the limiting spectral type for the companion to these stars.  We then calculated a 3$\sigma$ detection limit for each white dwarf by using the observed errors listed in UKIDSS DR8, and adding these to the predicted magnitudes. The distance to each white dwarf was calculated using the Bergeron models (Section~\ref{modelling}) using the effective temperatures and surface gravities produced from the automated fit of EIS06. For the single DA white dwarfs only found in MS99, temperature estimates were obtained from the literature or an automated fit to the stars optical photometry. For stars without a measured surface gravity, we assumed a log\,$g=8.0$. This applied to only 11 white dwarfs in the sample, which is not a large enough number to create a distance bias that could later affect the statistics. Subsequently, the grid of M, L and T dwarf absolute magnitudes were scaled to the estimated distance of each white dwarf and added to the predicted white dwarf $K$-band magnitudes until a match was made with the 3$\sigma$ detection limit. The spectral types of all the companion limits were summed (Table~\ref{limits_table}) and plotted as a histogram and a cumulative histogram in Figure~\ref{limits}.

\begin{table*}
\caption{Table of results for SDSS white dwarfs with NIR excess indicative of a companion: most likely companion spectral types and mergedClass stats indicating the possibility that a system is a widely separated or a close detached binary.}
\begin{center}
\begin{tabular}{ccccccccc}
\hline
SDSS\,J & WD &  Spec. Types & MergedClass & M$_{\rm WD}$ (M$_{\rm s}$) &  d (pc) & Age (Gyr) & M$_{\rm COMP}$ (M$_{\rm Jup}$) & a (AU) \\
\hline\hline
101642.93$+$044317.7    & -                     & DA$+$dM3      & -2    & $0.40\pm0.01$  &  $660\pm30$      & $>0.01$                   & $350\pm150$ & $<530$ \\
110826.47$+$092721.5    & -                     & DA$+$dM5      & -1    & $1.18\pm0.02$  &  $90\pm5$        & $2.6_{-0.1}^{+0.1}$         & $113\pm23$  &  $<63$  \\
162514.88$+$302610.8    & -                     & DA$+$dM5      & -1    & $0.42\pm0.02$  &  $680\pm90$      & $>0.0004$                &  $113\pm23$  & $<545$ \\
085956.47$+$082607.5    & -                     & DA$+$dM6      & -1    & $0.36\pm0.02$  &  $400\pm20$      & $>0.1$                    & $108\pm18$  & $<650$ \\
113416.09$+$055227.2    & -                     & DA$+$dM6      & -1    & $0.57\pm0.04$  &  $580\pm50$      & $1.6_{-0.4}^{+0.7}$       & $108\pm18$  & $<510$ \\
012032.27$-$001351.1    & -                     & DA$+$dM7      & -1    & $0.56\pm0.12$  &  $410\pm70$      & $2.2_{-1.1}^{+5.8}$       & $93\pm8$    & $\approx910$ \\
103736.75$+$013912.2    & -                     & DA$+$dM7      & -1    & $0.49\pm0.02$  &  $330\pm10$      & $3.7_{-0.8}^{+1.1}$       & $93\pm8$    & $<330$ \\
115808.44$-$012312.9    & 1155$-$011            & DA$+$dM7      & -1    & $\approx0.60^{1}$  &  $280\pm10$  & $\approx1.8^{1}$          & $93\pm8$    & $<270$ \\
134154.39$+$005600.8    & 1339$+$011            & DA$+$dM7      & +1    & $0.58\pm0.02$  &  $370\pm20$      & $1.5_{-0.2}^{+0.3}$        & $93\pm8$    & $<480$  \\
132925.21$+$123025.4 	& -          		& DA$+$dM7  	& -1 	& $0.36\pm0.01$  &  $210\pm10$      & $>0.2$                    & $93\pm8$    & $<235$ \\
233345.97$-$000843.0    & -                     & DA$+$dM7      & -1    & $0.83\pm0.18$  &  $390\pm100$     & $1.1_{-0.4}^{+1.2}$       & $93\pm8$    & $<350$ \\
092648.84$+$102828.8 	& -          		& DA$+$dM8  	& -1    & $0.68\pm0.17$  &  $480\pm110$     & $1.0_{-0.5}^{+2.3}$	& $85\pm5$    & $<800$ \\
125044.42$+$154957.4    & 1248$+$161            & DAH$+$dM8     & -1    & $\approx0.60^{1}$ &  $150\pm10$   & $>0.6$                    & $85\pm5$    & $<245$ \\
003923.04$+$003534.7 	& -          		& DA$+$dM9  	& -1	& $0.60\pm0.06$  &  $230\pm20$      & $1.8_{-0.5}^{+0.9}$       & $80\pm3$    & $<263$ \\
003902.47$-$003000.3 	& -           		& DA$+$dL0  	& -1	& $0.34\pm0.02$  &  $390\pm30$	    & $>0.3$	                & $>80$       & $<484$ \\ 
100259.88$+$093950.0 	& -            		& DA$+$dL0  	& -1	& $0.57\pm0.03$  &  $540\pm30$	    & $1.6_{-0.3}^{+0.5}$       & $78\pm7$    & $<420$ \\
131955.04$+$015259.5    & 1317$+$021            & DA$+$dL1      & -1    & $\approx0.60^{1}$ & $290\pm20$    &  $\approx1.3^{1}$       & $71\pm5$    & $<130$   \\
220841.63$-$000514.5    & -                     & DA$+$dL1      & +1    & $0.78\pm0.08$  &  $170\pm20$      & $1.5_{-0.3}^{+0.6}$       & $71\pm5$    & $\approx290$ \\
090759.59$+$053649.7 	& -          		& DA$+$dL4  	& -1	& $0.54\pm0.02$  &  $360\pm20$	    & $2.1_{-0.4}^{+0.5}$       & $58\pm2$    & $<600$ \\
013532.98$+$144555.8 	& 0132$+$145            & DA$+$dL5  	& -1	& $0.65\pm0.03$  &  $60\pm10$	    & $2.1_{-0.2}^{+0.3}$       & $58\pm2$    & $<57$  \\ 
093821.34$+$342035.6    & -                     & DA$+$dL5      & -1    & $0.61\pm0.08$  &  $160\pm20$      & $2.2_{-0.7}^{+1.4}$       & $59\pm5$    & $<154$  \\
103448.92$+$005201.4    & 1032$+$011            & DA$+$dL5      & -1    & $0.68\pm0.10$  &  $200\pm30$      & $1.4_{-0.4}^{+0.9}$       & $55\pm4$    & $<150$ \\
222551.65$+$001637.7 	& -          		& DA$+$dL7 	& +1	& $0.70\pm0.06$  &  $190\pm20$	    & $1.2_{-0.3}^{+0.4}$       & $47\pm3$    & $<350$ \\
121209.31$+$013627.7    & -                     & DAH$+$dL8     & -1    & $\approx0.60^{2}$ & $\approx150^{2}$ & $1-5^{2}$              & $\approx50^{2}$ & $\approx0.003$ \\  
222030.68$-$004107.9 	& -          		& DA$+$dL8  	& +1 	& $0.77\pm0.04$  &  $60\pm10$	    & $2.1_{-0.3}^{+0.4}$       & $49\pm3$    & 55 \\ 
154431.47$+$060104.3 	& -          		& DA$+$dT3  	& -1 	& $0.85\pm0.09$  &  $90\pm10$	    & $2.6_{-0.8}^{+0.8}$	& $46\pm2$    & $<190$ \\   
\hline
\end{tabular}
\end{center}
$^{1}$Surface gravity estimated from mass distribution of white dwarfs.\\
$^{2}$\citet{farihi08}
\label{results}
\end{table*}

\begin{table}
\caption{Number of each spectral type obtained as limits to the 'single' DA white dwarf population in UKIDSS DR8.}
\begin{center}
\begin{tabular}{cc|cc|cc}
\hline
Spectral Type & No. & Spectral Type & No. & Spectral Type & No. \\
\hline\hline
-      &   -  & L0  & 1  & T0      & 56 \\
-      &   -  & L1  & 7  & T1      & 41 \\
-      &   -  & L2  & 4  & T2      & 48 \\
-      &   -  & L3  & 8  & T3      & 33 \\
$<$M5  &   0  & L4  & 21 & T4      & 33 \\
M5     &   1  & L5  & 33 & T5      & 26 \\
M6     &   1  & L6  & 21 & T6      & 16  \\
M7     &   2  & L7  & 37 & T7      & 15  \\
M8     &   2  & L8  & 48 & T8      & 8  \\
M9     &   1  & L9  & 47 & $>$T8   & 11  \\ 
\hline
\end{tabular}
\end{center}
\label{limits_table}
\end{table}

It should be noted that for the isolated white dwarf sample, the distances might be under-estimated if they are, for example, unresolved double degenerate binaries. Thus, for this fraction ($\approx5$\% of all WDs are unresolved WD$+$WDs, \citealt{holberg09}), the sensitivity to ultracool companions may have been overestimated. It is also important to spectroscopically confirm the candidate secondary in each candidate binary system for the same reason i.e. in case the distance to the white dwarf has been under-estimated, and thus the temperature and mass of the companion are under-estimated.

\section{Survey Sensitivity to Unresolved Companions}
\label{DISCUSSION:sensitivity}
\begin{figure*}
\begin{center}
\psfig{file=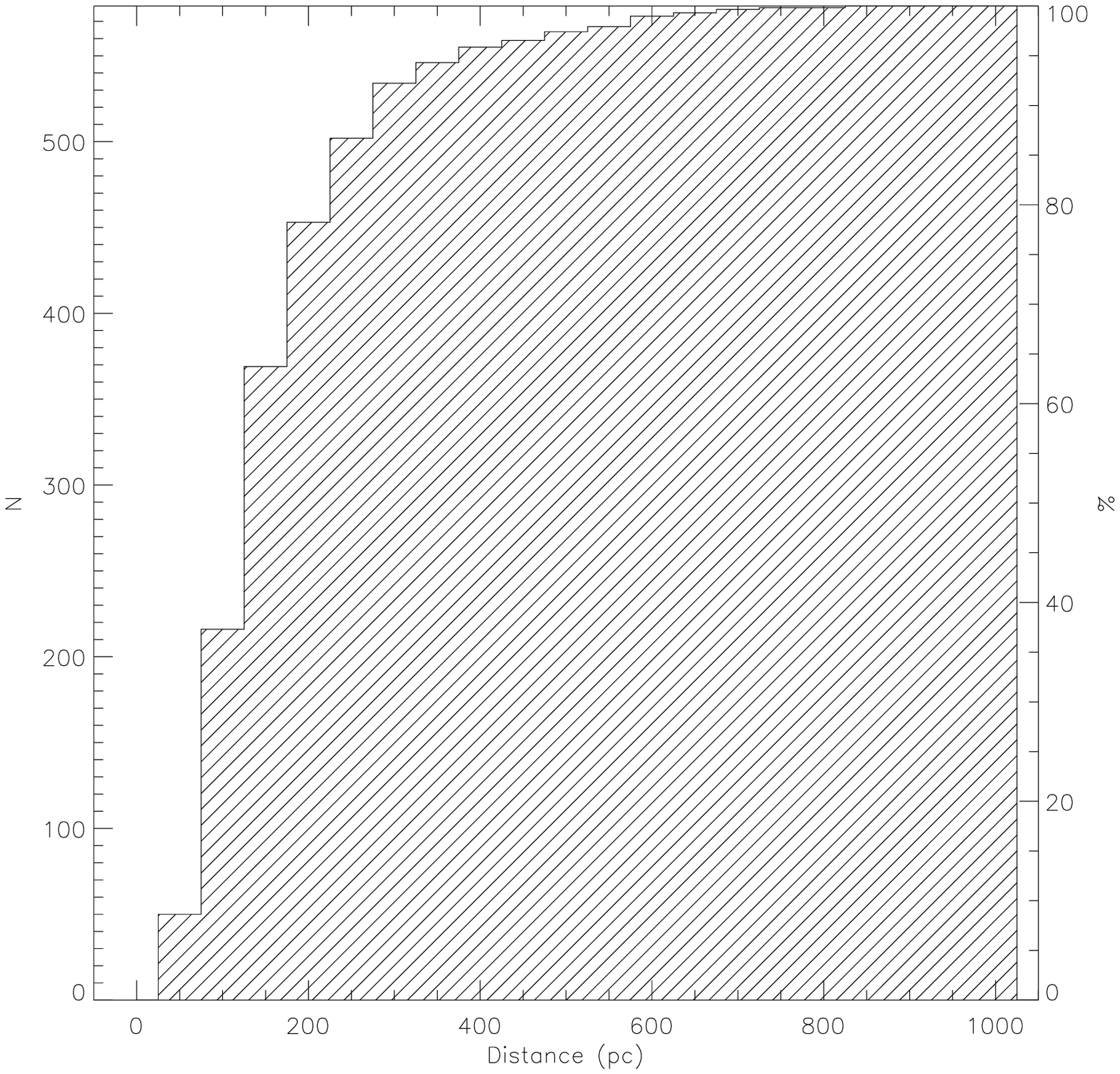,width=7.5cm,angle=0}
\psfig{file=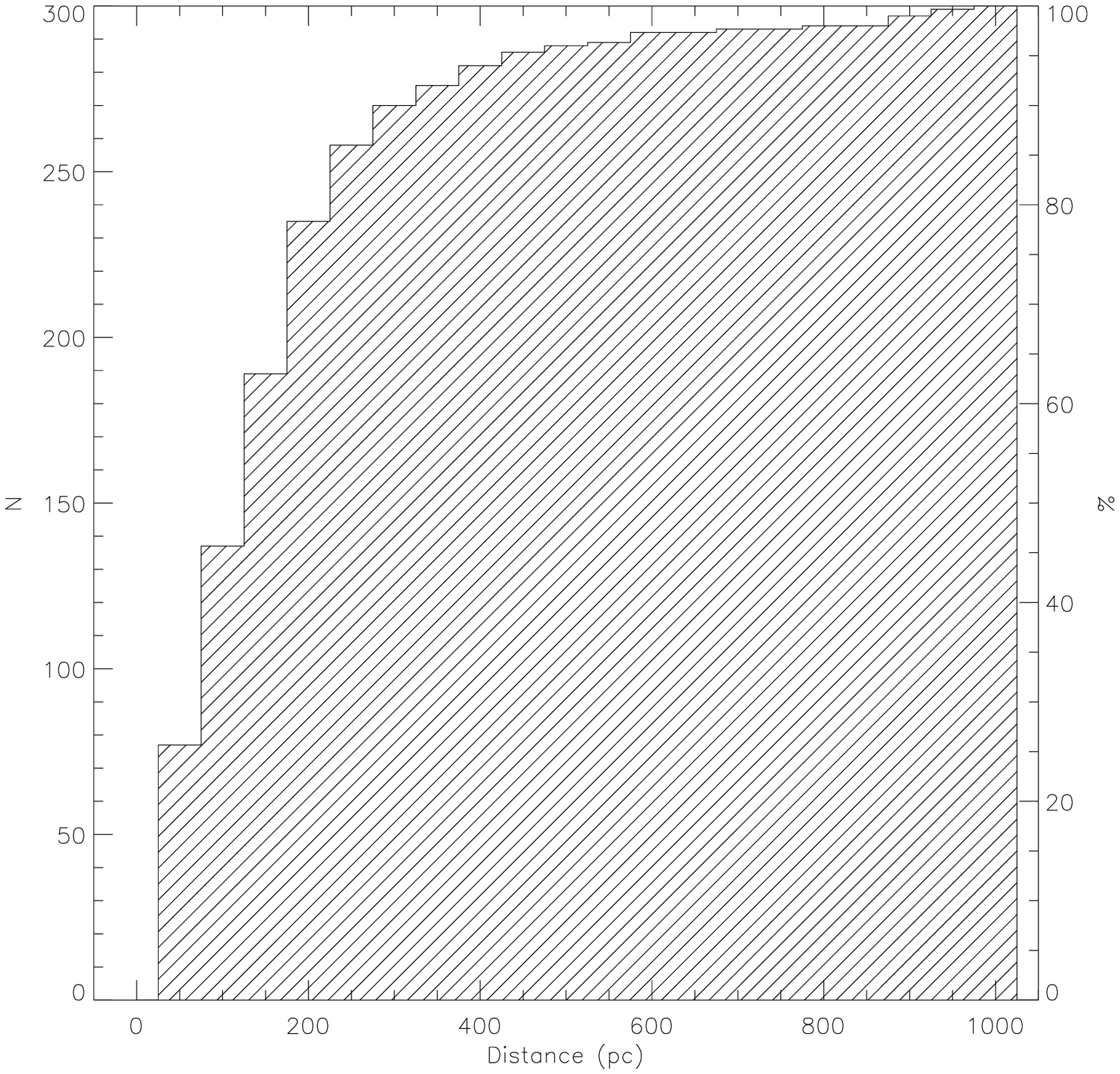,width=7.5cm,angle=0}
\end{center}
\caption{Left: Cumulative distribution of distances to the isolated DA white dwarfs in this work. Right: Cumulative distribution of distances to the DA$+$dM binaries in this work.}
\label{distance_histo}
\end{figure*}

Figure~\ref{limits} shows that the sensitivity of this survey for the isolated DA white dwarf sample peaks roughly at the L-T borderline, at approximately L8-T2. Figure~\ref{limits} also demonstrates that the sensitivity function then falls off as the limits approach a spectral type of late T, with only 43\% of the sample sensitive to the detection of a companion earlier than or equal to spectral type T0. In contrast 99\% of the sample is sensitive to the detection of a companion of spectral type earlier than or equal to L0, dropping to 51\% by L9. Figure~\ref{distance_histo}, shows a histogram of the distances to these single white dwarfs. 98\% of the sample lies within 500\,pc of the Sun.

In considering the total survey sensitivity, we must take the the 274 DA$+$dM binaries that fell onto our model grid into account. These must be included in the final sample for statistical analysis as many would have been sensitive to the detection of L and T-type companions, if the M dwarf companion were to be removed. However, the presence of the M dwarf companions may bias this population towards greater distances, as the secondary is far brighter in the near-infrared. Therefore, the fraction of DA$+$dM binaries that lie within 500\,pc must be determined so a direct comparison can be made against the isolated white dwarf population. The distances to these were estimated using the effective temperatures and surface gravities assigned by EIS06. We plot this data as a histogram in Figure~\ref{distance_histo}, with 93\% of the DA$+$Ms lying within 500\,pc. Therefore, there is not a significant difference between the isolated and white dwarf $+$ M dwarf distance distributions.

\section{Binary Statistics for White Dwarfs with Unresolved Substellar Companions}
\label{DISCUSSION:stats}
In this section we are going to estimate the fractions of DA white dwarfs with unresolved L dwarf, T dwarf and brown dwarf companions using the results of this survey.

We assume that the white dwarf $+$ L and T dwarf candidates, and the candidate disk systems, follow the same distance distribution as the single white dwarfs, as shown in the previous section. The new candidate white dwarf $+$ M dwarf systems can be added to the previously identified WD$+$dM binary sample. Therefore, the complete sample size of DA white dwarfs at any distance, with or without a near-infrared excess is 794. The number of these stars that lie within 500\,pc is 98\% of the total sample size giving 778 DA white dwarfs. 

We can now calculate an upper limit for the unresolved L dwarf, T dwarf and brown dwarf companion fractions to DA white dwarfs in UKIDSS. This will be done firstly for the DA $+$ L dwarfs, followed by the DA $+$ T dwarfs and then finally for the DA $+$ brown dwarf population.

The candidate L and T dwarfs have been identified through NIR photometry with only PHL\,5038 and SDSS\,J1212$+$013 confirmed as harbouring brown dwarf companions. PHL\,5038B has been confirmed spectroscopically \citep{steele09}, and the substellar status of secondary to SDSS\,1212$+$013 is supported by a number of convincing arguments \citep{farihi08}. However, due to its high proper motion, we have determined that the source of the near-infrared excess to NLTT\,5306 is directly associated with the star, and not due to a background object (Section~\ref{NLTT5306}). Therefore, we will assume that at least 3 of the detections in this survey are confirmed white dwarf $+$ brown dwarf binary systems containing L dwarf secondaries. These confirmed brown dwarfs can be used to set a lower limit on the unresolved L dwarf and brown dwarf companion fractions. 

Table~\ref{stats_table} shows the percentage of the sample sensitive to the detection of L dwarfs equal to or earlier than the specified spectral type, and the number of each spectral type potentially detected in this survey. In order to calculate the white dwarf $+$ L dwarf binary fraction, we are going to determine the effective number of detections. This value represents the number of each spectral type detected if the survey were 100\% efficient at detecting companions of all spectral types. The effective number of detections can then be calculated by dividing the actual number of detections by the sensitivity to each spectral type. For example, if the survey sensitivity to companions of spectral types earlier or equal to L0 was 50\%, and 1 L0 was detected, then the effective number of detections is 2. This is then summed for all spectral types to give the total effective number of L dwarfs detected:

$$n_{\rm eff}=\frac{2}{0.99}+\frac{2}{0.98}+\frac{0}{0.97}+\frac{0}{0.96}+\frac{1}{0.95}+\frac{3}{0.91}+\frac{0}{0.84}$$
$$+\frac{1}{0.80}+\frac{2}{0.73}+\frac{0}{0.64}=12.4$$\\

\noindent The L dwarf companion fraction is then the effective number of detections divided by the sample total. An error can be estimated by taking the square root of the reciprocal number of effective detections. Thus, an upper limit to the fraction of DA white dwarfs with unresolved L dwarf companions is $f_{\rm WD+dL}\leq1.6\pm0.5$\%. A lower limit can be estimated by assuming only 3 candidates (PHL\,5038, SDSS\,1212$+$013 and NLTT\,5306) are real. This gives a final range of  $0.4\leq f_{\rm WD+dL}\leq2.1$\%. 

Table~\ref{stats_table} also gives the percentage of the sample sensitive to the detection of T dwarfs equal to or earlier than the specified spectral type, and the number of each spectral type potentially detected in this survey. An upper limit to the fraction of white dwarf with T dwarf companions is then calculated as detailed previously giving $f_{\rm WD+dT}\leq0.8$\%. Although only one candidate T dwarf has been detected, an average over the entire T dwarf spectral range is a good start under these circumstances but ultimately, a more complex Bayesian analysis would yield a better result. Additional candidates from future surveys with greater sensitivity and a larger search field will greatly improve the statistics.

In order to determine the fraction of white dwarfs with brown dwarf companions, we must first estimate the spectral type where the secondary becomes truly substellar. This depends on the age of the secondary, which can be estimated using the total age (main-sequence lifetime $+$ cooling age) of the white dwarf primary. The average age of the white dwarf sample is 1.9$\pm0.7$\,Gyr. After $\approx0.1$\,Gyr brown dwarfs cool very slowly, so the average sample age can be used to estimate a spectral type where the stellar/substellar borderline occurs. Taking the upper limit on brown dwarf mass as 80\mjup, then an upper limit for the brown dwarf spectral type can be taken as L0 (Figure~\ref{mass_dusty}). 

An upper limit to the fraction of white dwarf with brown dwarf companions is then calculated as detailed previously giving $f_{\rm WD+BD}\leq2.1\pm0.5$\%. A lower limit can be estimated by assuming only 3 candidates (PHL\,5038, SDSS\,1212$+$013 and NLTT\,5306) are real. This gives a final range of  $0.4\leq f_{\rm WD+BD}\leq2.6$\%.  


Although these statistics are not particularly robust, they are suitable as a first approximation using the small numbers available. In order to improve upon these numbers, the sample needs to be enlarged. In the first instance, this will be done by UKIDSS, which is set to be completed by 2012. Looking ahead, future infrared surveys with VISTA and WISE will add to the white dwarf $+$ ultracool companion sample significantly.

It should also be noted that the calculations presented here underestimates the total fraction of white dwarfs with ultracool companions, as this search does not include widely orbiting, resolved companions. Searches for these systems are already being undertaken \citep{dayjones08}. Due to the aperture size used in analysing UKIDSS photometry ($2\arcsec$), we have found a small number of such systems as part of this work. This is due to the \it{mergedClass} \rm statistic which,  as noted previously, provides an indication of whether a source is a point or extended source (Section~\ref{ukidss:wideorclose}). In the first instance this value can be used to remove contaminating galaxies from a search, but it can also be an indication that a system is partially resolved (e.g. PHL\,5038). Thus, for this survey wide companions have only been discovered if they happen to be partially resolved in the UKIDSS database. 

\begin{table*}
\caption{Actual number and effective number of detections for DA white dwarf $+$ dL/dT binaries in the UKIDSS DR8 sample based on sensitivity estimates. No candidate companions later than spectral type T3 were detected.}
\begin{center}
\begin{tabular}{cccccc}
\hline
Spectral  & No. Isolated WDs       & \% Detectable of            & Actual No.     & Effective No. & Sample \%  \\
Type        & Sensitive to Spectral  &      Spectral Type           & Candidates   & Detected        &                      \\
                 & Type or Earlier            &       or Earlier                   & Detected        &                         &               \\
\hline\hline
L0 & 514 &  99\% & 2 & 2.0 & 0.3\% \\ 
L1 & 513 &  98\% & 2 & 2.0 & 0.3\% \\ 
L2 & 506 &  97\% & 0 & 0.0 & 0.0\%  \\
L3 & 502 &  96\% & 0 & 0.0 & 0.0\%  \\
L4 & 494 &  95\% & 1 & 1.1 & 0.1\% \\ 
L5 & 473 &  91\% & 3 & 3.3 & 0.4\% \\ 
L6 & 440 &  84\% & 0 & 0.0 & 0.0\%  \\
L7 & 419 &  80\% & 1 & 1.3 & 0.2\% \\
L8 & 382 &  73\% & 2 & 2.7 & 0.3\% \\
L9 & 334 &  64\% & 0 & 0.0 & 0.0\%  \\
T0 &  287 & 55\% & 0 & 0.0 & 0.0\% \\
T1 &  231 & 44\% & 0 & 0.0 & 0.0\% \\
T2 &  190 & 37\% & 0 & 0.0 & 0.0\% \\
T3 &  142 & 27\% & 1 & 3.7 & 0.5\% \\
T4 &  109 & 21\% & 0 & 0.0 & 0.0\% \\
T5 &  76  & 15\% & 0 & 0.0 & 0.0\% \\
T6 &  50  & 10\% & 0 & 0.0 & 0.0\% \\
T7 &  34  & 7\%  & 0 & 0.0 & 0.0\% \\
T8 &  19  & 4\%  & 0 & 0.0 & 0.0\% \\

\hline
\end{tabular}
\end{center}
\label{stats_table}
\end{table*}

\section{Correcting for Completeness}
Some of the white dwarfs with a near-infrared excess may not have been detected in UKIDSS if the companion were to be removed. We can correct for this bias by removing those stars with predicted $K$-band magnitudes which are outside the $5\sigma$ detection limits of the survey. We can then repeat the statistics using this new, magnitude limited sample. After applying this constraint, we are left with 413 ``single'' DA white dwarfs and 151 previousley identified DA+dM binaries, giving a total sample size of 564. Only 3 of the white dwarf $+$ brown dwarf candidates (SDSS\,J0135$+$144, PHL\,5038 and SDSS\,J1544+060) are within the magnitude limited sample, giving the following binary fractions; $f_{\rm WD+dL}\geq0.4\pm0.3$\%, $f_{\rm WD+dT}\geq0.2$\% and $f_{\rm WD+BD}\geq0.5\pm0.3$.

\section{Comparison to Previous Estimates}
\subsection{The White Dwarf $+$ L dwarf Binary Fraction}

\begin{figure}
\begin{center}
\psfig{file=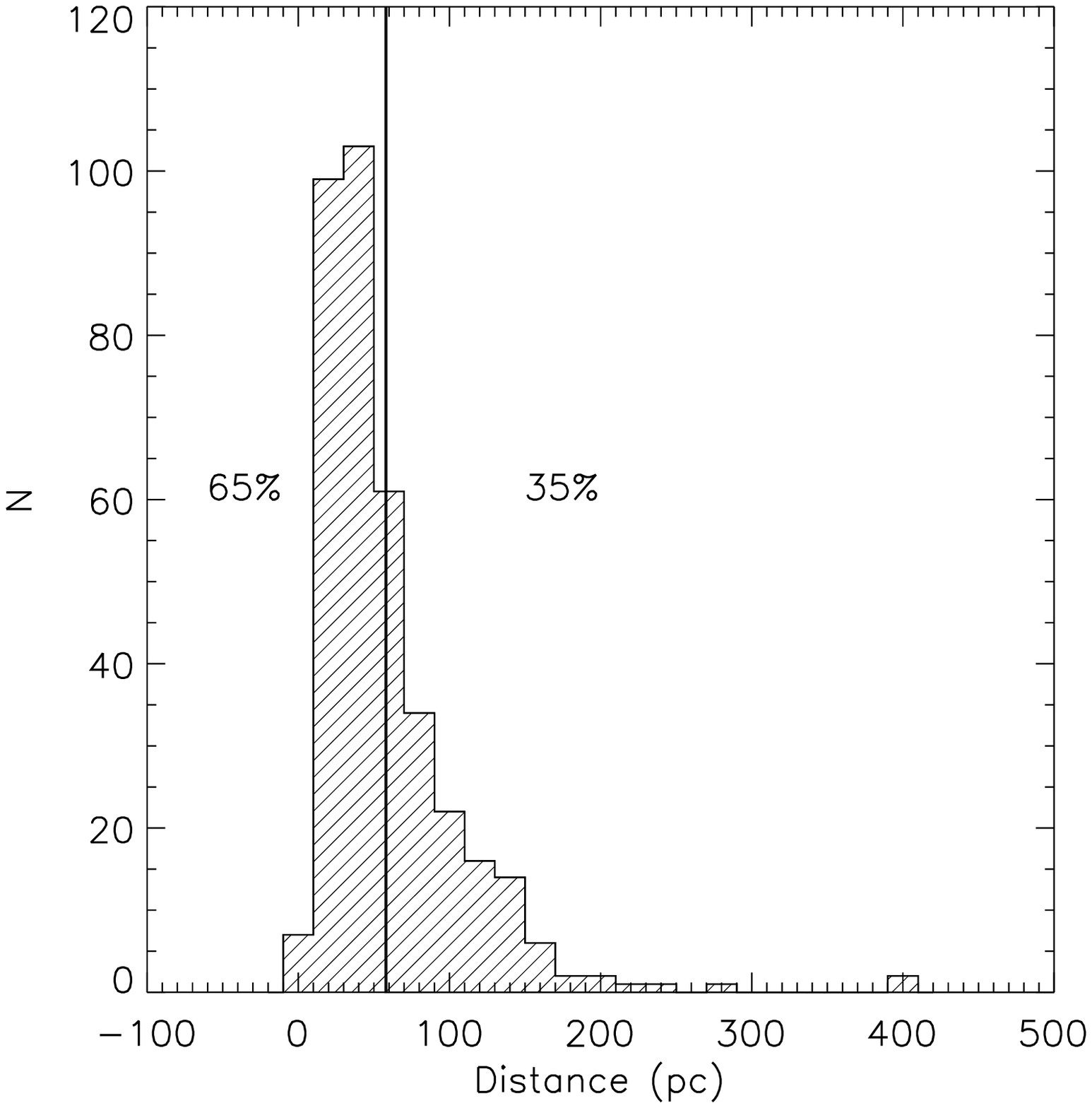,width=8.0cm,angle=0}
\end{center}
\caption{Distance distribution of the white dwarfs presented in Farihi et al. (2005). The mean distance is shown as a solid line.}
\label{jay_dist}
\begin{center}
\psfig{file=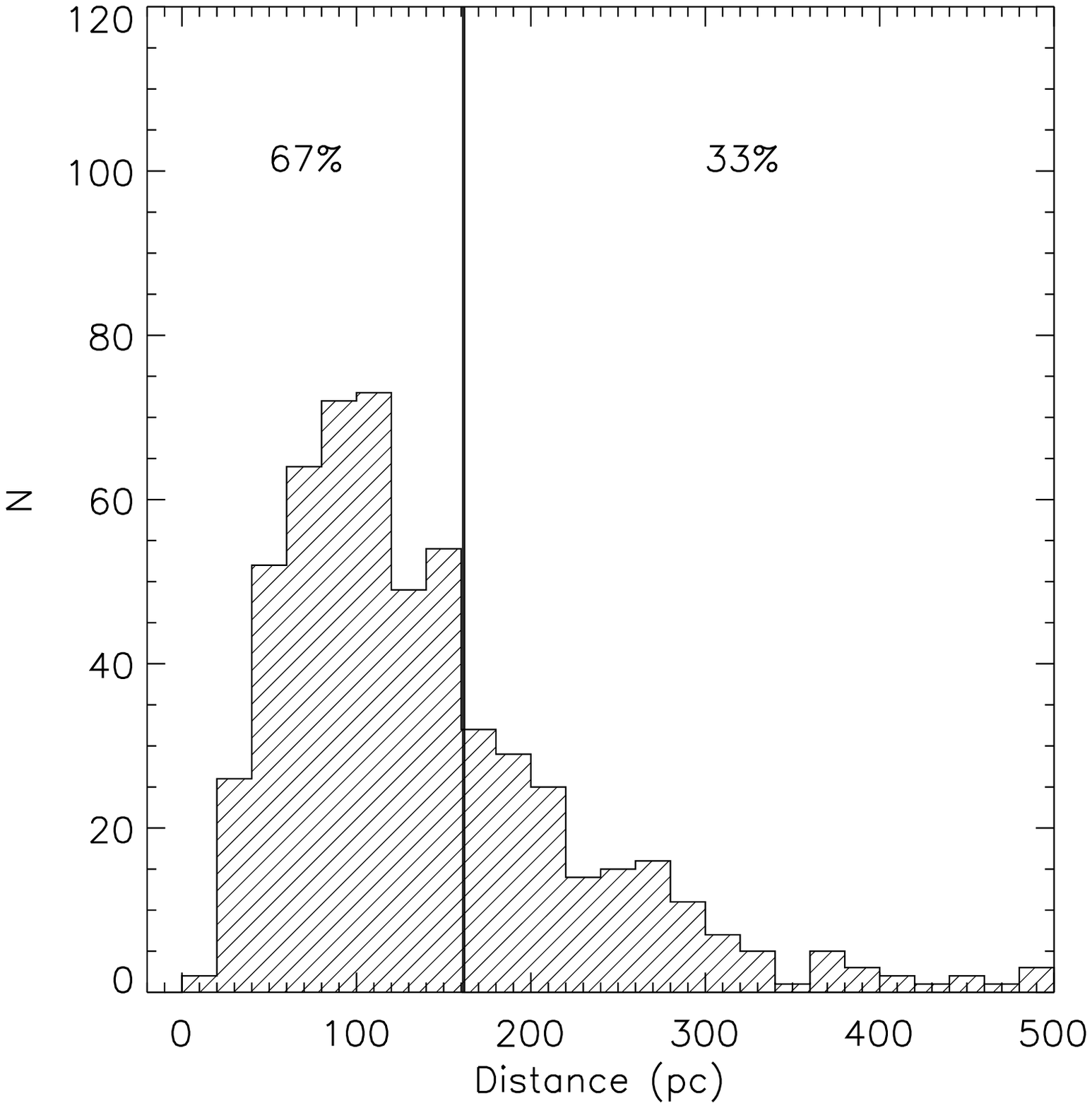,width=8.0cm,angle=0}
\end{center}
\caption{Distance distribution of the single DA white dwarfs present in the UKIDSS sample. The mean distance is shown as a solid line.}
\label{ukidss_dist}
\end{figure}

\citet{farihi05} (hereafter FBZ05) estimated the white dwarf $+$ L dwarf binary fraction as $f_{\rm WD+dL}\leq0.5$\% from a NIR study of 371 white dwarfs with 2MASS photometry. FBZ05 estimated their average sensitive to unresolved companions to be L8, the same as the peak of the sensitivity in the UKIDSS data. 
In order for the L dwarf fraction from UKIDSS ($0.4\leq f_{\rm WD+dL}\leq2.1$\%) to agree with the FBZ05 result either (1) All of the remaining photometric candidates presented here are not WD$+$dL binaries i.e. the lower limit of the calculated L dwarf fraction is more likely to be correct, or (2) FBZ05 have under-estimated the fraction of white dwarfs with L dwarfs. As the first candidate for which spectroscopy was obtained (PHL\,5038) was confirmed to be real, and in other cases the photometric classification remains quite convincing, then explanation (1) seems initially less likely than (2). To investigate this further, we will perform a detailed examination as to how FBZ05 arrived at their L dwarf companion fraction. We will then follow the same method for the UKIDSS sample, and make a direct comparison to the detailed sensitivity analysis discussed previously.

It should be noted that the FBZ05 sample includes white dwarfs not of spectral type DA. However, these are a minority in the sample, and their removal would not significantly affect the fraction calculated by FBZ05.

Figures~\ref{jay_dist} and ~\ref{ukidss_dist} show the distance distributions of both surveys, with the mean distances plotted (FBZ05 $=$ 57\,pc and UKIDSS $=$ 161\,pc). FBZ05 calculate the detection limit of their survey by determining the ``average'' white dwarf from their sample (found by calculating the average temperature, $T_{\rm eff}=13000$\,K, of their sample and assuming log$g=8.0$), and placing an upper limit on any putative companion's spectral type assuming this "average" white dwarf is situated at the mean sample distance. FBZ05's "average" white dwarf has absolute magnitudes $M_{H}=11.8$\,mags and $M_{K_{\rm s}}=11.9$\,mags. These magnitudes are beyond the detection limits of the 2MASS survey ($M_{H}\sim11.6$\,mags and $K_{\rm s}\sim10.8$\,mags), and so FBZ05 set the detection limit by assuming a companion has a magnitude equal to the difference between the 2MASS survey limits and this ``average'' white dwarf. This gives a limit of L8 and L4 in the 2MASS $H$ and $K_{\rm s}$-bands respectively. 

Following the FBZ05 method, an ``average'' white dwarf must be analysed using the UKIDSS sample, for which the average temperature is $T_{\rm eff}=14000$\,K. Assuming a surface gravity of log$g=8.0$ this equates to 2MASS absolute magnitudes of  $M_{H}=11.7$\,mags and $M_{K_{\rm s}}=11.8$\,mags. Using the average distance of 161\,pc, the 5$\sigma$ limits of the UKIDSS survey are $M_{H}=12.8$\,mags and $M_{K_{\rm s}}=12.2$\,mags.

 It should be noted at this point that these are fainter than the corresponding magnitudes of the ``average'' white dwarf, differing from the FBZ05 analysis. In FBZ05 the "average" white dwarf is in fact brighter in magnitude than the 2MASS limits, allowing for a limit to be set on the minimum luminosity of a putative companion by using the difference between those two values. As the opposite case is true for UKIDSS, it would make sense to use the addition of a typical 3$\sigma$ error as the limiting magnitude. An interesting point can be made here; if FBZ05 had added the 3$\sigma$ error at the 2MASS limit to their initial estimate, then the "average" companion would have been brighter, and therefore of an earlier spectral type. 
 
At 161\,pc the "average" UKIDSS white dwarf has magnitudes $H=17.7$\,mags and $K=17.8$\,mags, which have typical errors of $\pm$0.05 and $\pm$0.1 respectively. Assuming a companion is detected at the $>3\sigma$ level, this corresponds to a companion with apparent magnitudes of $H=20.0$\,mags and $K=18.8$\,mags. This sets a limiting spectral type for the UKIDSS survey of later than T9 in both bands.

Figure~\ref{limits} shows that it is indeed the case that the UKIDSS survey is sometimes sensitive to late ($>$T8) T dwarf companions. However, the sensitivity analysis for each white dwarf individually shows that this accounts for $<2$\% of the entire sample. Clearly it is not the case that this survey is capable of detecting late T dwarf companions around every white dwarf. Indeed, we have shown in Section~\ref{DISCUSSION:sensitivity} that the sensitivity of this survey peaks at the L-T border. The UKIDSS  "average" white dwarf is clearly not representative of either the mean or the median star in the sample. Indeed, it is only representative of the extreme of the survey sensitivity and therefore not representative at all. Therefore, we suspect that FBZ05's "average" white dwarf is unlikely to represent their mean or median star either. For example, if the distance distribution to each sample is inspected, 35\% and 33\% of white dwarfs lie beyond the average distance from FBZ05 and UKIDSS respectively. Hence, it is likely that \it{at least} \rm that fraction are not in fact as sensitive to low mass companions as the "average" star.

FBZ05 state "NIR excess detection requires photometric accuracy, not deep imaging". In fact, as UKIDSS is $\approx$3 magnitudes deeper than 2MASS, the UKIDSS photometric errors are much smaller at the fainter end of the 2MASS survey where many of the white dwarfs lie. In 2MASS the photometric signal-to-noise ratio is  $>10$ for $K_{S}<14.3$ mags, so a typical error is 0.1\,mags. In the equivalent UKIDSS bandpass, the typical error is 0.005\,mags. Clearly, UKIDSS will provide much better accuracy at the 2MASS limits, and continue this trend until the limits of the survey are reached. Indeed, it may be the case that 2MASS is not as sensitive as the ``average'' white dwarf would suggest.

In conclusion, the use of an ``average'' white dwarf to place sensitivity limits on a sample may be the problem. The detection of an ultracool companion requires that the companion exceed the white dwarf's luminosity by 3$\sigma$ at an infrared wavelength. The luminosity of a star is proportional to $R^{2}$ and $T^{4}$, hence an ``average'' white dwarf determined according to temperature is not representative of the sample. Unfortunately, FBZ05 do not detail the parameters needed for each white dwarf (Effective temperature and distance) to perform a similar analysis to that presented for the UKIDSS data. The reasons for the surprisingly similar sensitivities of both the 2MASS and UKIDSS surveys cannot be fully understood until such an analysis can be performed.


\subsection{The Main-sequence Star $+$ Brown Dwarf Binary Fraction}

\citet{mccarthy04} presented a coronographic NIR search for substellar companions to nearby main-sequence stars. They found that the frequency of brown dwarf companions to G, K and M stars orbiting at a distance between 75-300\,AU to be 1$\pm1$\%. They also calculated a companion fraction for massive brown dwarfs orbiting between 120-1200\,AU to be 0.7$\pm$0.7\%.  \citet{grether06} analysed the relative fraction of close (P$<5$\,yr) stellar, brown dwarf and planetary companions to nearby sun-like stars and found that close brown dwarf companions occur at a frequency of $<1$\%. Therefore, the brown dwarf companion fraction to white dwarfs ($0.4\leq f_{\rm WD+BD}\leq2.6$\%) and hence, their progenitors, from the UKIDSS survey agrees with the estimates of both \citet{mccarthy04} and \citet{grether06} within errors.

However, there exists a bias among white dwarfs in that they are generally descended from stars more massive than the Sun. It could easily be the case that such stars have a greater frequency of brown dwarf companions than solar-type stars.

The brown dwarf desert is a noted deficit in the frequency of brown dwarf companions relative to the frequency of less massive planetary companions \citep{marcy00} or relative to the frequency of more massive stellar companions to Solar-like hosts. \citet{grether06} verified the existence of this phenomenon, with the desert extending to $<3$\,AU.  Determining the separations of the WD$+$BD binaries in the UKIDSS sample will allow additional investigation of the brown dwarf desert. It may even be possible to determine the companion fraction at different orbital radii.

\section{Summary}
We have cross correlated the UKIDSS LAS DR8 with both the \citet{eisenstein06} and the McCook \& Sion online catalogues of spectroscopically identified white dwarfs, in order to search for stars displaying near-infrared excesses indicative of low mass companions or debris disks. The sample included 811 hydrogen atmosphere DA white dwarfs with $1500<T_{\rm eff}<100,000$\,K, $6.5<$log\,$g<9.5$, and both $H$ and $K$-band photometry present in the UKIDSS archive.

We identified 313 white dwarfs with near-infrared excesses, with 274 of these recovering previously identified binaries. This left 39 candidate white dwarfs for further analysis. Twenty seven of these had multiple excesses indicative of a low mass companion, with 9-11 of these having a predicted mass in the range associated with brown dwarfs, and a further 14-16 likely very low mass stellar companions. Seven stars in the sample showed evidence of contamination by a foreground or background object. The remaining 3 each had a $K$-band excess indicative of a debris disk including the already discovered DAd SDSS\,J1228$+$104. Two magnetic white dwarfs were found with a near-infrared excess; (1) SDSS\,J1212$+$01 - a previously identified DAH$+$dL8 binary, and (2) SDSS\,J1250$+$154 - potentially explained by the presence of an M8 companion and additional cyclotron emission.

We have placed limits on the latest spectral type of unresolved companion detectable around each of the apparently single DA white dwarfs detected in the UKIDSS database. We found that this survey has a sensitivity of $\approx99$\% to companions of spectral type L0 or earlier and $\approx43$\% to companions of spectral type T0 or earlier. These sensitivities were used in conjunction with the photometrically detected L and T dwarf companions to estimate the unresolved brown dwarf companion fraction to DA white dwarfs in UKIDSS as $0.4\leq f_{\rm WD+BD}\leq2.6$\%. Correcting for the completeness of the survey gives a final binary fraction of $f_{\rm WD+BD}\geq0.5\pm0.3$\%.

The WD+dL binary fraction in UKIDSS is significantly higher than that of \citet{farihi05}. One possible explanation for this discrepancy is that  \citet{farihi05} have used an ``average'' white dwarf to set the limits of their 2MASS survey. By utilising the same method for the UKIDSS white dwarfs, we have shown that this "average" white dwarf may not be representative of the mean survey sensitivity. A more thorough analysis of the 2MASS survey sensitivity is required, using the same method employed here, before the differences can be fully understood. This will require each white dwarf to be analysed individually in order to place a limit on the coolest detectable companion around each star.

 We found that the fraction of ultracool companions to white dwarfs was found to be consistent with estimates of the main-sequence $+$ brown dwarf binary fraction.

\section{Notes}

An updated verion of Pierre Bergeron's synthetic model grid was made available in February 2011 \citep*{tremblay11}. These revisions should have little effect on the magnitudes and distances calculate in this work, so will be used for the analysis of future UKIDSS data releases. 

Before submission the authors became aware of the study of Girven et al. (2011, submitted) who have carried out an independent study to identify photometric white dwarf candidates in SDSS DR7, and also search for infrared excesses. A comparison of results shows each study to be broadly consistent.

\section{Acknowledgements}

PRS is supported by RoPACs, a Marie Curie Initial Training Network funded by the European Commissions's Seventh Framework Programme. MRB acknowledges receipt of an STFC Advanced Fellowship. Balmer lines in the models were calculated with the modified Stark broadening profiles of Tremblay \& Bergeron (2009), kindly made available by the authors.

\bibliographystyle{mn}
\bibliography{refs}

\begin{thebibliography}{}

\bibitem[\protect\citeauthoryear{{Baraffe} \& {Chabrier}}{{Baraffe} \&
  {Chabrier}}{1996}]{baraffe96}
{Baraffe} I.,  {Chabrier} G.,  1996, ApJL, 461, L51+

\bibitem[\protect\citeauthoryear{{Baraffe}, {Chabrier}, {Allard} \&
  {Hauschildt}}{{Baraffe} et~al.}{2002}]{baraffe02}
{Baraffe} I.,  {Chabrier} G.,  {Allard} F.,    {Hauschildt} P.~H.,  2002, A\&A,
  382, 563

\bibitem[\protect\citeauthoryear{{Becklin}, {Farihi}, {Jura}, {Song},
  {Weinberger} \& {Zuckerman}}{{Becklin} et~al.}{2005}]{becklin05}
{Becklin} E.~E.,  {Farihi} J.,  {Jura} M.,  {Song} I.,  {Weinberger} A.~J.,
  {Zuckerman} B.,  2005, ApJL, 632, L119

\bibitem[\protect\citeauthoryear{{Becklin} \& {Zuckerman}}{{Becklin} \&
  {Zuckerman}}{1988}]{becklin88}
{Becklin} E.~E.,  {Zuckerman} B.,  1988, Nature, 336, 656

\bibitem[\protect\citeauthoryear{{Bergeron}, {Saffer} \& {Liebert}}{{Bergeron}
  et~al.}{1992}]{bergeron92}
{Bergeron} P.,  {Saffer} R.~A.,    {Liebert} J.,  1992, ApJ, 394, 228

\bibitem[\protect\citeauthoryear{{Bergeron}, {Wesemael} \&
  {Beauchamp}}{{Bergeron} et~al.}{1995}]{bergeron95}
{Bergeron} P.,  {Wesemael} F.,    {Beauchamp} A.,  1995, PASP, 107, 1047

\bibitem[\protect\citeauthoryear{{Brinkworth}, {G{\"a}nsicke}, {Marsh}, {Hoard}
  \& {Tappert}}{{Brinkworth} et~al.}{2009}]{brinkworth09}
{Brinkworth} C.~S.,  {G{\"a}nsicke} B.~T.,  {Marsh} T.~R.,  {Hoard} D.~W.,
  {Tappert} C.,  2009, ApJ, 696, 1402

\bibitem[\protect\citeauthoryear{Burleigh et~al.,}{Burleigh
  et~al.}{2006}]{burleigh06b}
Burleigh M.~R.,  et~al., 2006, MNRAS, 373, 1416

\bibitem[\protect\citeauthoryear{{Burleigh}, {Hogan}, {Dobbie}, {Napiwotzki} \&
  {Maxted}}{{Burleigh} et~al.}{2006}]{burleigh06}
{Burleigh} M.~R.,  {Hogan} E.,  {Dobbie} P.~D.,  {Napiwotzki} R.,    {Maxted}
  P.~F.~L.,  2006, MNRAS, 373, L55

\bibitem[\protect\citeauthoryear{{Caballero}, {Burgasser} \&
  {Klement}}{{Caballero} et~al.}{2008}]{caballero08}
{Caballero} J.~A.,  {Burgasser} A.~J.,    {Klement} R.,  2008, A\&A, 488, 181

\bibitem[\protect\citeauthoryear{{Carpenter}}{{Carpenter}}{2001}]{carpenter01}
{Carpenter} J.~M.,  2001, AJ, 121, 2851

\bibitem[\protect\citeauthoryear{{Casali} et~al.,}{{Casali}
  et~al.}{2007}]{casali07}
{Casali} M.,  et~al., 2007, A\&A, 467, 777

\bibitem[\protect\citeauthoryear{{Chabrier}, {Baraffe}, {Allard} \&
  {Hauschildt}}{{Chabrier} et~al.}{2000}]{chabrier00}
{Chabrier} G.,  {Baraffe} I.,  {Allard} F.,    {Hauschildt} P.,  2000, ApJ,
  542, 464

\bibitem[\protect\citeauthoryear{{Chiu}, {Fan}, {Leggett}, {Golimowski},
  {Zheng}, {Geballe}, {Schneider} \& {Brinkmann}}{{Chiu} et~al.}{2006}]{chiu06}
{Chiu} K.,  {Fan} X.,  {Leggett} S.~K.,  {Golimowski} D.~A.,  {Zheng} W.,
  {Geballe} T.~R.,  {Schneider} D.~P.,    {Brinkmann} J.,  2006, AJ, 131, 2722

\bibitem[\protect\citeauthoryear{{Cushing}, {Rayner} \& {Vacca}}{{Cushing}
  et~al.}{2005}]{cushing05}
{Cushing} M.~C.,  {Rayner} J.~T.,    {Vacca} W.~D.,  2005, ApJ, 623, 1115

\bibitem[\protect\citeauthoryear{{Day-Jones}, {Pinfield}, {Jones},
  {Napiwotzki}, {Burningham}, {Jenkins} \& {UKIDSS Cool Dwarf Science Working
  Group}}{{Day-Jones} et~al.}{2008}]{dayjones08}
{Day-Jones} A.,  {Pinfield} D.~J.,  {Jones} H.~R.~A.,  {Napiwotzki} R.,
  {Burningham} B.,  {Jenkins} J.~S.,    {UKIDSS Cool Dwarf Science Working
  Group} 2008, in American Astronomical Society Meeting Abstracts Vol.~211 of
  American Astronomical Society Meeting Abstracts, {Hunting For Wild Brown
  Dwarf Companions To White Dwarfs In UKIDSS And SDSS}.
pp 103.30--+

\bibitem[\protect\citeauthoryear{{Day-Jones}, {Pinfield}, {Ruiz}, {Beaumont},
  {Burningham}, {Gallardo}, {Gianninas}, {Bergeron}, {Napiwotzki}, {Jenkins},
  {Zhang}, {Murray}, {Catal{\'a}n} \& {Gomes}}{{Day-Jones}
  et~al.}{2011}]{dayjones11}
{Day-Jones} A.~C.,  {Pinfield} D.~J.,  {Ruiz} M.~T.,  {Beaumont} H.,
  {Burningham} B.,  {Gallardo} J.,  {Gianninas} A.,  {Bergeron} P.,
  {Napiwotzki} R.,  {Jenkins} J.~S.,  {Zhang} Z.~H.,  {Murray} D.~N.,
  {Catal{\'a}n} S.,    {Gomes} J.,  2011, MNRAS, 410, 705

\bibitem[\protect\citeauthoryear{{Debes}, {L{\'o}pez-Morales}, {Bonanos} \&
  {Weinberger}}{{Debes} et~al.}{2006}]{debes06}
{Debes} J.~H.,  {L{\'o}pez-Morales} M.,  {Bonanos} A.~Z.,    {Weinberger}
  A.~J.,  2006, ApJL, 647, L147

\bibitem[\protect\citeauthoryear{{Dobbie}, {Burleigh}, {Levan}, {Barstow},
  {Napiwotzki}, {Holberg}, {Hubeny} \& {Howell}}{{Dobbie}
  et~al.}{2005}]{dobbie05}
{Dobbie} P.~D.,  {Burleigh} M.~R.,  {Levan} A.~J.,  {Barstow} M.~A.,
  {Napiwotzki} R.,  {Holberg} J.~B.,  {Hubeny} I.,    {Howell} S.~B.,  2005,
  MNRAS, 357, 1049

\bibitem[\protect\citeauthoryear{{Dobbie}, {Napiwotzki}, {Burleigh}, {Barstow},
  {Boyce}, {Casewell}, {Jameson}, {Hubeny} \& {Fontaine}}{{Dobbie}
  et~al.}{2006}]{dobbie06}
{Dobbie} P.~D.,  {Napiwotzki} R.,  {Burleigh} M.~R.,  {Barstow} M.~A.,  {Boyce}
  D.~D.,  {Casewell} S.~L.,  {Jameson} R.~F.,  {Hubeny} I.,    {Fontaine} G.,
  2006, MNRAS, 369, 383

\bibitem[\protect\citeauthoryear{{Dye} et~al.,}{{Dye}  et~al.}{2006}]{dye06}
{Dye} S.,  et~al., 2006, MNRAS, 372, 1227

\bibitem[\protect\citeauthoryear{{Eisenstein} et~al.,}{{Eisenstein}
  et~al.}{2006}]{eisenstein06}
{Eisenstein} D.~J.,  et~al., 2006, ApJS, 167, 40

\bibitem[\protect\citeauthoryear{{Farihi}, {Becklin} \& {Zuckerman}}{{Farihi}
  et~al.}{2005}]{farihi05}
{Farihi} J.,  {Becklin} E.~E.,    {Zuckerman} B.,  2005, ApJS, 161, 394

\bibitem[\protect\citeauthoryear{{Farihi}, {Burleigh} \& {Hoard}}{{Farihi}
  et~al.}{2008}]{farihi08}
{Farihi} J.,  {Burleigh} M.~R.,    {Hoard} D.~W.,  2008, ApJ, 674, 421

\bibitem[\protect\citeauthoryear{{Farihi} \& {Christopher}}{{Farihi} \&
  {Christopher}}{2004}]{farihi04}
{Farihi} J.,  {Christopher} M.,  2004, AJ, 128, 1868

\bibitem[\protect\citeauthoryear{{Farihi}, {Hoard} \& {Wachter}}{{Farihi}
  et~al.}{2006}]{farihi05b}
{Farihi} J.,  {Hoard} D.~W.,    {Wachter} S.,  2006, ApJ, 646, 480

\bibitem[\protect\citeauthoryear{{Farihi}, {Hoard} \& {Wachter}}{{Farihi}
  et~al.}{2010}]{farihi10}
{Farihi} J.,  {Hoard} D.~W.,    {Wachter} S.,  2010, ApJS, 190, 275

\bibitem[\protect\citeauthoryear{{G{\"a}nsicke}, {Marsh}, {Southworth} \&
  {Rebassa-Mansergas}}{{G{\"a}nsicke} et~al.}{2006}]{gansicke06}
{G{\"a}nsicke} B.~T.,  {Marsh} T.~R.,  {Southworth} J.,    {Rebassa-Mansergas}
  A.,  2006, Science, 314, 1908

\bibitem[\protect\citeauthoryear{{Girardi}, {Bressan}, {Bertelli} \&
  {Chiosi}}{{Girardi} et~al.}{2000}]{girardi00}
{Girardi} L.,  {Bressan} A.,  {Bertelli} G.,    {Chiosi} C.,  2000, AAS, 141,
  371

\bibitem[\protect\citeauthoryear{{Golimowski} et~al.,}{{Golimowski}
  et~al.}{2004}]{golimowski04}
{Golimowski} D.~A.,  et~al., 2004, AJ, 128, 1733

\bibitem[\protect\citeauthoryear{{Green}, {Ali} \& {Napiwotzki}}{{Green}
  et~al.}{2000}]{green00}
{Green} P.~J.,  {Ali} B.,    {Napiwotzki} R.,  2000, ApJ, 540, 992

\bibitem[\protect\citeauthoryear{{Grether} \& {Lineweaver}}{{Grether} \&
  {Lineweaver}}{2006}]{grether06}
{Grether} D.,  {Lineweaver} C.~H.,  2006, ApJ, 640, 1051

\bibitem[\protect\citeauthoryear{{Hewett}, {Warren}, {Leggett} \&
  {Hodgkin}}{{Hewett} et~al.}{2006}]{hewett06}
{Hewett} P.~C.,  {Warren} S.~J.,  {Leggett} S.~K.,    {Hodgkin} S.~T.,  2006,
  MNRAS, 367, 454

\bibitem[\protect\citeauthoryear{{Hoard}, {Wachter}, {Sturch}, {Widhalm},
  {Weiler}, {Pretorius}, {Wellhouse} \& {Gibiansky}}{{Hoard}
  et~al.}{2007}]{hoard07}
{Hoard} D.~W.,  {Wachter} S.,  {Sturch} L.~K.,  {Widhalm} A.~M.,  {Weiler}
  K.~P.,  {Pretorius} M.~L.,  {Wellhouse} J.~W.,    {Gibiansky} M.,  2007, AJ,
  134, 26

\bibitem[\protect\citeauthoryear{{Holberg}}{{Holberg}}{2009}]{holberg09}
{Holberg} J.~B.,  2009, Journal of Physics Conference Series, 172, 012022

\bibitem[\protect\citeauthoryear{{Holberg} \& {Bergeron}}{{Holberg} \&
  {Bergeron}}{2006}]{holberg06}
{Holberg} J.~B.,  {Bergeron} P.,  2006, AJ, 132, 1221

\bibitem[\protect\citeauthoryear{{Holberg} \& {Magargal}}{{Holberg} \&
  {Magargal}}{2005}]{holberg05}
{Holberg} J.~B.,  {Magargal} K.,  2005, in 14th European Workshop on White
  Dwarfs, ed. D. Koester \& S. Moehler, ASP Conf. Ser. (San Francisco ASP)
  Vol.~334, {}.
p.~419

\bibitem[\protect\citeauthoryear{{Hubeny}}{{Hubeny}}{1988}]{hubeny88}
{Hubeny} I.,  1988, Computer Physics Communications, 52, 103

\bibitem[\protect\citeauthoryear{{Hubeny} \& {Lanz}}{{Hubeny} \&
  {Lanz}}{1995}]{hubeny95}
{Hubeny} I.,  {Lanz} T.,  1995, ApJ, 439, 875

\bibitem[\protect\citeauthoryear{{Jura}}{{Jura}}{2003}]{jura03}
{Jura} M.,  2003, ApJ, 584, L91

\bibitem[\protect\citeauthoryear{{Jura}, {Farihi} \& {Zuckerman}}{{Jura}
  et~al.}{2007}]{jura07}
{Jura} M.,  {Farihi} J.,    {Zuckerman} B.,  2007, ApJ, 663, 1285

\bibitem[\protect\citeauthoryear{{Kalirai}, {Hansen}, {Kelson}, {Reitzel},
  {Rich} \& {Richer}}{{Kalirai} et~al.}{2008}]{kalirai08}
{Kalirai} J.~S.,  {Hansen} B.~M.~S.,  {Kelson} D.~D.,  {Reitzel} D.~B.,  {Rich}
  R.~M.,    {Richer} H.~B.,  2008, ApJ, 676, 594

\bibitem[\protect\citeauthoryear{{Kawka}, {Vennes}, {Schmidt}, {Wickramasinghe}
  \& {Koch}}{{Kawka} et~al.}{2007}]{kawka07}
{Kawka} A.,  {Vennes} S.,  {Schmidt} G.~D.,  {Wickramasinghe} D.~T.,    {Koch}
  R.,  2007, ApJ, 654, 499

\bibitem[\protect\citeauthoryear{{Kilic} \& {Redfield}}{{Kilic} \&
  {Redfield}}{2007}]{kilic07}
{Kilic} M.,  {Redfield} S.,  2007, ApJ, 660, 641

\bibitem[\protect\citeauthoryear{{Kilic}, {von Hippel}, {Leggett} \&
  {Winget}}{{Kilic} et~al.}{2005}]{kilic05}
{Kilic} M.,  {von Hippel} T.,  {Leggett} S.~K.,    {Winget} D.~E.,  2005, ApJ,
  632, L115

\bibitem[\protect\citeauthoryear{{Kilic}, {von Hippel}, {Leggett} \&
  {Winget}}{{Kilic} et~al.}{2006}]{kilic06}
{Kilic} M.,  {von Hippel} T.,  {Leggett} S.~K.,    {Winget} D.~E.,  2006, ApJ,
  646, 474

\bibitem[\protect\citeauthoryear{Knapp et~al.,}{Knapp  et~al.}{2004}]{knapp04}
Knapp G.~R.,  et~al., 2004, AJ, 127, 3553

\bibitem[\protect\citeauthoryear{{Knutson} et~al.,}{{Knutson}
  et~al.}{2007}]{knutson07}
{Knutson} H.~A.,  et~al., 2007, Nature, 447, 183

\bibitem[\protect\citeauthoryear{{Koen} \& {Maxted}}{{Koen} \&
  {Maxted}}{2006}]{koen06}
{Koen} C.,  {Maxted} P.~F.~L.,  2006, MNRAS, 371, 1675

\bibitem[\protect\citeauthoryear{{Koester}}{{Koester}}{2008}]{koester08}
{Koester} D.,  2008, ArXiv e-prints

\bibitem[\protect\citeauthoryear{{Lawrence} et~al.,}{{Lawrence}
  et~al.}{2007}]{lawrence07}
{Lawrence} A.,  et~al., 2007, MNRAS, 379, 1599

\bibitem[\protect\citeauthoryear{{L{\'e}pine} \& {Shara}}{{L{\'e}pine} \&
  {Shara}}{2005}]{lepine05}
{L{\'e}pine} S.,  {Shara} M.~M.,  2005, AJ, 129, 1483

\bibitem[\protect\citeauthoryear{{Liebert} et~al.,}{{Liebert}
  et~al.}{2005}]{liebert05}
{Liebert} J.,  et~al., 2005, AJ, 129, 2376

\bibitem[\protect\citeauthoryear{{Marcy} \& {Butler}}{{Marcy} \&
  {Butler}}{2000}]{marcy00}
{Marcy} G.~W.,  {Butler} R.~P.,  2000, PASP, 112, 137

\bibitem[\protect\citeauthoryear{{Maxted}, {Napiwotzki}, {Dobbie} \&
  {Burleigh}}{{Maxted} et~al.}{2006}]{maxted06}
{Maxted} P.~F.~L.,  {Napiwotzki} R.,  {Dobbie} P.~D.,    {Burleigh} M.~R.,
  2006, Nature, 442, 543

\bibitem[\protect\citeauthoryear{{McCarthy} \& {Zuckerman}}{{McCarthy} \&
  {Zuckerman}}{2004}]{mccarthy04}
{McCarthy} C.,  {Zuckerman} B.,  2004, AJ, 127, 2871

\bibitem[\protect\citeauthoryear{{McCook} \& {Sion}}{{McCook} \&
  {Sion}}{1999}]{mccook99}
{McCook} G.~P.,  {Sion} E.~M.,  1999, ApJS, 121, 1

\bibitem[\protect\citeauthoryear{{Patten}, {Stauffer}, {Burrows}, {Marengo},
  {Hora}, {Luhman}, {Sonnett}, {Henry}, {Raghavan}, {Megeath}, {Liebert} \&
  {Fazio}}{{Patten} et~al.}{2006}]{patten06}
{Patten} B.~M.,  {Stauffer} J.~R.,  {Burrows} A.,  {Marengo} M.,  {Hora} J.~L.,
   {Luhman} K.~L.,  {Sonnett} S.~M.,  {Henry} T.~J.,  {Raghavan} D.,  {Megeath}
  S.~T.,  {Liebert} J.,    {Fazio} G.~G.,  2006, ApJ, 651, 502

\bibitem[\protect\citeauthoryear{{Probst}}{{Probst}}{1983}]{probst83}
{Probst} R.~G.,  1983, ApJS, 53, 335

\bibitem[\protect\citeauthoryear{{Reach}, {Kuchner}, {von Hippel}, {Burrows},
  {Kilic}, {Mullally} \& {Winget}}{{Reach} et~al.}{2005}]{reach05}
{Reach} W.~T.,  {Kuchner} M.~J.,  {von Hippel} T.,  {Burrows} A.,  {Kilic} M.,
  {Mullally} F.,    {Winget} D.~E.,  2005, ApJ, 635, L161

\bibitem[\protect\citeauthoryear{{Schmidt}, {Szkody}, {Silvestri}, {Cushing},
  {Liebert} \& {Smith}}{{Schmidt} et~al.}{2005}]{schmidt05}
{Schmidt} G.~D.,  {Szkody} P.,  {Silvestri} N.~M.,  {Cushing} M.~C.,  {Liebert}
  J.,    {Smith} P.~S.,  2005, ApJL, 630, L173

\bibitem[\protect\citeauthoryear{{Skrutskie} et~al.,}{{Skrutskie}
  et~al.}{2006}]{skrutskie06}
{Skrutskie} M.~F.,  et~al., 2006, AJ, 131, 1163

\bibitem[\protect\citeauthoryear{{Steele}, {Burleigh}, {Farihi},
  {G{\"a}nsicke}, {Jameson}, {Dobbie} \& {Barstow}}{{Steele}
  et~al.}{2009}]{steele09}
{Steele} P.~R.,  {Burleigh} M.~R.,  {Farihi} J.,  {G{\"a}nsicke} B.~T.,
  {Jameson} R.~F.,  {Dobbie} P.~D.,    {Barstow} M.~A.,  2009, A\&A, 500, 1207

\bibitem[\protect\citeauthoryear{{Su} et~al.,}{{Su}  et~al.}{2007}]{su07}
{Su} K.~Y.~L.,  et~al., 2007, ApJL, 657, L41

\bibitem[\protect\citeauthoryear{{Tremblay} \& {Bergeron}}{{Tremblay} \&
  {Bergeron}}{2007}]{tremblay07}
{Tremblay} P.-E.,  {Bergeron} P.,  2007, ApJ, 657, 1013

\bibitem[\protect\citeauthoryear{{Tremblay}, {Bergeron} \&
  {Gianninas}}{{Tremblay} et~al.}{2011}]{tremblay11}
{Tremblay} P.-E.,  {Bergeron} P.,    {Gianninas} A.,  2011, ApJ, 730, 128

\bibitem[\protect\citeauthoryear{{Vanlandingham}, {Schmidt}, {Eisenstein},
  {Harris}, {Anderson}, {Hall}, {Liebert}, {Schneider}, {Silvestri}, {Stinson}
  \& {Wolfe}}{{Vanlandingham} et~al.}{2005}]{vanland05}
{Vanlandingham} K.~M.,  {Schmidt} G.~D.,  {Eisenstein} D.~J.,  {Harris} H.~C.,
  {Anderson} S.~F.,  {Hall} P.~B.,  {Liebert} J.,  {Schneider} D.~P.,
  {Silvestri} N.~M.,  {Stinson} G.~S.,    {Wolfe} M.~A.,  2005, AJ, 130, 734

\bibitem[\protect\citeauthoryear{{Vennes}, {Smith}, {Boyle}, {Croom}, {Kawka},
  {Shanks}, {Miller} \& {Loaring}}{{Vennes} et~al.}{2002}]{vennes02}
{Vennes} S.,  {Smith} R.~J.,  {Boyle} B.~J.,  {Croom} S.~M.,  {Kawka} A.,
  {Shanks} T.,  {Miller} L.,    {Loaring} N.,  2002, MNRAS, 335, 673

\bibitem[\protect\citeauthoryear{{von Hippel}, {Kuchner}, {Kilic}, {Mullally}
  \& {Reach}}{{von Hippel} et~al.}{2007}]{vonhippel07}
{von Hippel} T.,  {Kuchner} M.~J.,  {Kilic} M.,  {Mullally} F.,    {Reach}
  W.~T.,  2007, ApJ, 662, 544

\bibitem[\protect\citeauthoryear{{Vrba}, {Henden}, {Luginbuhl}, {Guetter},
  {Munn}, {Canzian}, {Burgasser}, {Kirkpatrick}, {Fan}, {Geballe},
  {Golimowski}, {Knapp}, {Leggett}, {Schneider} \& {Brinkmann}}{{Vrba}
  et~al.}{2004}]{vrba04}
{Vrba} F.~J.,  {Henden} A.~A.,  {Luginbuhl} C.~B.,  {Guetter} H.~H.,  {Munn}
  J.~A.,  {Canzian} B.,  {Burgasser} A.~J.,  {Kirkpatrick} J.~D.,  {Fan} X.,
  {Geballe} T.~R.,  {Golimowski} D.~A.,  {Knapp} G.~R.,  {Leggett} S.~K.,
  {Schneider} D.~P.,    {Brinkmann} J.,  2004, AJ, 127, 2948

\bibitem[\protect\citeauthoryear{{Wachter}, {Hoard}, {Hansen}, {Wilcox},
  {Taylor} \& {Finkelstein}}{{Wachter} et~al.}{2003}]{wachter03}
{Wachter} S.,  {Hoard} D.~W.,  {Hansen} K.~H.,  {Wilcox} R.~E.,  {Taylor}
  H.~M.,    {Finkelstein} S.~L.,  2003, ApJ, 586, 1356

\bibitem[\protect\citeauthoryear{{Warren} et~al.,}{{Warren}
  et~al.}{2007a}]{warren07b}
{Warren} S.~J.,  et~al., 2007a, ArXiv Astrophysics e-prints

\bibitem[\protect\citeauthoryear{{Warren} et~al.,}{{Warren}
  et~al.}{2007b}]{warren07}
{Warren} S.~J.,  et~al., 2007b, MNRAS, 375, 213

\bibitem[\protect\citeauthoryear{{Wellhouse}, {Hoard}, {Howell}, {Wachter} \&
  {Esin}}{{Wellhouse} et~al.}{2005}]{wellhouse05}
{Wellhouse} J.~W.,  {Hoard} D.~W.,  {Howell} S.~B.,  {Wachter} S.,    {Esin}
  A.~A.,  2005, PASP, 117, 1378

\bibitem[\protect\citeauthoryear{{York} et~al.,}{{York}  et~al.}{2000}]{york00}
{York} D.~G.,  et~al., 2000, AJ, 120, 1579

\bibitem[\protect\citeauthoryear{{Zuckerman} \& {Becklin}}{{Zuckerman} \&
  {Becklin}}{1987}]{zuckerman87}
{Zuckerman} B.,  {Becklin} E.~E.,  1987, ApJL, 319, L99

\bibitem[\protect\citeauthoryear{{Zuckerman}, {Koester}, {Reid} \&
  {H{\"u}nsch}}{{Zuckerman} et~al.}{2003}]{zuckerman03}
{Zuckerman} B.,  {Koester} D.,  {Reid} I.~N.,    {H{\"u}nsch} M.,  2003, ApJ,
  596, 477

\end{thebibliography}

\clearpage

\appendix

\section{White Dwarfs Model Spectra and Photometry}
\label{spectra}
\clearpage

\begin{figure}
\begin{center}
\psfig{file=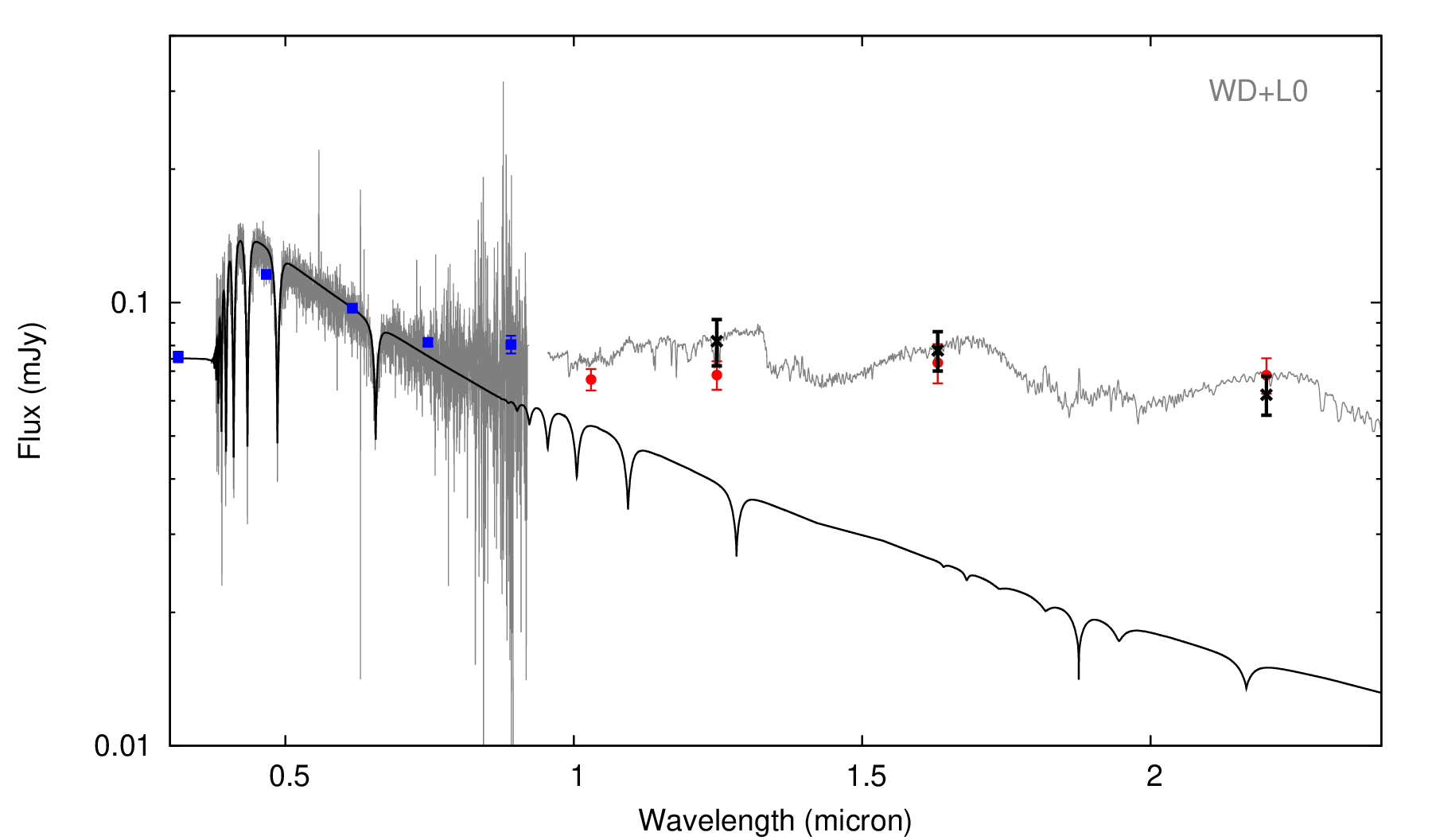,width=5.0cm,angle=-90}
\end{center}
\caption{SDSS\,J003902.47$-$003000.3 model spectrum (black solid) with SDSS $ugriz'$ (squares), UKIDSS $YJHK$ (circles) and NTT $JHK$ photometry (crosses). Also shown are the SDSS spectrum (light grey) and a composite WD$+$L0 spectrum (dark grey).}
\label{003902}
\end{figure}

\begin{figure}
\begin{center}
\psfig{file=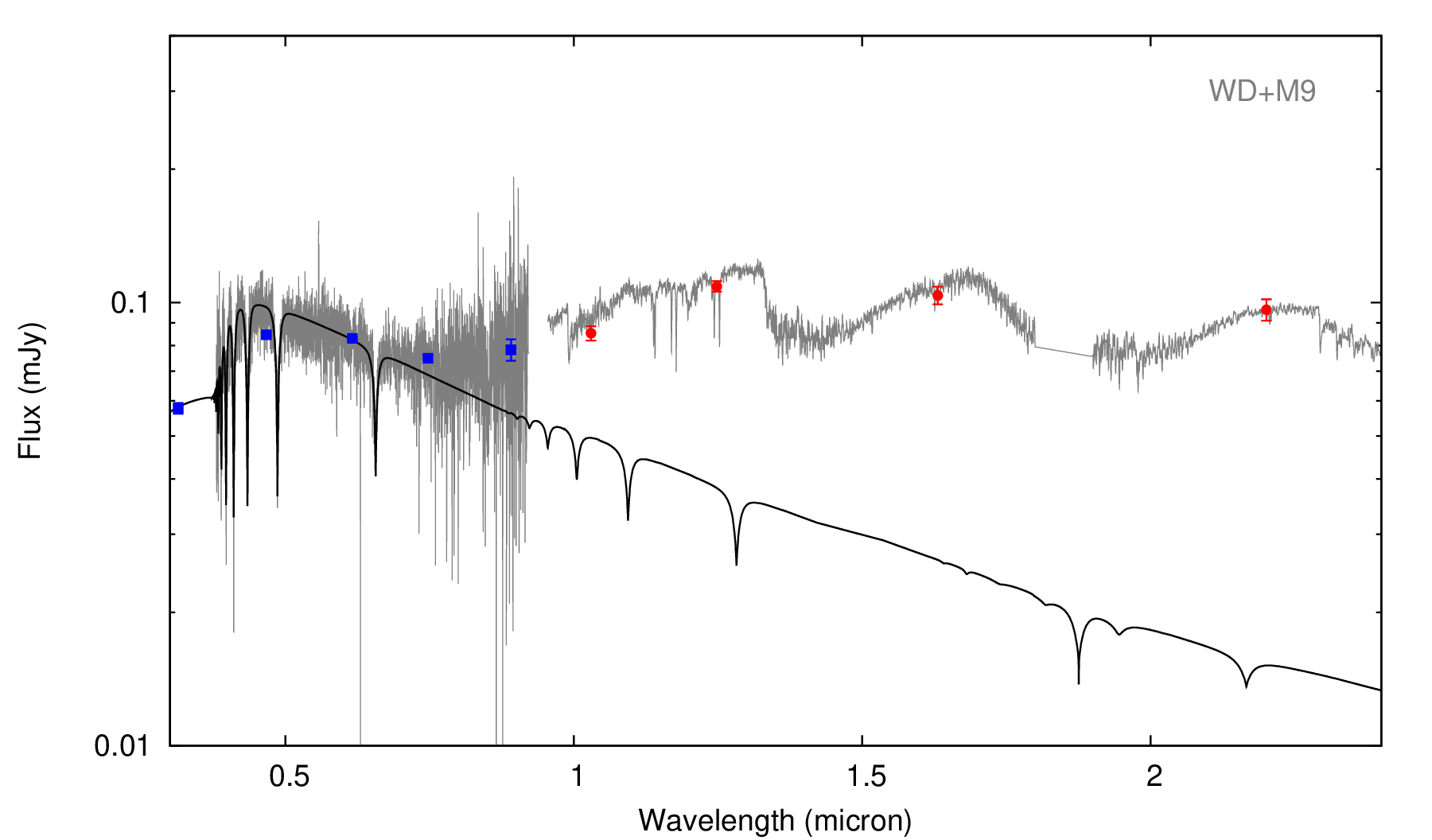,width=5.0cm,angle=-90}
\end{center}
\caption{SDSS\,J003923.04$+$003534.7 model spectrum (black solid) with SDSS $ugriz'$ (squares) and UKIDSS $YJHK$ photometry (circles). Also shown are the SDSS spectrum (light grey) and a composite WD$+$M9 spectrum (dark grey).}
\label{003923}
\end{figure}

\begin{figure}
\begin{center}
\psfig{file=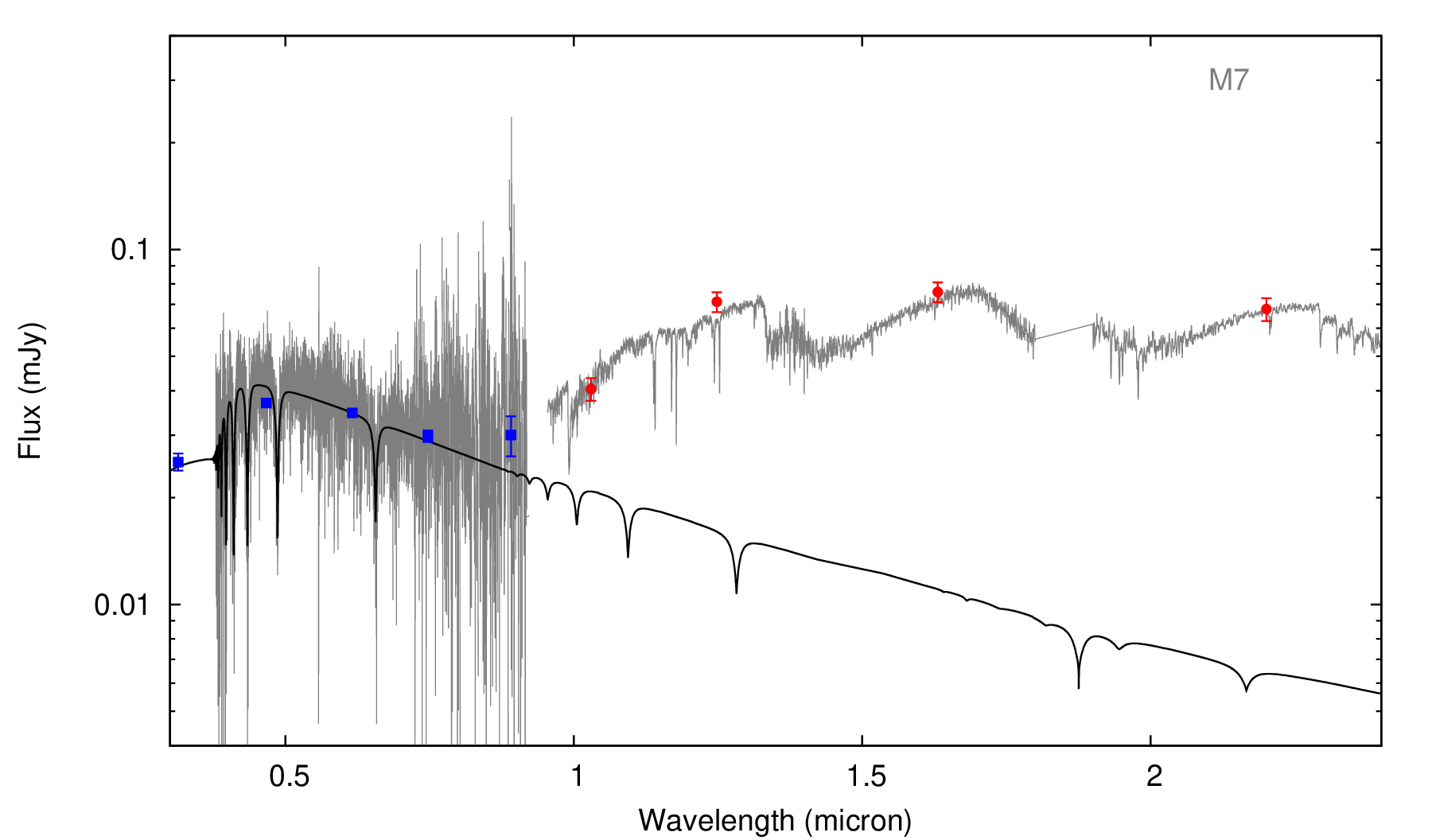,width=5.0cm,angle=-90}
\end{center}
\caption{SDSS\,J012032.27$-$001351.1 model spectrum (black solid) with SDSS $ugriz'$ (squares) and UKIDSS $YJHK$ photometry (circles). Also shown is the SDSS spectrum (light grey) and an M7 spectrum scaled to the $K$-band magnitude of the putative secondary.}
\label{012032}
\end{figure}

\begin{figure}
\begin{center}
\psfig{file=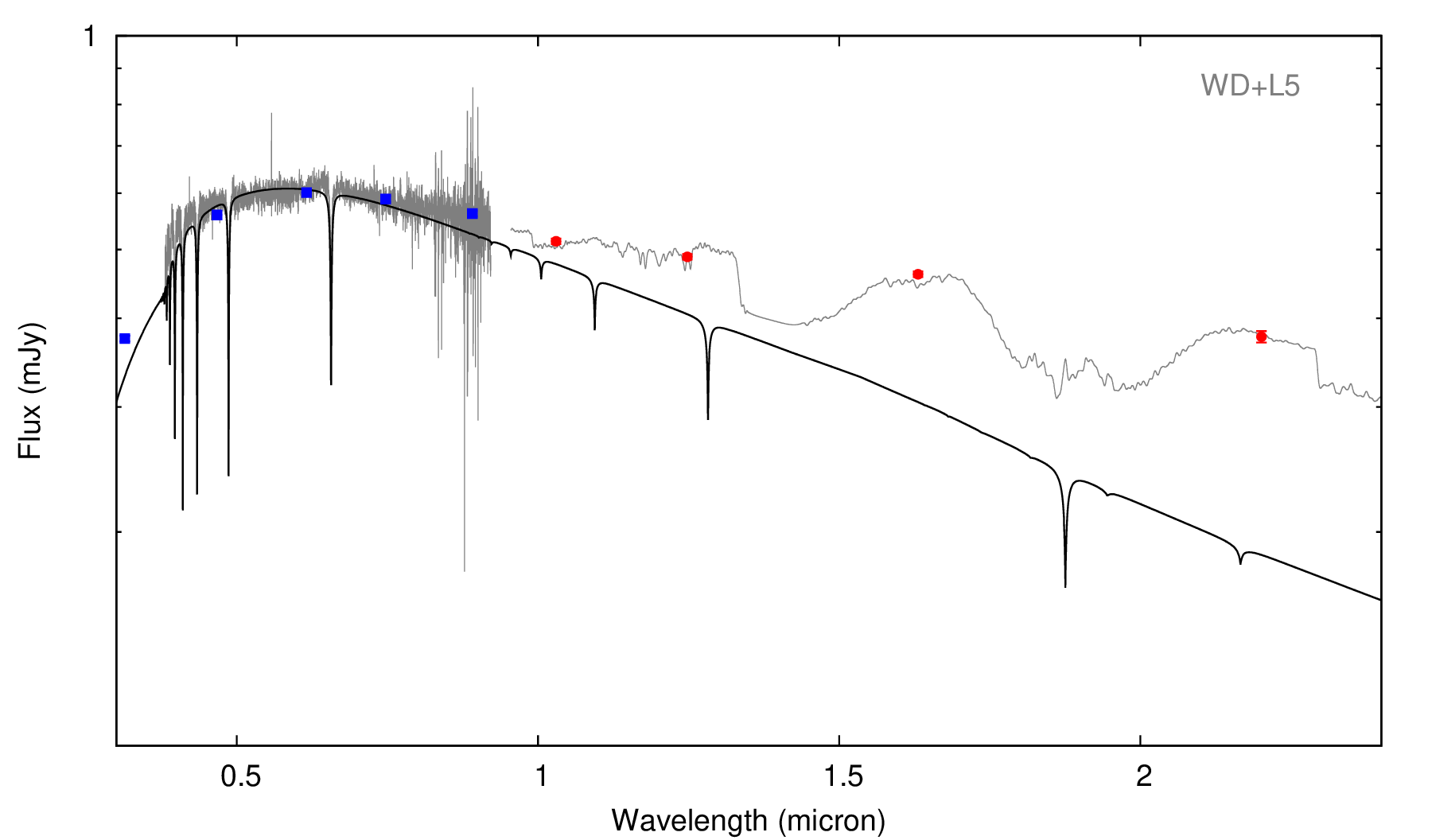,width=5.0cm,angle=-90}
\end{center}
\caption{SDSS\,J013532.98$+$144555.8 model spectrum (black solid) with SDSS $griz'$ (squares) and UKIDSS $YJHK$ photometry (circles). Also shown are the SDSS spectrum (light grey) and a composite WD$+$L5 spectrum (dark grey).}
\label{013532}
\end{figure}

\begin{figure}
\begin{center}
\psfig{file=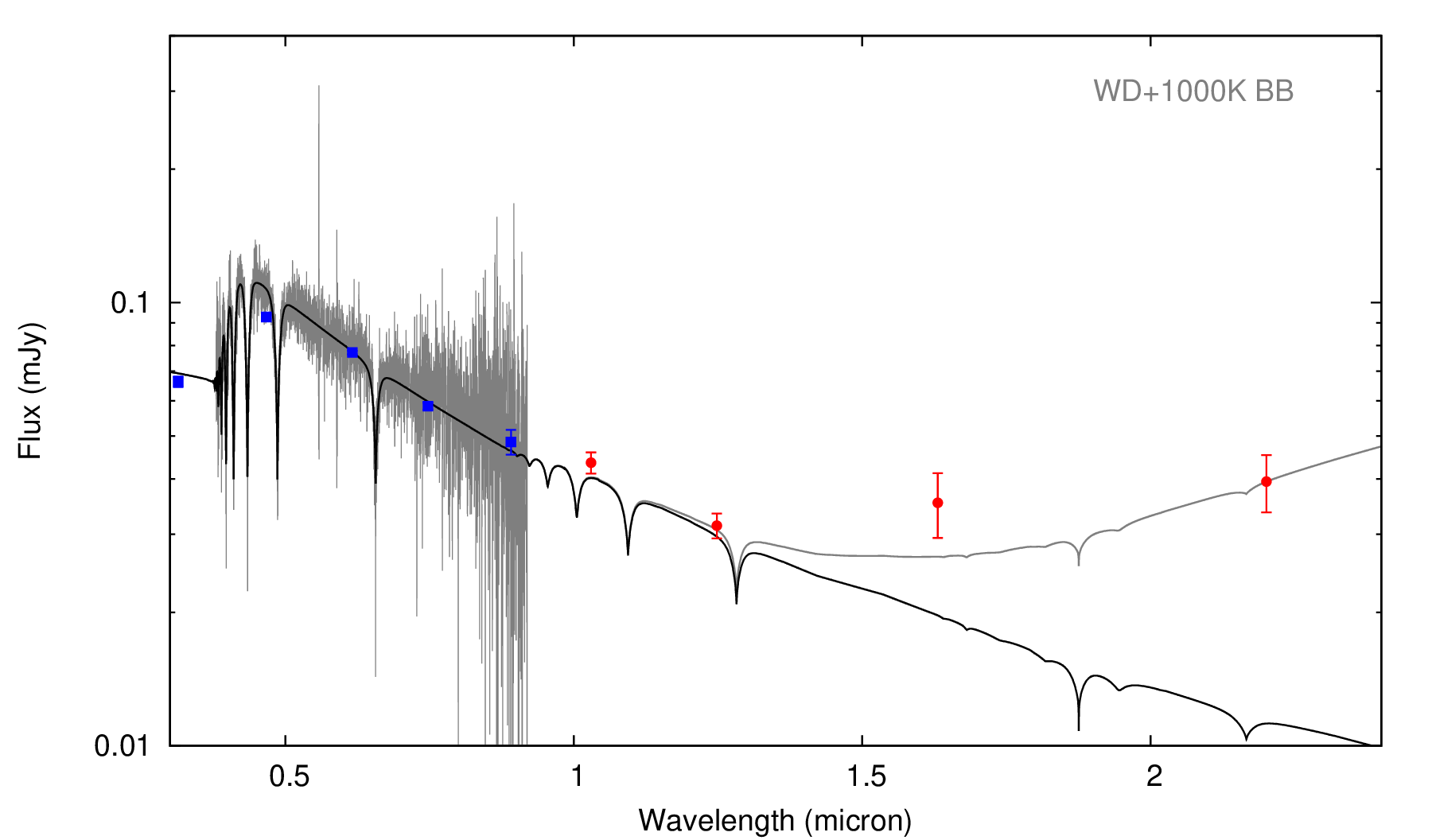,width=5.0cm,angle=-90}
\end{center}
\caption{SDSS\,J075327.05$+$244705.5 model spectrum (black solid) with SDSS $ugriz'$ (squares) and UKIDSS $YJHK$ photometry (circles). Also shown is the SDSS spectrum (light grey)}
\label{075327}
\end{figure}

\begin{figure}
\begin{center}
\psfig{file=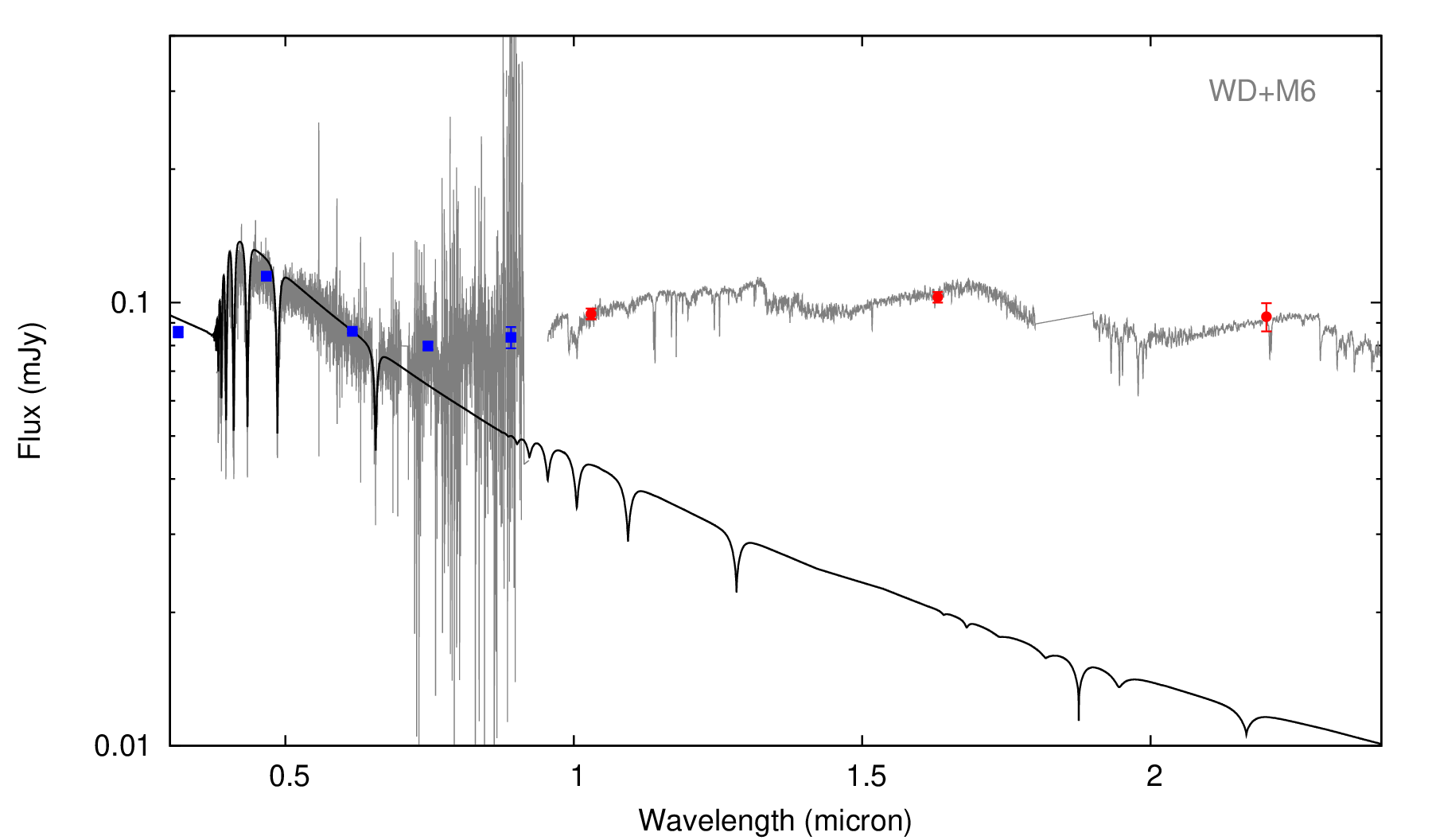,width=5.0cm,angle=-90}
\end{center}
\caption{SDSS\,J085956.47$+$082607.5 model spectrum (solid black) with SDSS $ugriz'$ (squares) and UKIDSS $YHK$ photometry (circles). Also shown are the SDSS spectrum (light grey) and a composite WD$+$M6 spectrum (dark grey).}
\label{085956}
\end{figure}

\clearpage

\begin{figure}
\begin{center}
\psfig{file=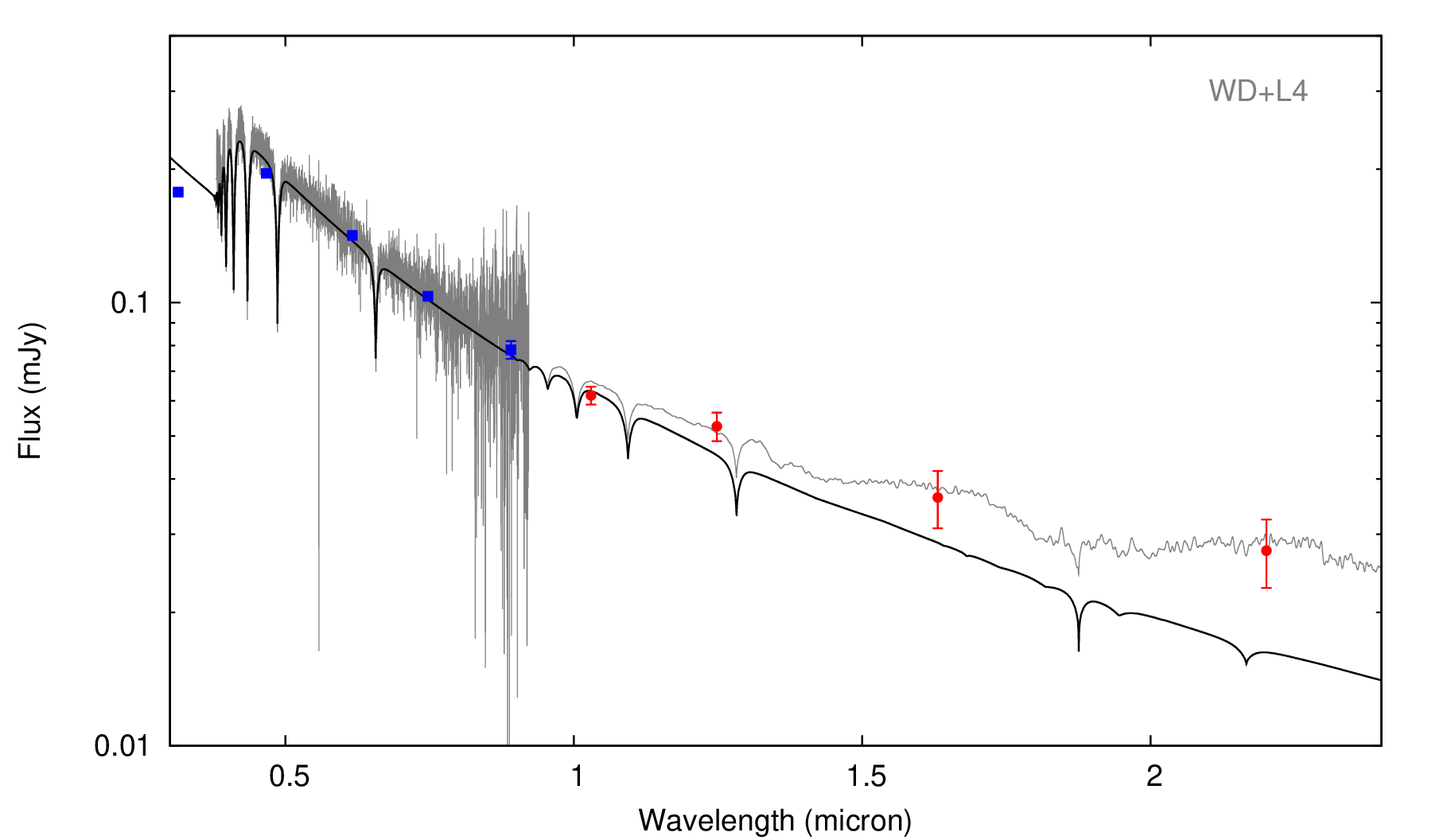,width=5.0cm,angle=-90}
\end{center}
\caption{SDSS\,J090759.59$+$053649.7 model spectrum (solid black) with SDSS $ugriz'$ (squares) and UKIDSS $YJHK$ photometry (circles). Also shown are the SDSS spectrum (light grey) and a composite WD$+$L4 spectrum (dark grey).}
\label{090759}
\end{figure}

\begin{figure}
\begin{center}
\psfig{file=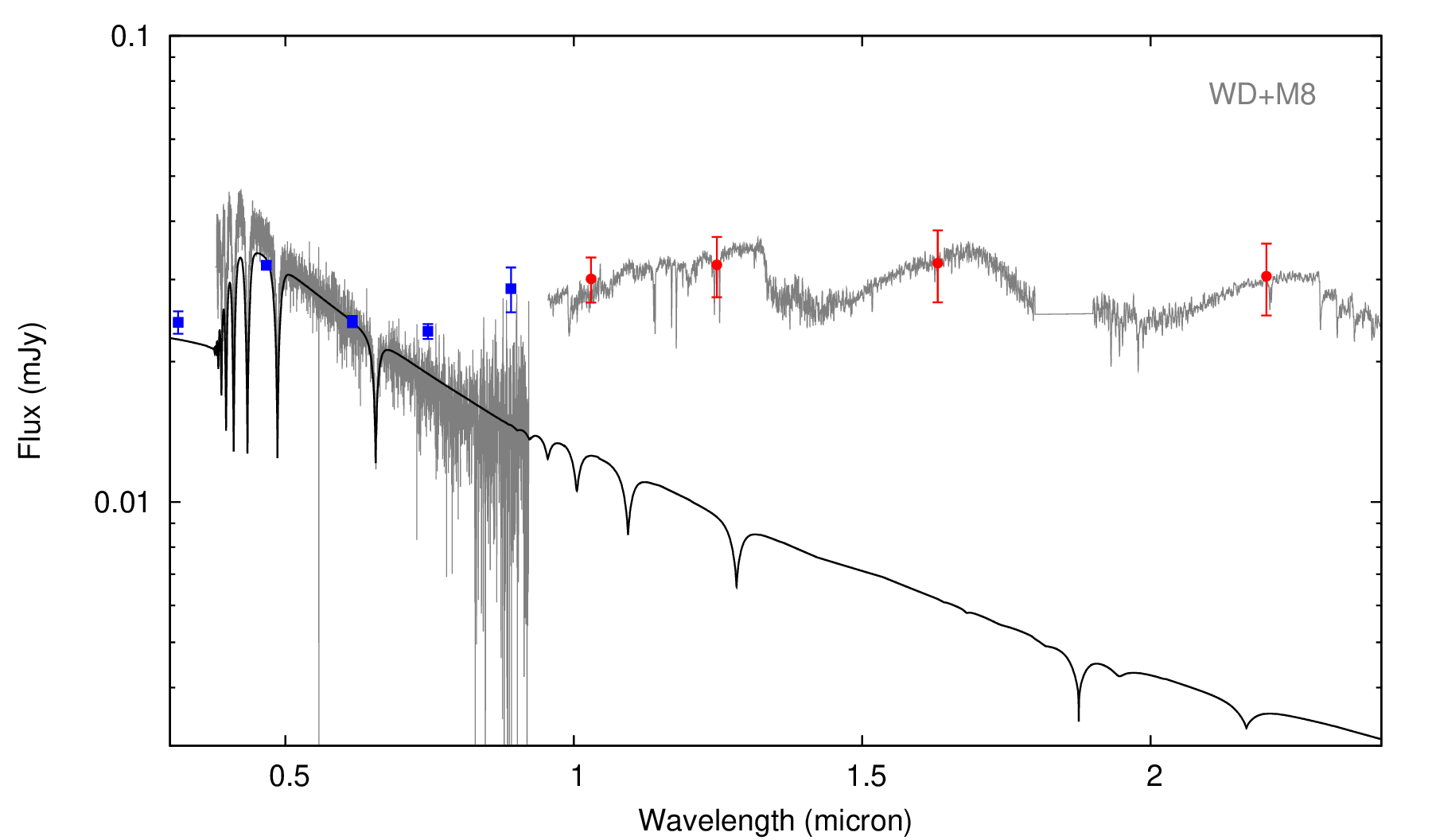,width=5.0cm,angle=-90}
\end{center}
\caption{SDSS\,J092648.48$+$102828.8 model spectrum (solid black) with SDSS $ugriz'$ (squares) and UKIDSS $YJHK$ photometry (circles). Also shown are the SDSS spectrum (light grey) and a composite WD$+$M8 spectrum. (dark grey).}
\label{092648}
\end{figure}

\begin{figure}
\begin{center}
\psfig{file=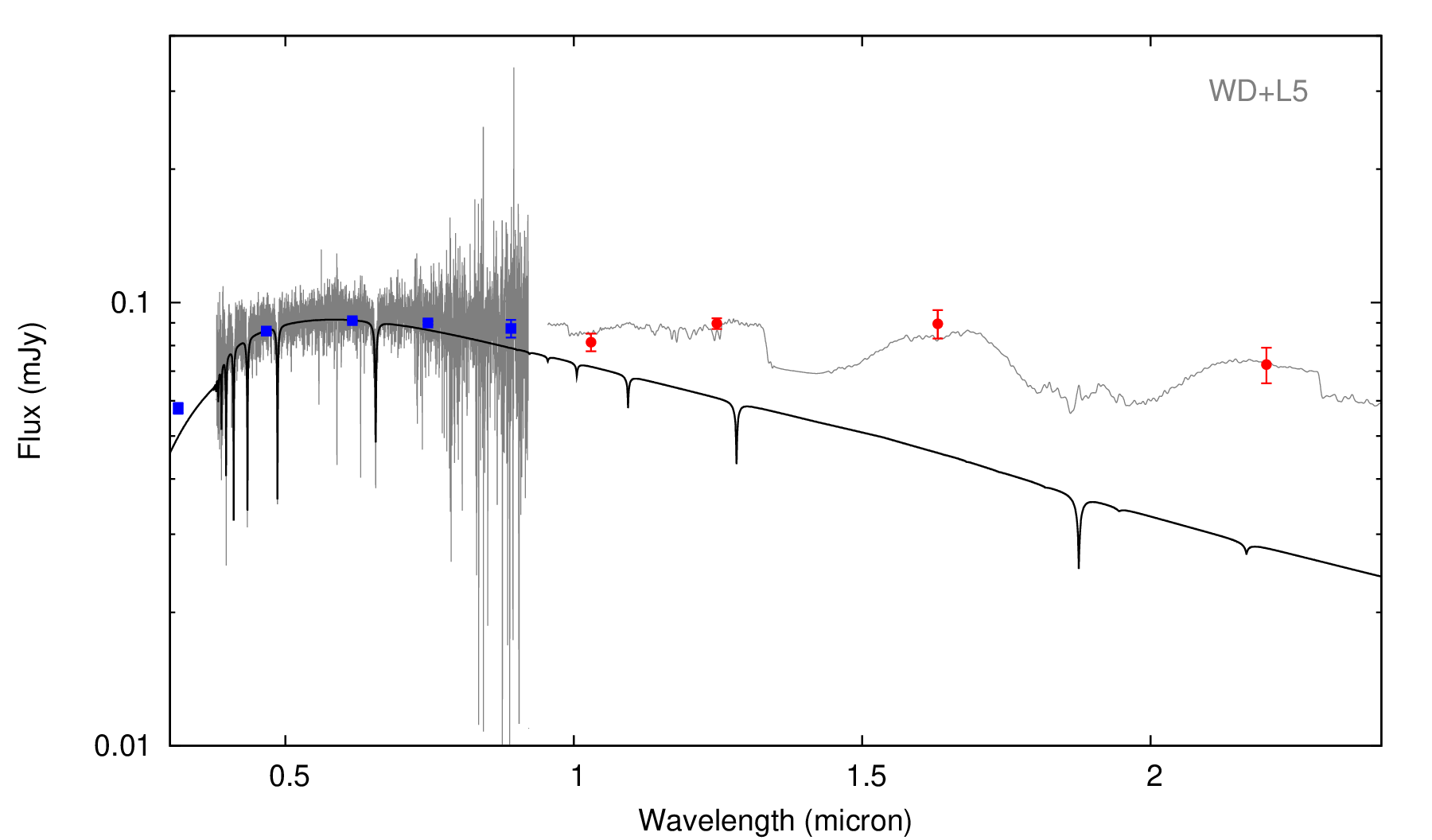,width=5.0cm,angle=-90}
\end{center}
\caption{SDSS\,J093821.34$+$342035.6 model spectrum (solid black) with SDSS $ugriz'$ (squares) and UKIDSS $YJHK$ photometry (circles). Also shown are the SDSS spectrum (light grey) and a composite WD$+$L5 spectrum. (dark grey).}
\label{093821}
\end{figure}

\begin{figure}
\begin{center}
\psfig{file=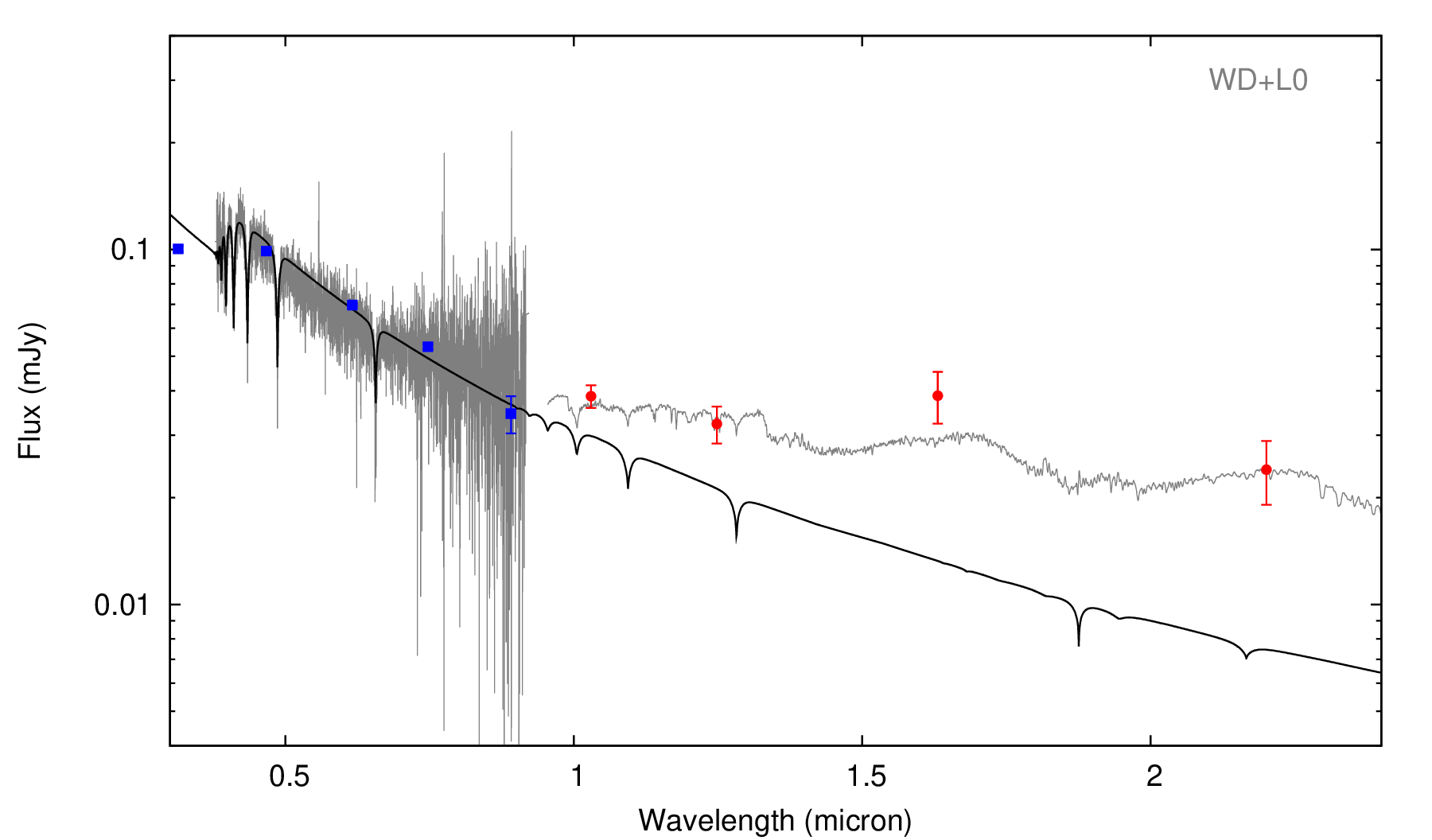,width=5.0cm,angle=-90}
\end{center}
\caption{SDSS\,J100259.88$+$093950.0 model spectrum (solid black) with SDSS $ugriz'$ (squares) and UKIDSS $YJHK$ photometry (circles). Also shown are the SDSS spectrum (light grey) and a composite WD$+$L0 dwarf spectrum (dark grey).}
\label{100259}
\end{figure}

\begin{figure}
\begin{center}
\psfig{file=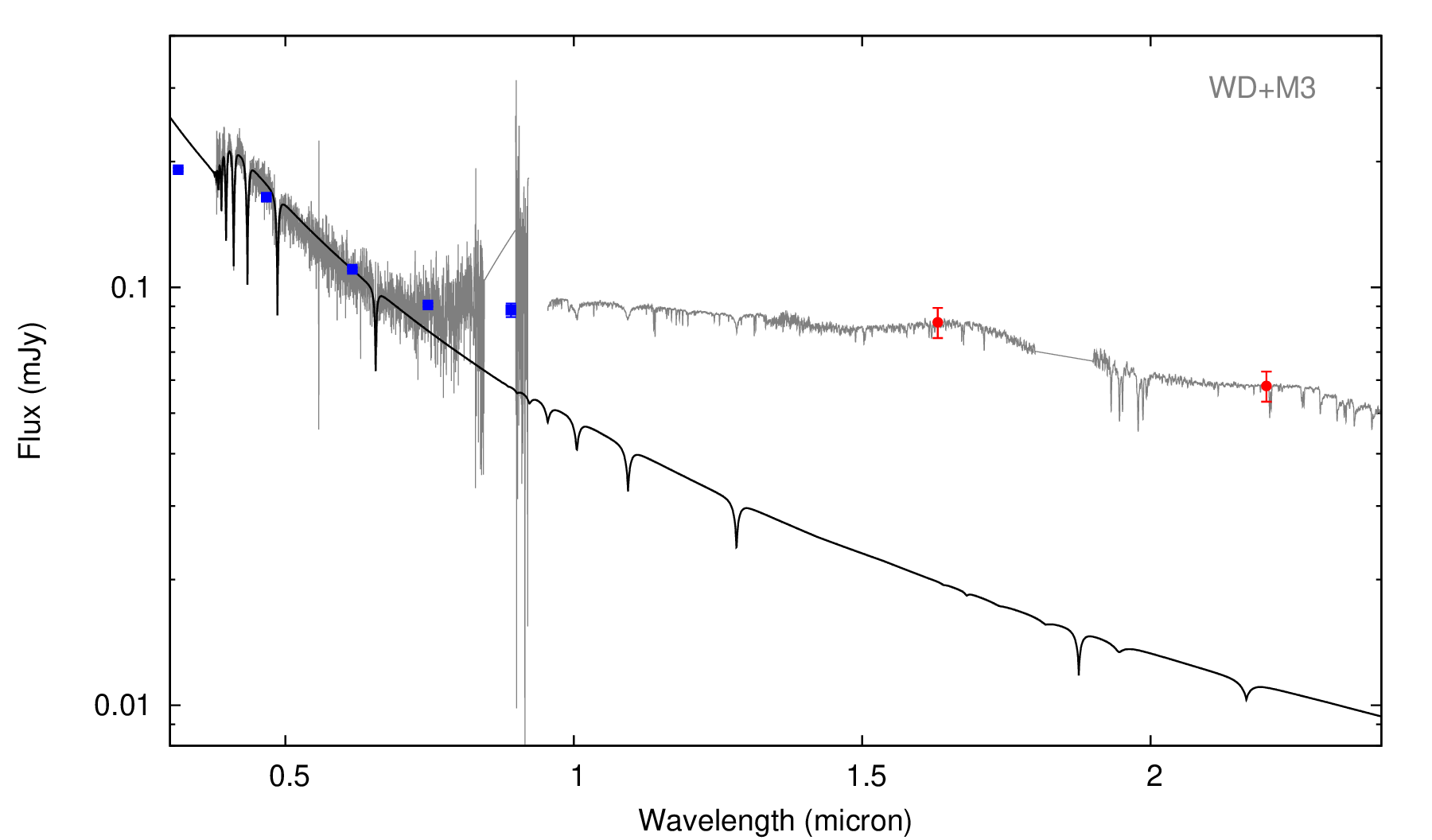,width=5.0cm,angle=-90}
\end{center}
\caption{SDSS\,J101642.93$+$044317.7 model spectrum (solid black) with SDSS $ugriz'$ (squares) and UKIDSS $HK$ photometry (circles). Also shown are the SDSS spectrum (light grey) and a composite WD$+$M3 dwarf spectrum (dark grey).}
\label{101642}
\end{figure}

\begin{figure}
\begin{center}
\psfig{file=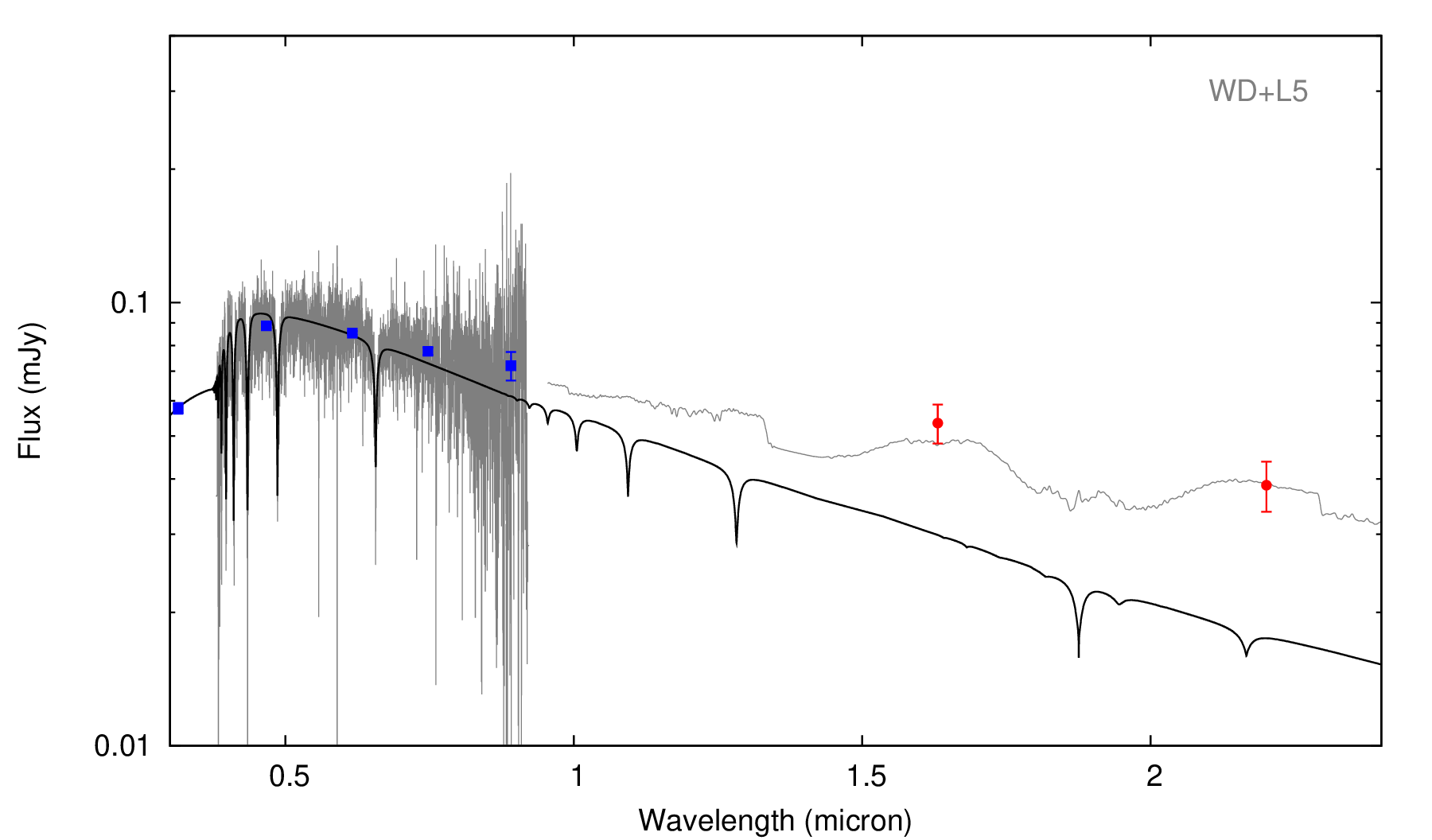,width=5.0cm,angle=-90}
\end{center}
\caption{SDSS\,J103448.92$+$005201.4 model spectrum (solid black) with SDSS $ugriz'$ (squares) and UKIDSS $HK$ photometry (circles). Also shown are the SDSS spectrum (light grey) and a composite WD$+$L5 dwarf spectrum (dark grey).}
\label{103448}
\end{figure}

\clearpage

\begin{figure}
\begin{center}
\psfig{file=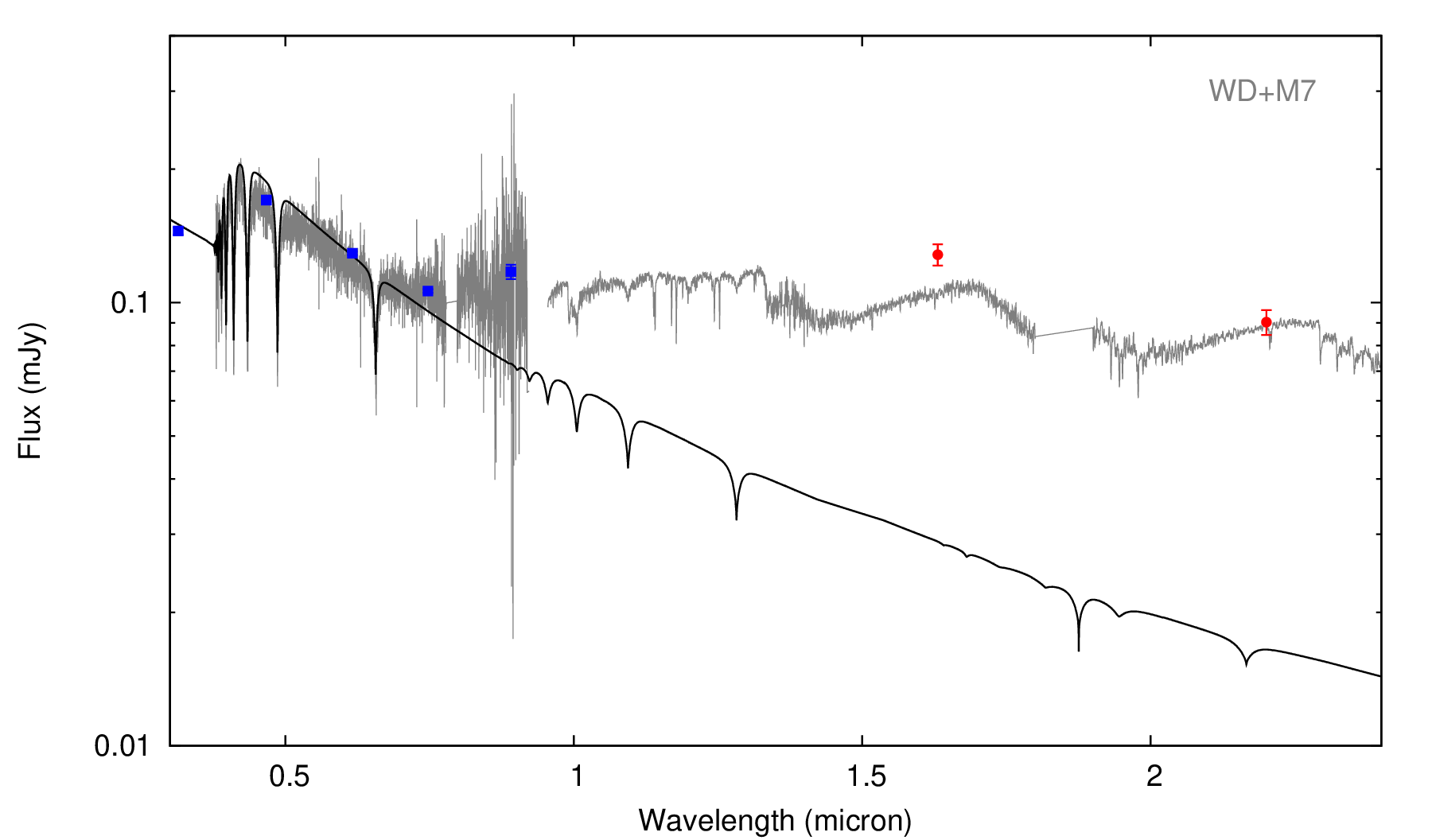,width=5.0cm,angle=-90}
\end{center}
\caption{SDSS\,J103736.75$+$013912.2 model spectrum (solid black) with SDSS $ugriz'$ (squares) and UKIDSS $HK$ photometry (circles). Also shown are the SDSS spectrum (light grey) and a composite WD$+$M7 dwarf spectrum (dark grey).}
\label{103736}
\end{figure}

\begin{figure}
\begin{center}
\psfig{file=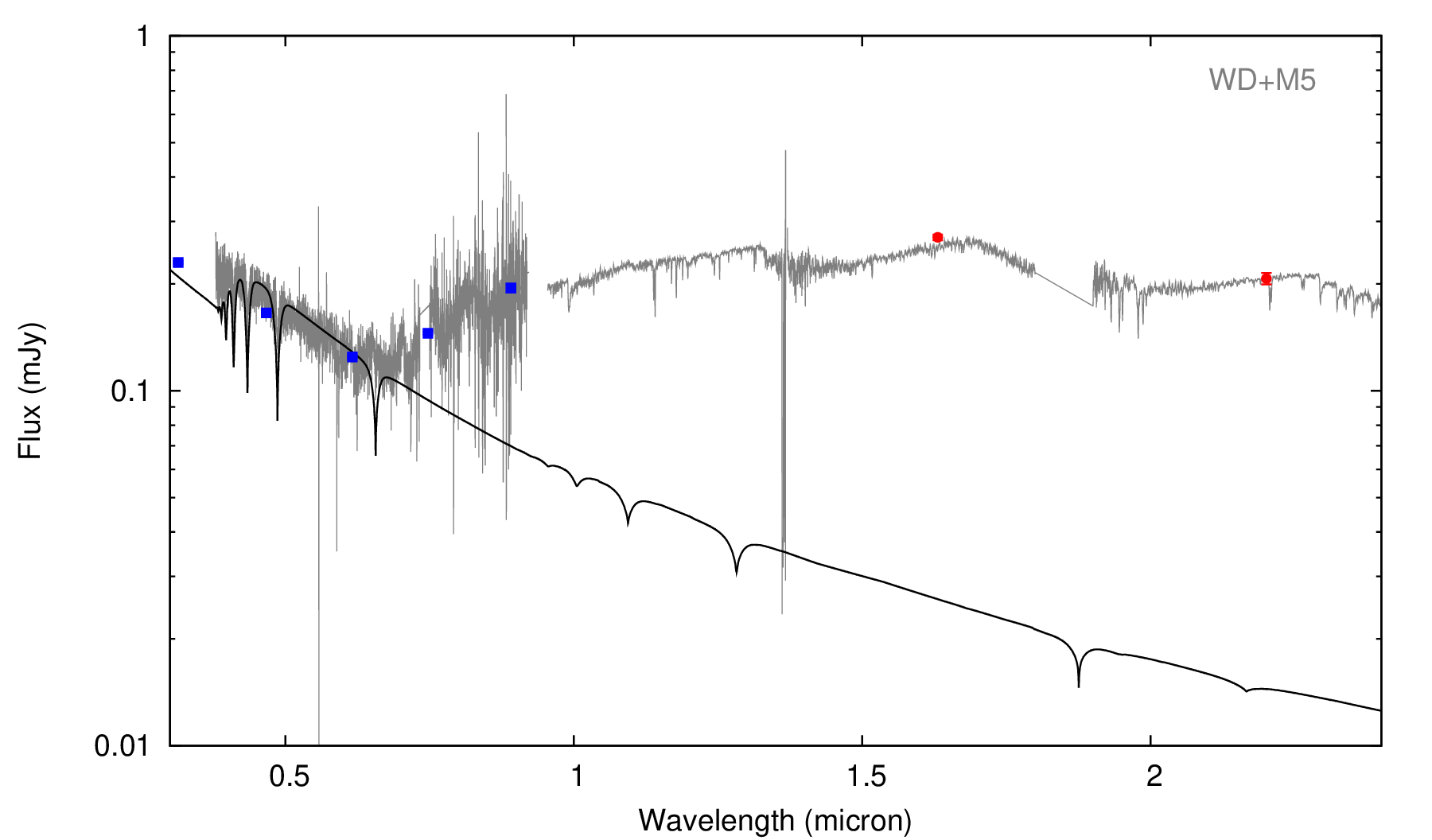,width=5.0cm,angle=-90}
\end{center}
\caption{SDSS\,J110826.47$+$092721.5 model spectrum (solid black) with SDSS $ugriz'$ (squares) and UKIDSS $HK$ photometry (circles). Also shown are the SDSS spectrum (light grey) and a composite WD$+$M5 dwarf spectrum (dark grey).}
\label{110826}
\end{figure}

\begin{figure}
\begin{center}
\psfig{file=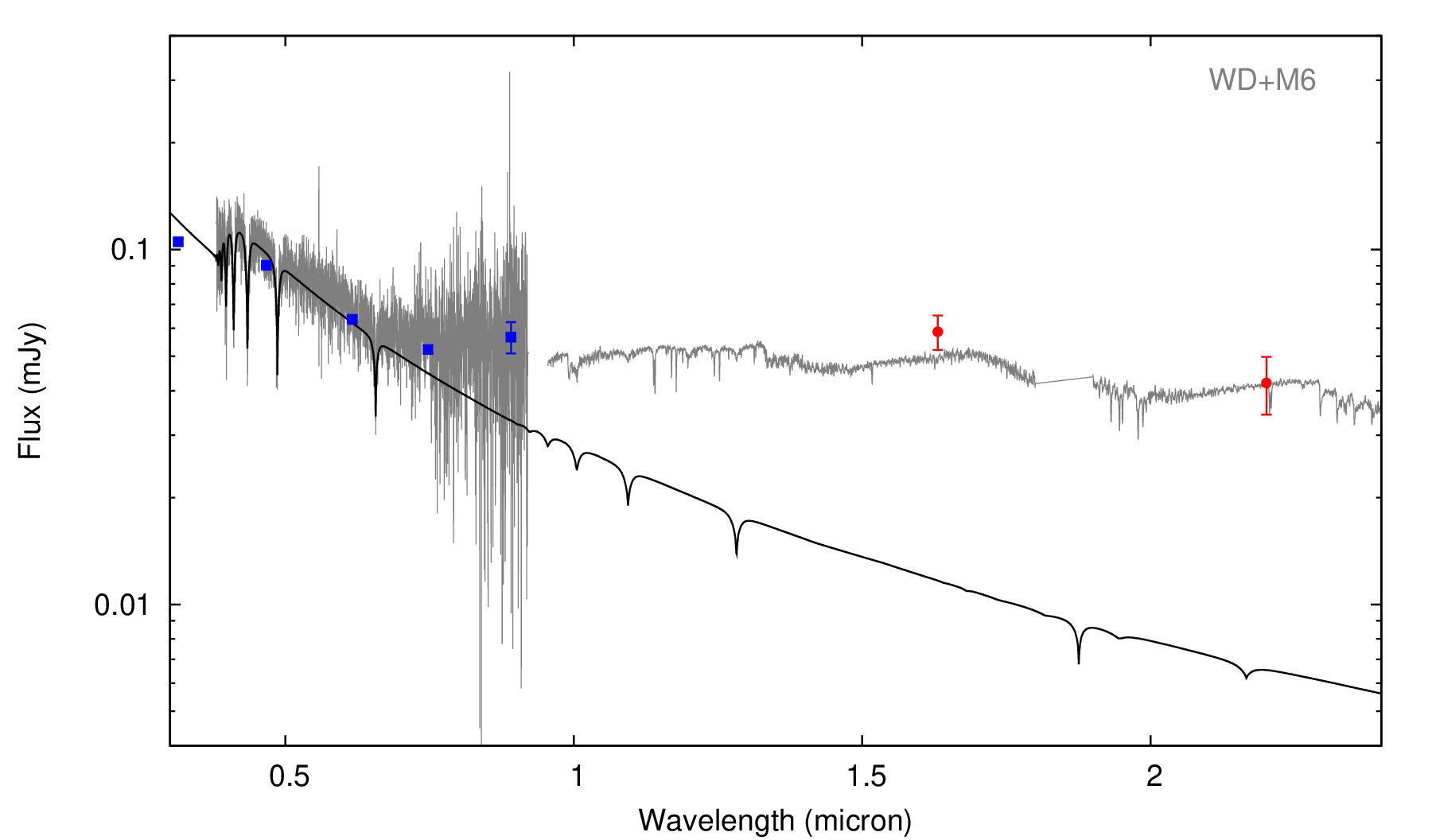,width=5.0cm,angle=-90}
\end{center}
\caption{SDSS\,J113416.09$+$055227.2 model spectrum (solid black) with SDSS $ugriz'$ (squares) and UKIDSS $HK$ photometry (circles). Also shown are the SDSS spectrum (light grey) and a composite WD$+$L3 dwarf spectrum (dark grey).}
\label{113416}
\end{figure}

\begin{figure}
\begin{center}
\psfig{file=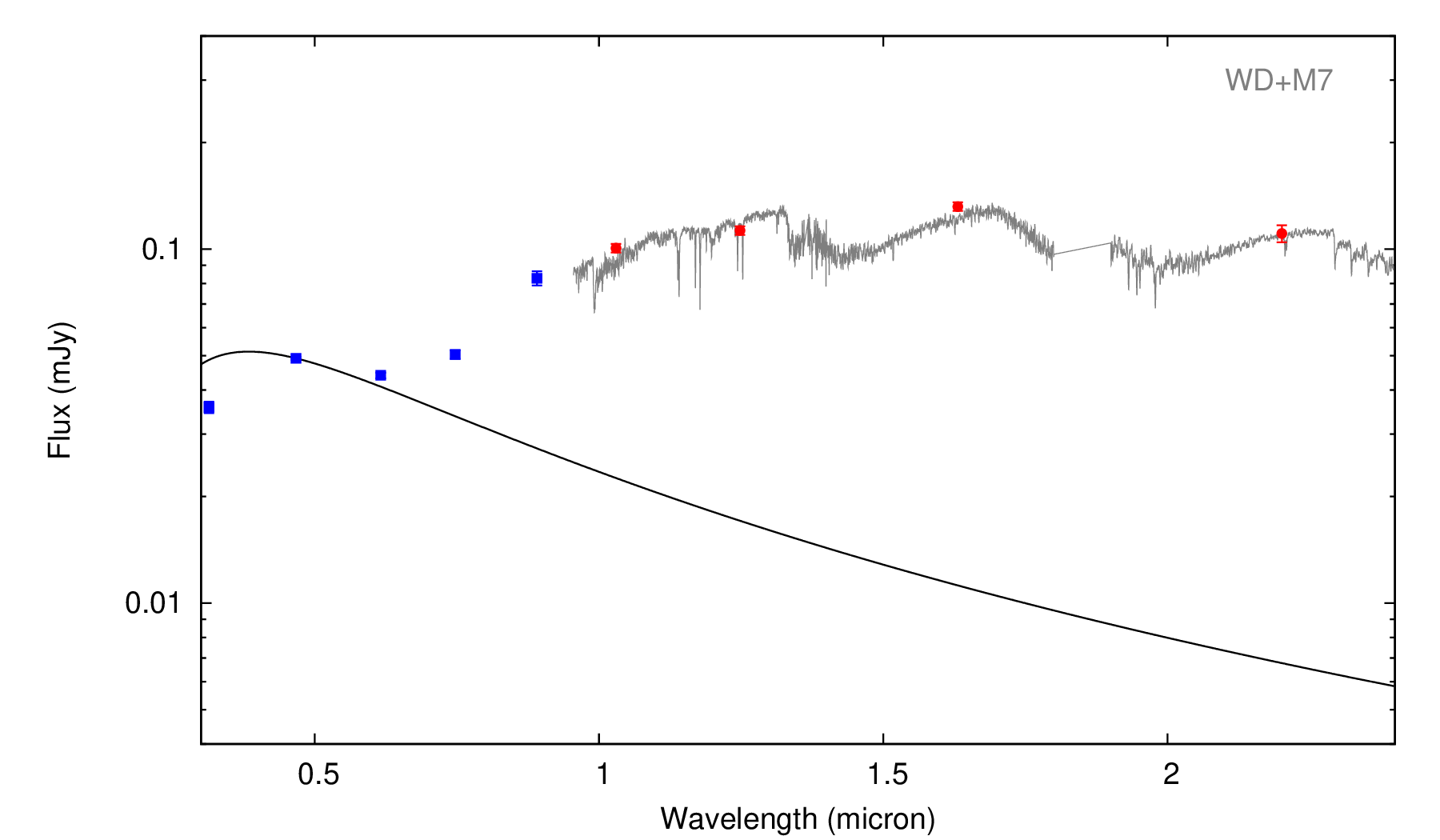,width=5.0cm,angle=-90}
\end{center}
\caption{WD1155$-$011 model spectrum (solid black) with SDSS $ugriz'$ (squares) and UKIDSS $YJHK$ photometry (circles). Also shown is a composite WD$+$M7 dwarf spectrum (dark grey).}
\label{1155}
\end{figure}

\begin{figure}
\begin{center}
\psfig{file=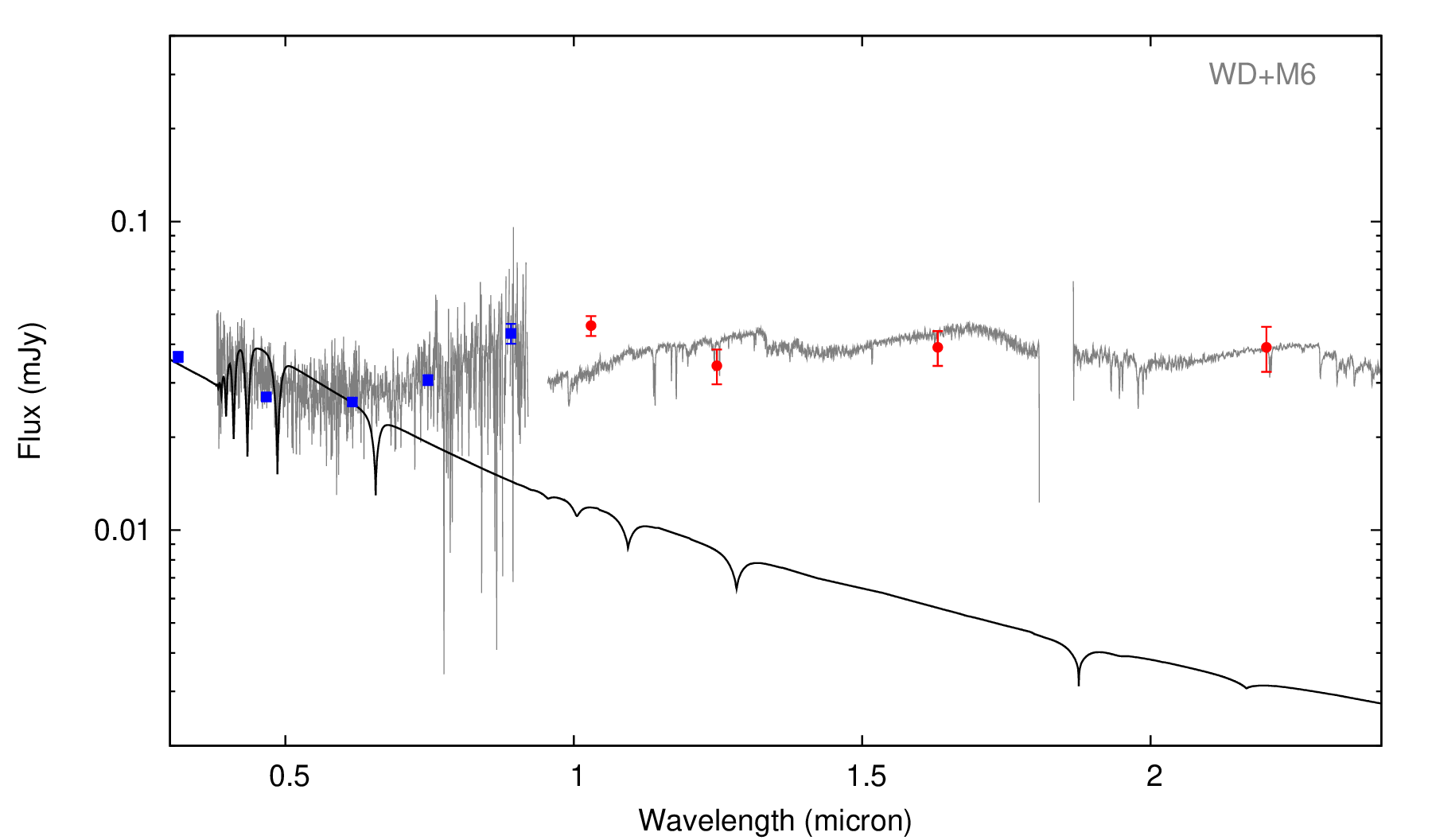,width=5.0cm,angle=-90}
\end{center}
\caption{SDSS\,J122625.93$+$002700.2 model spectrum (solid black) with SDSS $ugriz'$ (squares) and UKIDSS $HK$ photometry (circles). Also shown are the SDSS spectrum (light grey) and a composite WD$+$M6 dwarf spectrum (dark grey).}
\label{122625}
\end{figure}

\begin{figure}
\begin{center}
\psfig{file=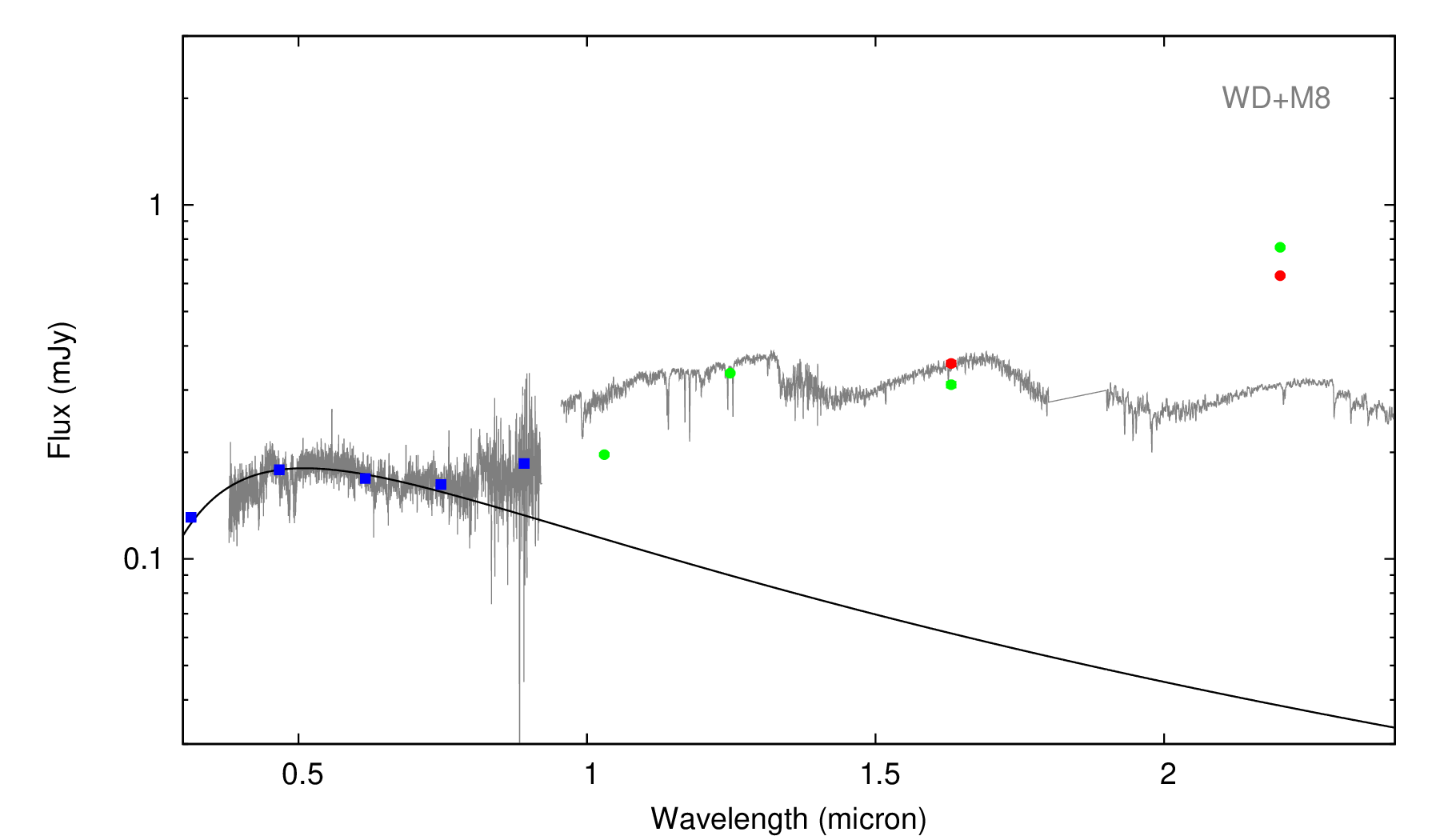,width=5.0cm,angle=-90}
\end{center}
\caption{WD\,1248$+$161 model spectrum (solid black) with SDSS $ugriz'$ (squares) and UKIDSS $JHK$ (circles) photometry. Also shown are the SDSS spectrum (light grey) and a composite WD$+$M8 dwarf spectrum (dark grey). The single $J$-band observation was classified as noise in the DR8 database and should be treated with caution.}
\label{125044}
\end{figure}

\clearpage

\begin{figure}
\begin{center}
\psfig{file=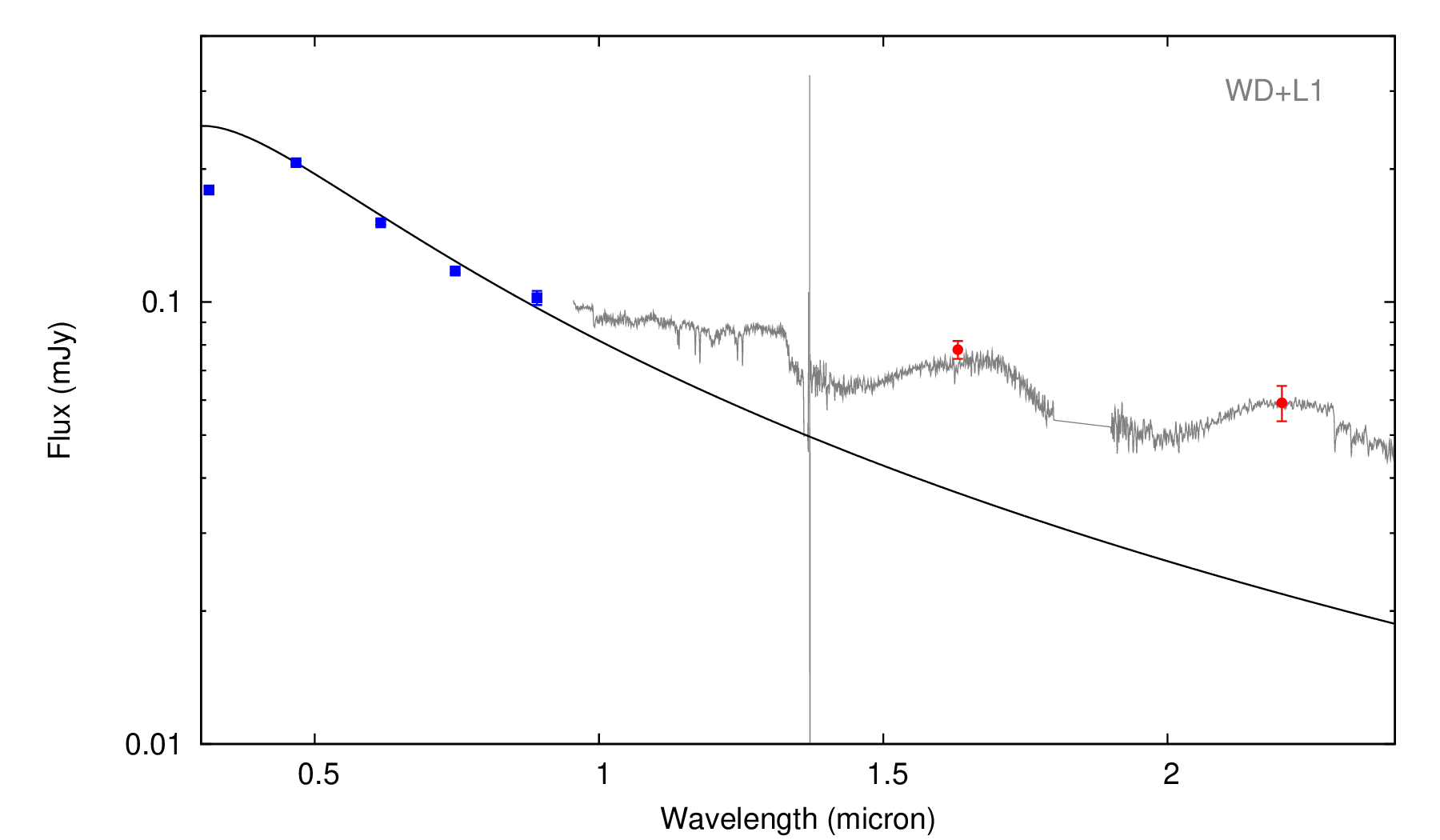,width=5.0cm,angle=-90}
\end{center}
\caption{WD1317$+$021 model spectrum (solid black) with SDSS $ugriz'$ (squares) and UKIDSS $YJHK$ photometry (circles). Also shown is a composite WD$+$L1 dwarf spectrum (dark grey).}
\label{1317}
\end{figure}

\begin{figure}
\begin{center}
\psfig{file=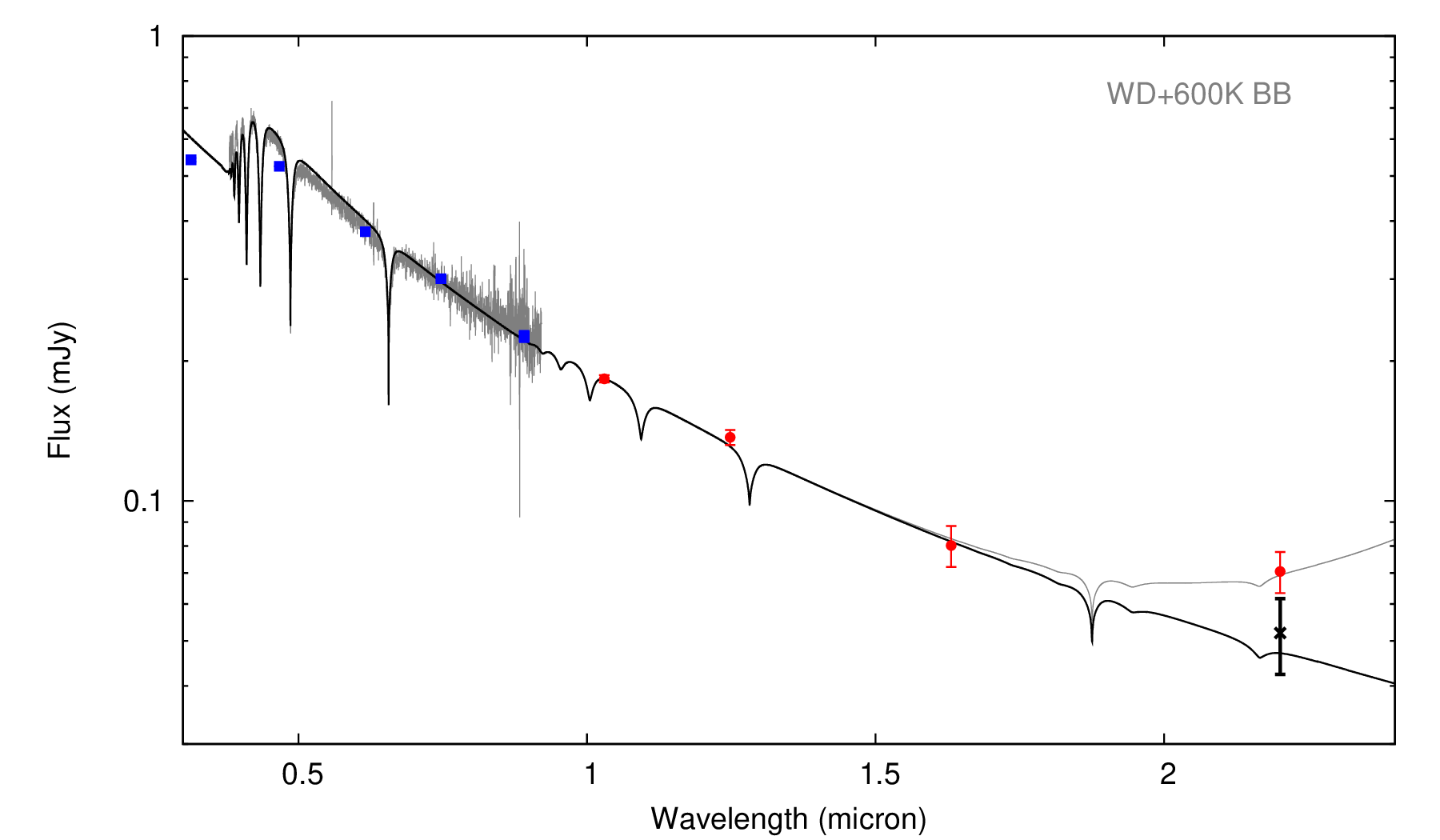,width=5.0cm,angle=-90}
\end{center}
\caption{WD\,1318$+$005 model spectrum (solid black) with SDSS $ugriz'$ (squares), UKIDSS $YJHK$ (circles) and AAO IRIS2 $K$-band photometry (cross). Also shown are the SDSS spectrum (light grey) and a combined WD $+$ 600\,K blackbody model (dark grey).}
\label{132044}
\end{figure}

\begin{figure}
\begin{center}
\psfig{file=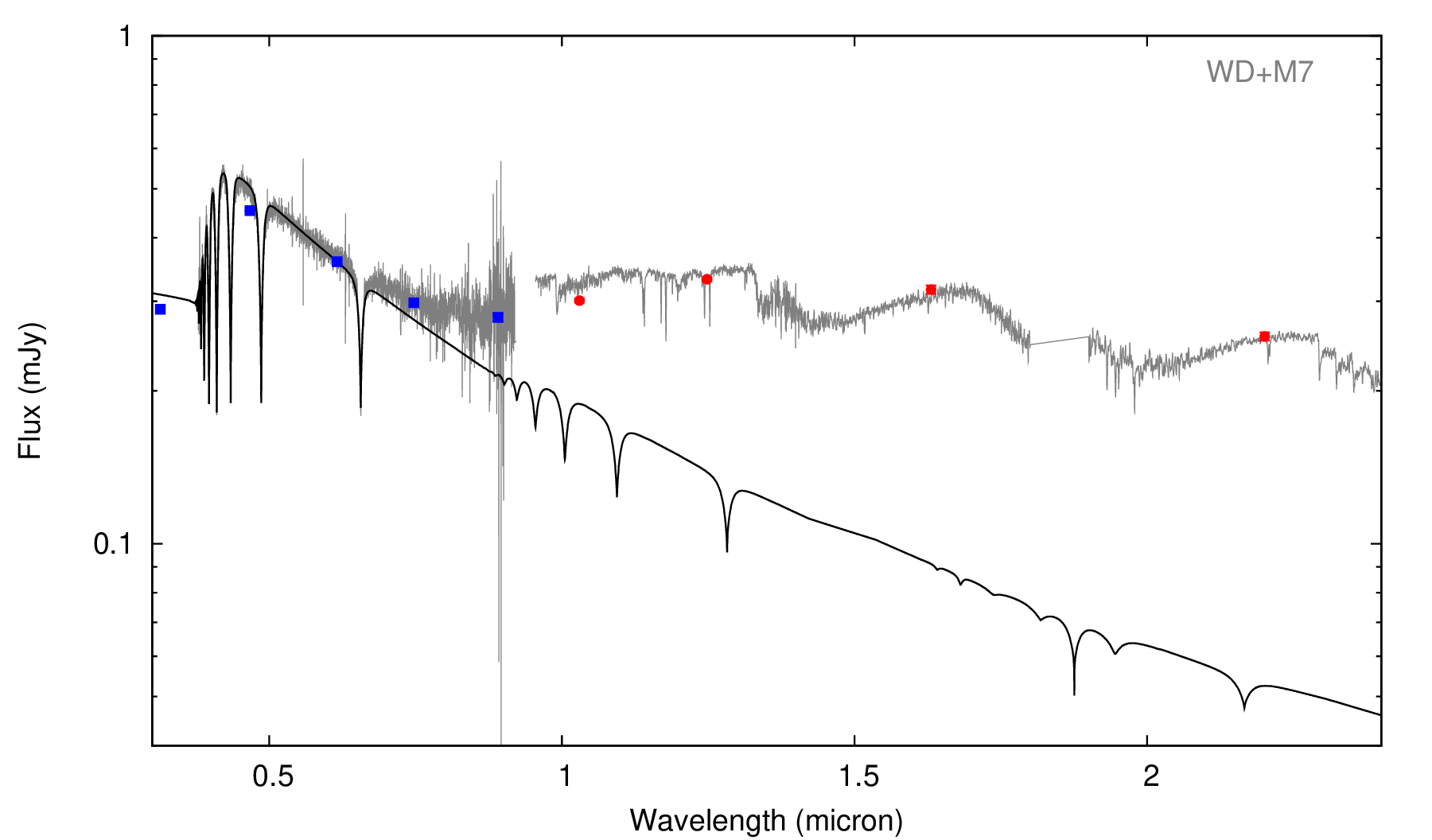,width=5.0cm,angle=-90}
\end{center}
\caption{SDSS\,J132925.21$+$123025.4 model spectrum (solid black) with SDSS $ugriz'$ (squares) and UKIDSS $YJHK$ photometry (circles). Also shown are the SDSS spectrum (light grey) and a composite WD$+$M7 dwarf spectrum (dark grey).}
\label{132925}
\end{figure}

\begin{figure}
\begin{center}
\psfig{file=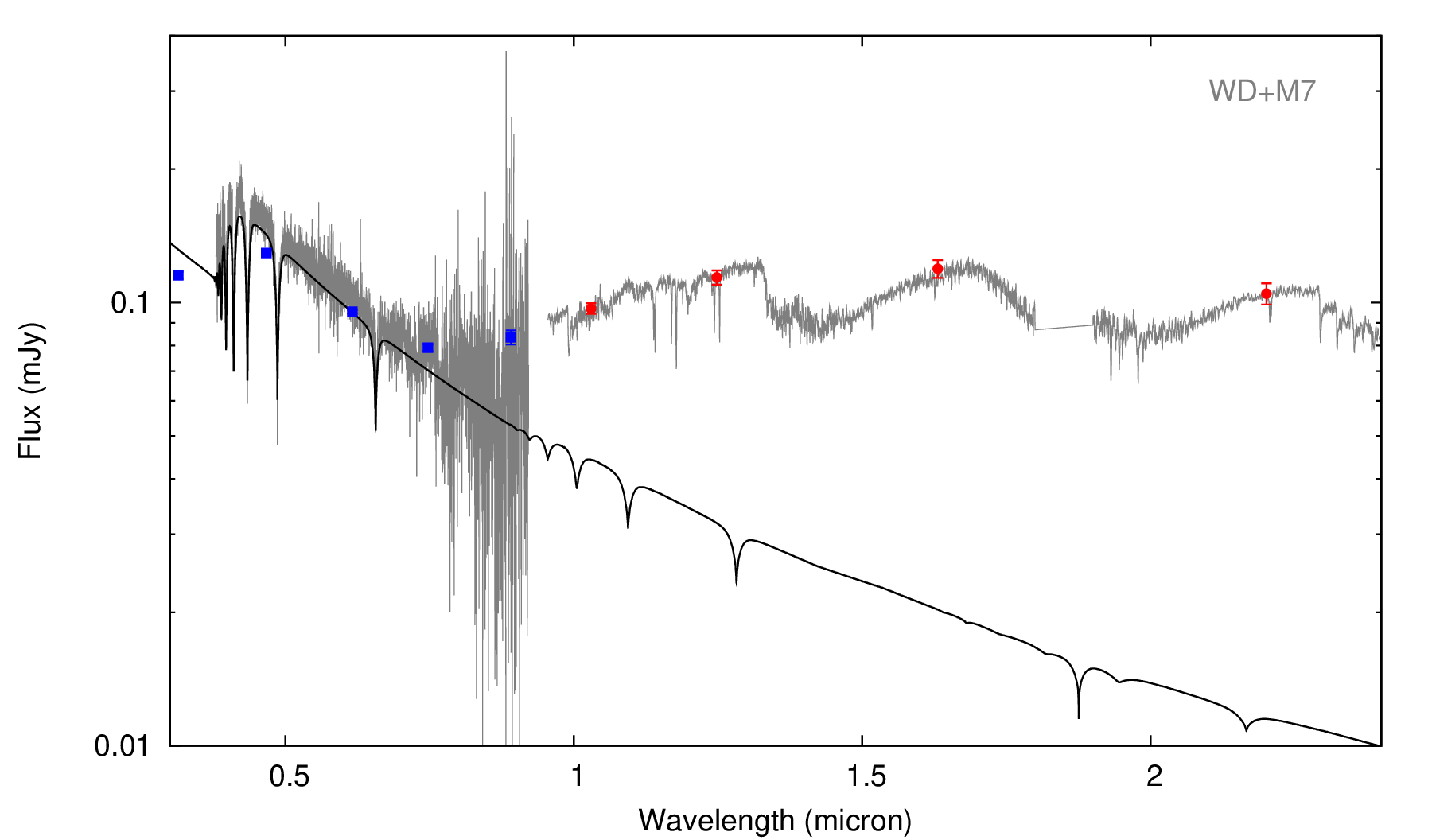,width=5.0cm,angle=-90}
\end{center}
\caption{SDSS\,J134154.29$+$005600.8 model spectrum (solid black) with SDSS $ugriz'$ (squares) and UKIDSS $YJHK$ photometry (circles). Also shown are the SDSS spectrum (light grey) and a composite WD$+$M7 dwarf spectrum (dark grey).}
\label{134154}
\end{figure}

\begin{figure}
\begin{center}
\psfig{file=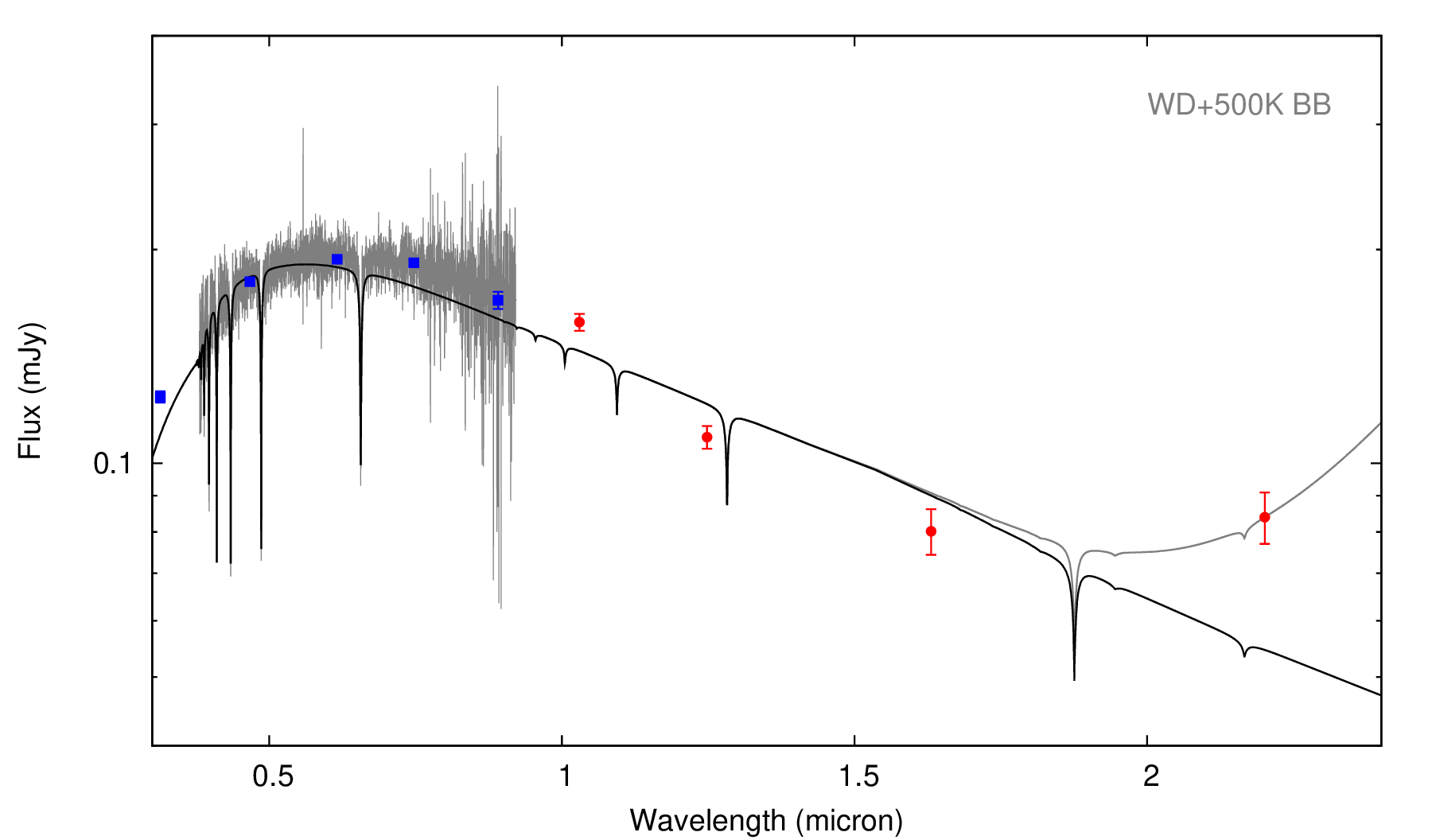,width=5.0cm,angle=-90}
\end{center}
\caption{SDSS\,J141448.25$+$021257.7 model spectrum (solid black) with SDSS $ugriz'$ (squares) and UKIDSS $YJHK$ photometry (circles). Also shown are the SDSS spectrum (light grey) and a combined WD$+$700\,K blackbody model (dashed grey).}
\label{141448}
\end{figure}

\begin{figure}
\begin{center}
\psfig{file=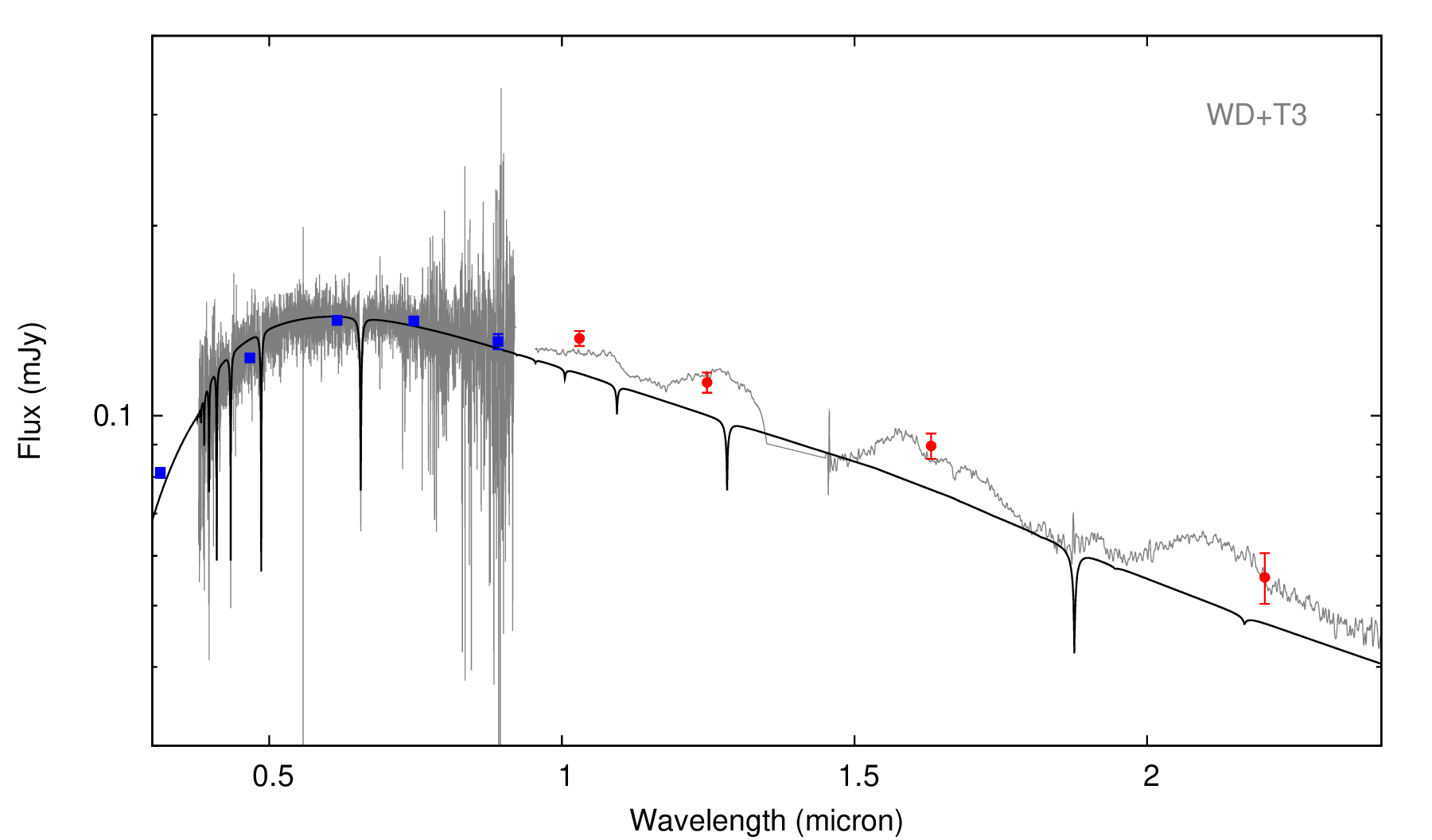,width=5.0cm,angle=-90}
\end{center}
\caption{SDSS\,J154431.47$+$060104.3 model spectrum (solid black) with SDSS $ugriz'$ (squares) and UKIDSS $YJHK$ photometry (circles). Also shown are the SDSS spectrum (light grey) and a composite WD$+$T3 dwarf spectrum (dark grey).}
\label{154431}
\end{figure}

\clearpage


\begin{figure}
\begin{center}
\psfig{file=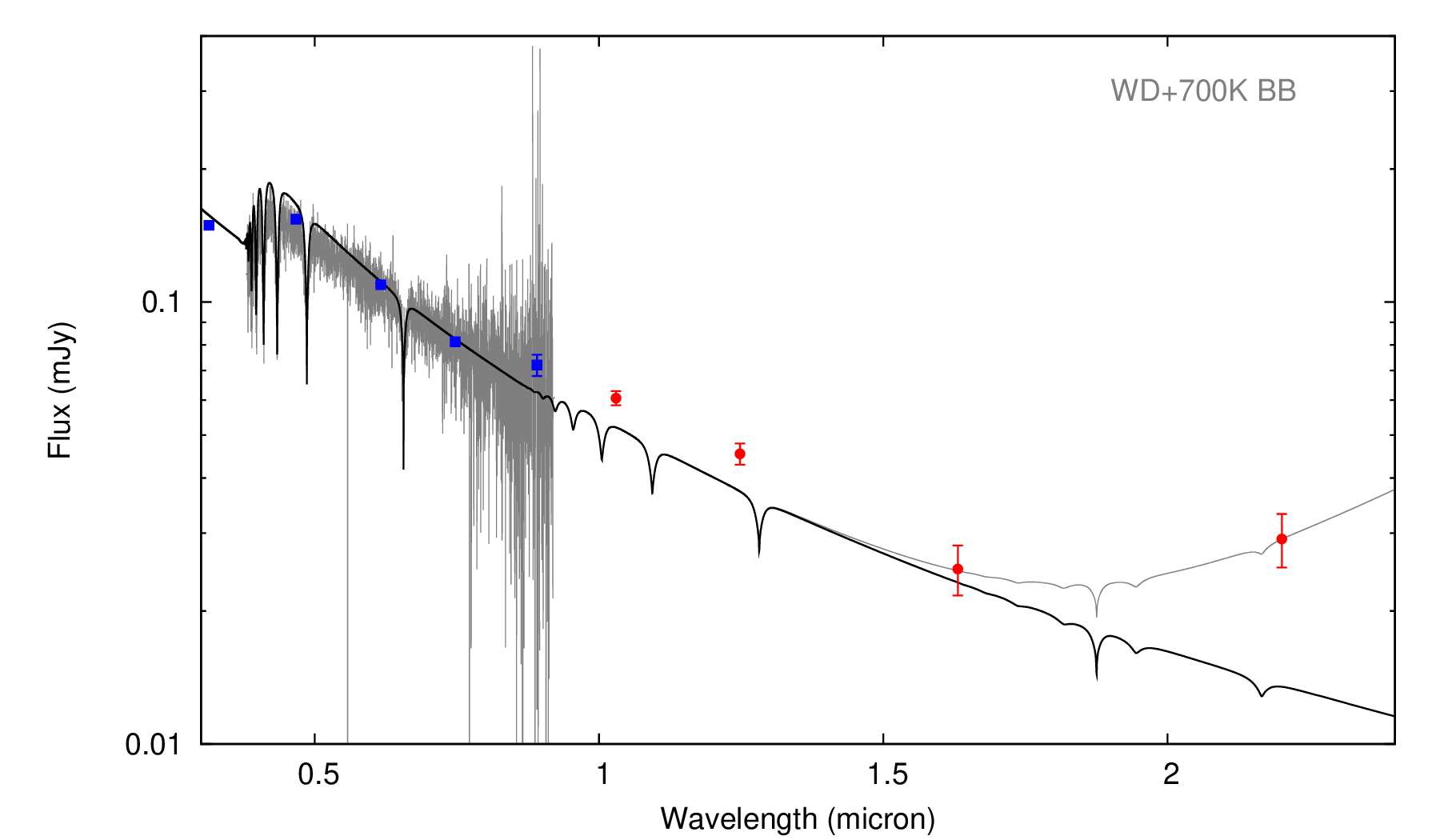,width=5.0cm,angle=-90}
\end{center}
\caption{SDSS\,J155720.77$+$091624.7 model spectrum (solid black) with SDSS $ugriz'$ (squares) and UKIDSS $YJHK$ photometry (circles). Also shown are the SDSS spectrum (light grey),  a combined WD$+$L4 dwarf model (dark grey), and a combined WD$+$700\,K blackbody model (dashed grey).}
\label{155720}
\end{figure}

\begin{figure}
\begin{center}
\psfig{file=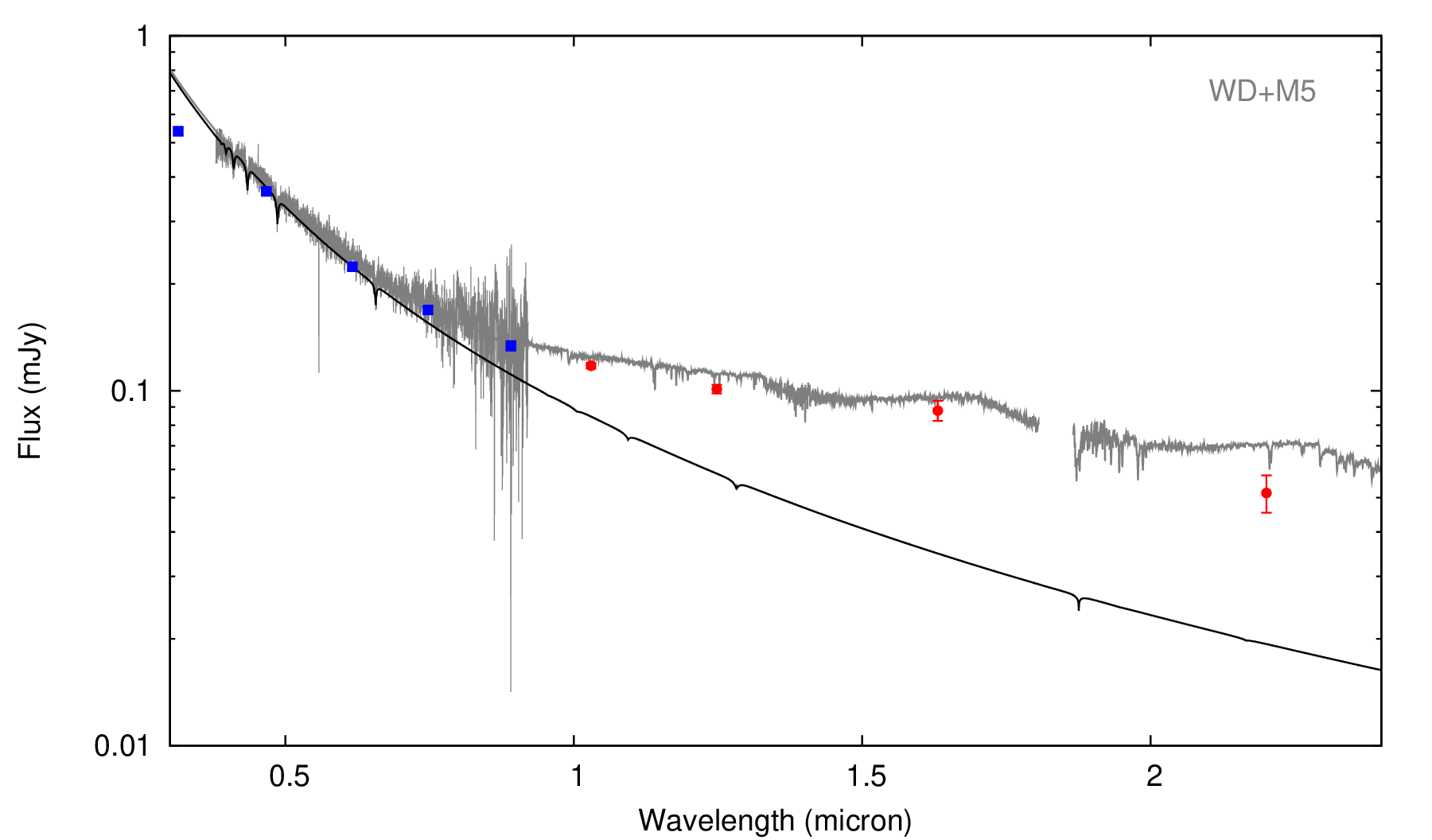,width=5.0cm,angle=-90}
\end{center}
\caption{SDSS\,J162514.88$+$302610.8 model spectrum (solid black) with SDSS $ugriz'$ (squares) and UKIDSS $YJHK$ photometry (circles). Also shown are the SDSS spectrum (light grey) and a composite WD$+$M5 dwarf spectrum (dark grey).}
\label{162514}
\end{figure}

\begin{figure}
\begin{center}
\psfig{file=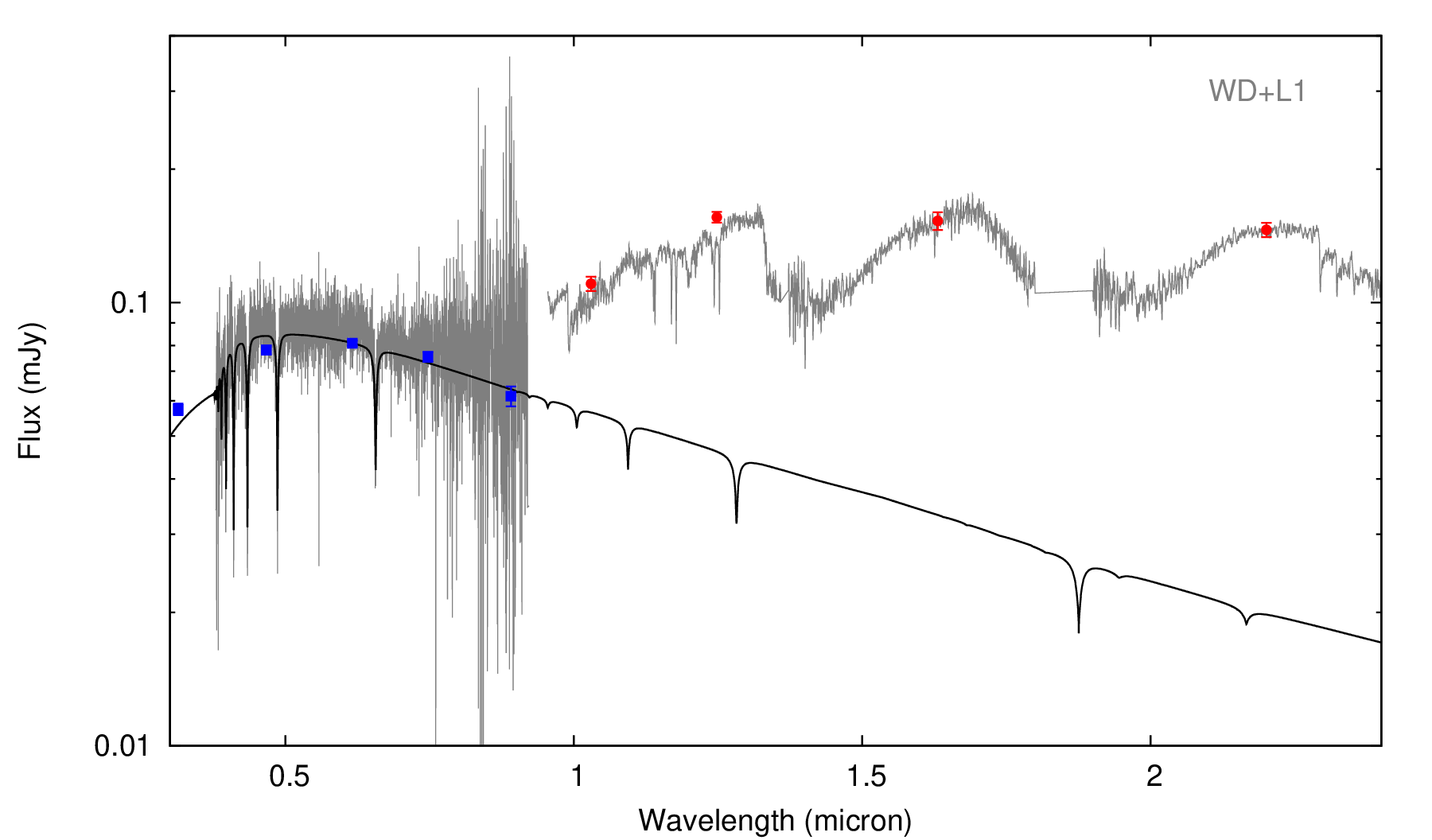,width=5.0cm,angle=-90}
\end{center}
\caption{SDSS\,J220841.63$-$000514.5 model spectrum (solid black) with SDSS $ugriz'$ (squares) and UKIDSS $YJHK$ photometry (circles). Also shown are the SDSS spectrum (light grey) and a composite WD$+$L1 dwarf spectrum (dark grey).}
\label{220841}
\end{figure}

\begin{figure}
\begin{center}
\psfig{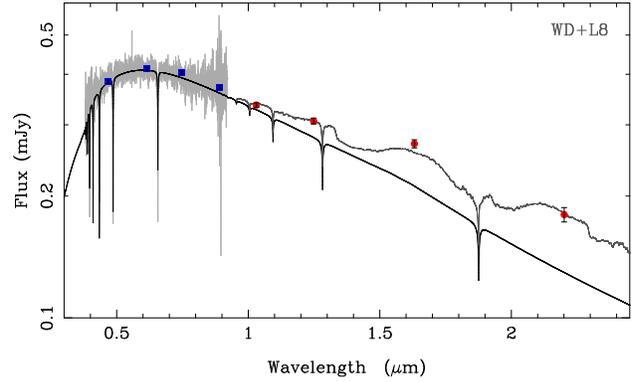}
\caption{PHL\,5038 model spectrum (solid black) with SDSS $ugriz'$ (squares) and UKIDSS $YJHK$ photometry (circles). Also shown are the SDSS spectrum (light grey) and a composite WD$+$L8 dwarf spectrum (dark grey).}
\end{center}
\label{phl5038_ukidss}
\end{figure}

\begin{figure}
\begin{center}
\psfig{file=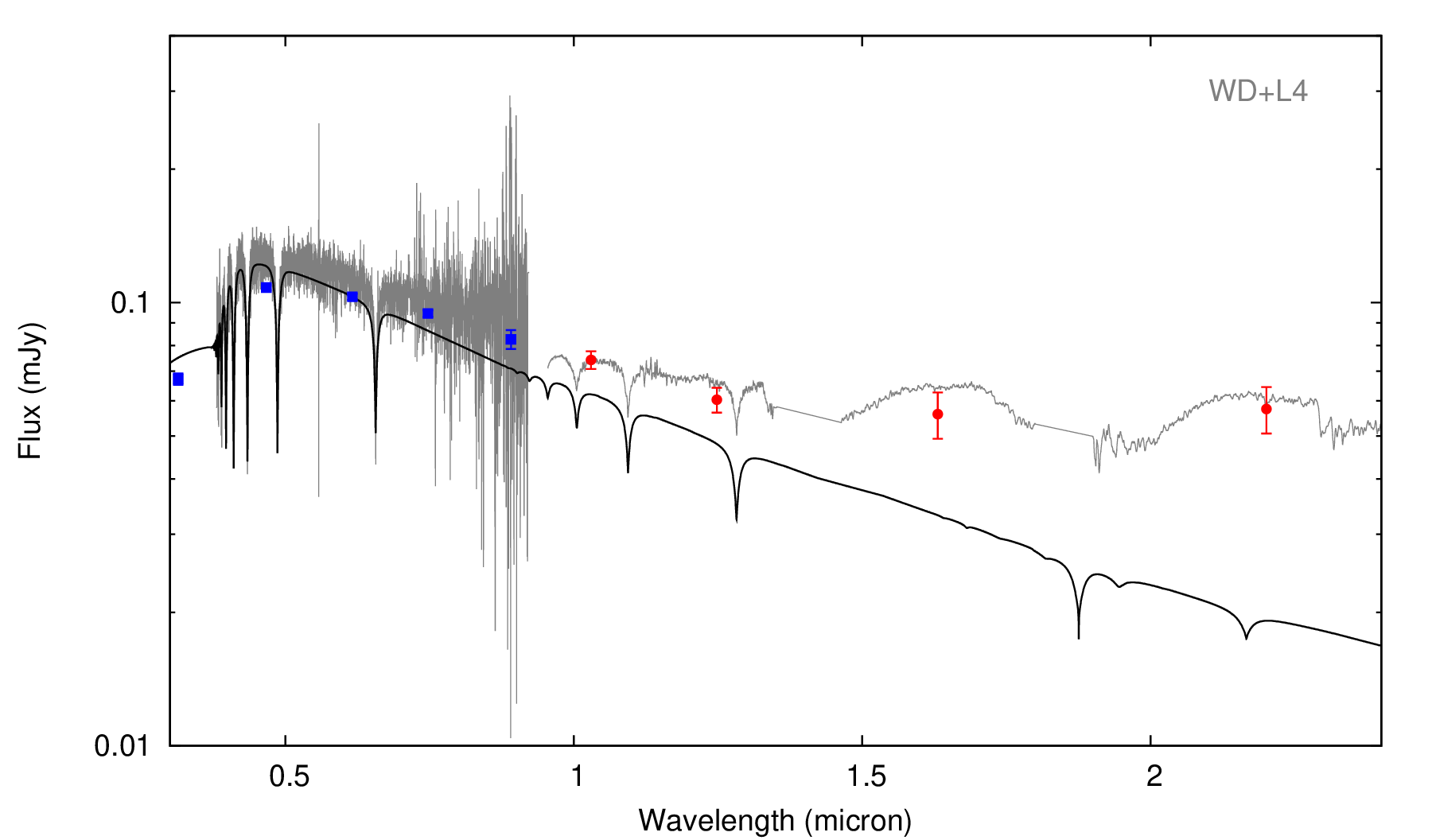,width=5.0cm,angle=-90}
\end{center}
\caption{SDSS\,J222551.65$+$001637.7 model spectrum (solid black) with SDSS $ugriz'$ (squares) and UKIDSS $YJHK$ photometry (circles). Also shown are the SDSS spectrum (light grey) and a composite WD$+$L7.5 dwarf spectrum (dark grey).}
\label{222551}
\end{figure}

\begin{figure}
\begin{center}
\psfig{file=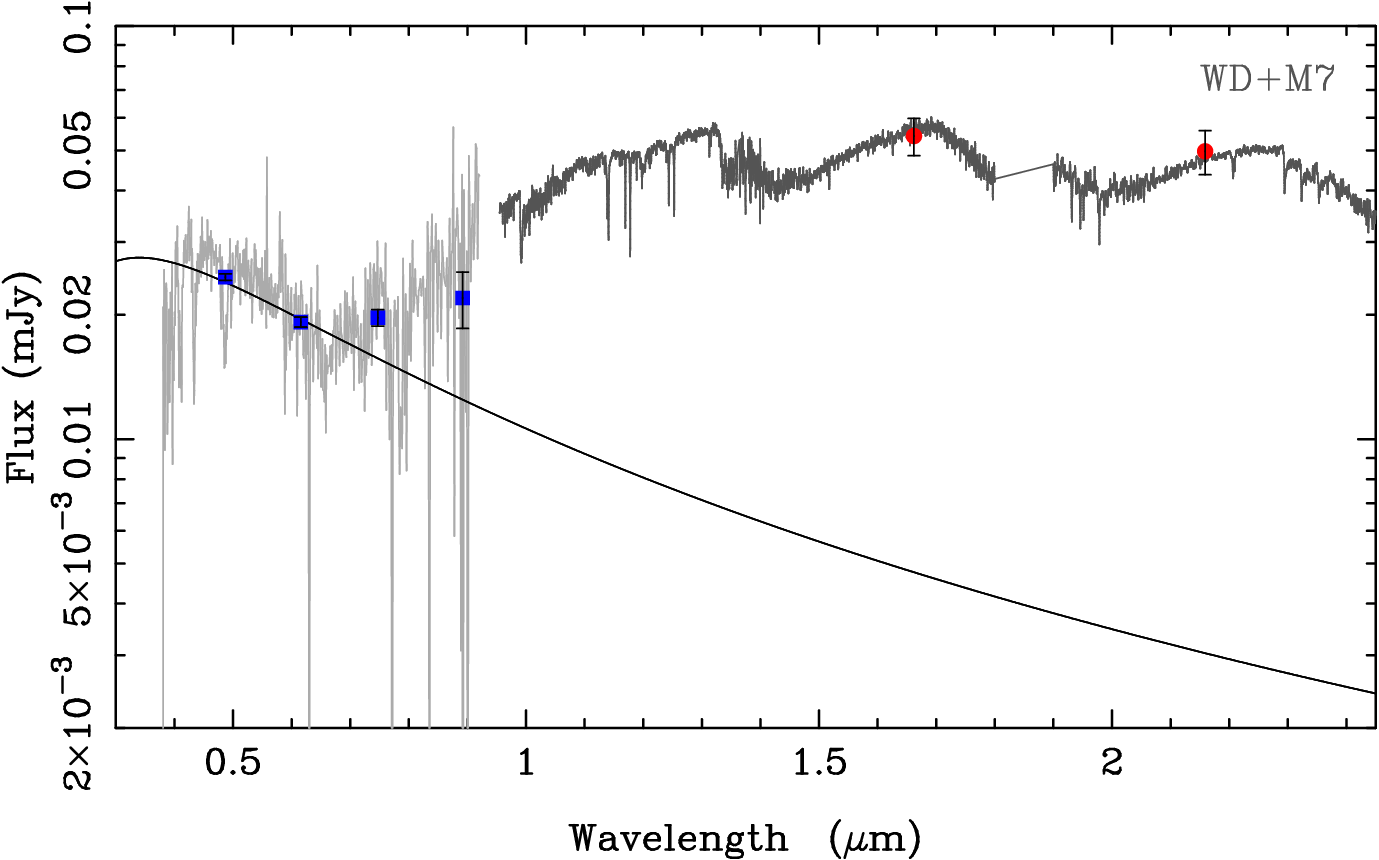,width=5.0cm,angle=-90}
\end{center}
\caption{SDSS\,J233345.97$-$000843.0 model spectrum (solid black) with SDSS $ugriz'$ (squares) and UKIDSS $HK$ photometry (circles). Also shown are the SDSS spectrum (light grey) and a composite WD$+$M7 dwarf spectrum (dark grey).}
\label{233345}
\end{figure}

\end{document}